\documentclass[twocolumn]{aastex631}
\usepackage{graphicx}
\pdfoutput=1 
\usepackage{amssymb}
\usepackage{gensymb}
\usepackage[caption=false]{subfig}
\usepackage{booktabs}
\usepackage{xcolor}
\usepackage{multirow}

\newcommand{\Teff}{$\rm{T_{eff}}$}
\newcommand{\Tspot}{$\rm{T_{spot}}$}
\newcommand{\Thot}{$\rm{T_{hot}}$}
\newcommand{\Tamb}{$\rm{T_{amb}}$}
\newcommand{\fspot}{$f_{\rm spot}$}
\newcommand{\fhot}{$f_{\rm hot}$}
\newcommand{\deltafspot}{$\Delta f_{\rm spot}$}

\newcommand{\Fallgp}{0.075$\pm$0.006}
\newcommand{\Fallrp}{0.071$\pm$0.006}
\newcommand{\Fallip}{0.041$\pm$0.007}
\newcommand{\Springgp}{0.075$\pm$0.003}
\newcommand{\Springrp}{0.075$\pm$0.003}

\newcommand{\Photfspot}{$0.57^{+0.22}_{-0.28}$}
\newcommand{\Photfhot}{$<0.40$}
\newcommand{\Photdeltafspot}{$0.06\pm0.03$}
\newcommand{\PhotTspot}{$3141^{+266}_{-389}$}
\newcommand{\PhotThot}{$4873^{+1922}_{-727}$}
\newcommand{\PhotTeff}{$3664\pm101$}
\newcommand{\PhotTamb}{$3719^{+302}_{-502}$}
\newcommand{\Photchisq}{$2.59~(2.59)$}

\newcommand{\Specfspot}{$0.39\pm0.04$}
\newcommand{\Specfhot}{$<0.01$}
\newcommand{\Specdeltafspot}{$<0.40$}
\newcommand{\SpecTspot}{$2974^{+72}_{-71}$}
\newcommand{\SpecThot}{$8681^{+868}_{-629}$}
\newcommand{\SpecTamb}{$4002^{+14}_{-15}$}
\newcommand{\SpecTeff}{$3783\pm37$}
\newcommand{\Specchisq}{$2124~(0.581)$}

\newcommand{\ensemblefspot}{$0.39\pm0.04$}
\newcommand{\ensemblefhot}{$<0.005$}
\newcommand{\ensembledeltafspot}{$0.05^{+0.004}_{-0.003}$}
\newcommand{\ensembleTspot}{$3003^{+63}_{-71}$}
\newcommand{\ensembleThot}{$8671^{+890}_{-629}$}
\newcommand{\ensembleTamb}{$4003^{+15}_{-14}$}
\newcommand{\ensembleTeff}{$3789\pm35$}
\newcommand{\ensemblechisq}{$2136~(0.584)$}

\newcommand{\PhotfspottwoT}{$0.61^{+0.22}_{-0.27}$}
\newcommand{\PhotdeltafspottwoT}{$0.08^{+0.06}_{-0.04}$}
\newcommand{\PhotTspottwoT}{$3454^{+155}_{-208}$}
\newcommand{\PhotTefftwoT}{$3649\pm98$}
\newcommand{\PhotTambtwoT}{$3876^{+234}_{-157}$}
\newcommand{\PhotchisqtwoT}{$1.9~(1.9)$}

\newcommand{\SpecfspottwoT}{$0.41\pm0.03$}
\newcommand{\SpecdeltafspottwoT}{$<0.4$}
\newcommand{\SpecTspottwoT}{$3083^{+31}_{-45}$}
\newcommand{\SpecTambtwoT}{$3998^{9}_{11}$}
\newcommand{\SpecTefftwoT}{$3704\pm24$}
\newcommand{\SpecchisqtwoT}{$3060~(0.837)$}

\newcommand{\ensemblefspottwoT}{$0.41\pm0.03$}
\newcommand{\ensembledeltafspottwoT}{$0.05\pm0.003$}
\newcommand{\ensembleTspottwoT}{$3093^{+29}_{-41}$}
\newcommand{\ensembleTambtwoT}{$3998^{+10}_{-12}$}
\newcommand{\ensembleTefftwoT}{$3707\pm24$}
\newcommand{\ensemblechisqtwoT}{$3058~(0.836)$}

\begin{document}

\title{Quantifying the Transit Light Source Effect: Measurements of Spot Temperature and Coverage on the Photosphere of AU Microscopii with High-Resolution Spectroscopy and Multi-Color Photometry}

% william.waalkes@colorado.edu
\author[0000-0002-8961-0352]{William~C.~Waalkes}
\altaffiliation{NSF Graduate Research Fellow}
\affiliation{Department of Astrophysical \& Planetary Sciences, University of Colorado Boulder, 2000 Colorado Ave, Boulder, CO 80309, USA}

% zachory.bertathompson@colorado.edu
\author[0000-0002-3321-4924]{Zachory~K.~Berta-Thompson}
\affiliation{Department of Astrophysical \& Planetary Sciences, University of Colorado Boulder, 2000 Colorado Ave, Boulder, CO 80309, USA}

% Elisabeth.R.Newton@dartmouth.edu
\author[0000-0003-4150-841X]{Elisabeth Newton}
\affiliation{Department of Physics and Astronomy, Dartmouth College, Hanover NH 03755, USA}

% awmann@unc.edu
\author[0000-0003-3654-1602]{Andrew W. Mann}
\affiliation{Department of Physics and Astronomy, The University of North Carolina at Chapel Hill, Chapel Hill, NC 27599-3255, USA}

% pgao@carnegiescience.edu
\author[0000-0002-8518-9601]{Peter Gao}
\affiliation{Earth and Planets Laboratory, Carnegie Institute of Washington, Washington DC, USA}

% hannah.wakeford@bristol.ac.uk
\author[0000-0003-4328-3867]{Hannah R. Wakeford}
\affiliation{School of Physics, University of Bristol, Bristol, UK}

% lili.alderson@bristol.ac.uk
\author[0000-0001-8703-7751]{Lili Alderson}
\affiliation{School of Physics, University of Bristol, Bristol, UK}

% pplavcha@gmu.edu
\author[0000-0002-8864-1667]{Peter Plavchan}
\affiliation{Department of Physics and Astronomy, George Mason University, Fairfax, VA 22030, USA}

\begin{abstract}
AU Mic is an active 24$\pm$3 Myr pre-main sequence M dwarf in the stellar neighborhood (d=9.7 pc) with a rotation period of 4.86 days.
The two transiting planets orbiting AU Mic, AU Mic b and c, are warm sub-Neptunes on 8.5 and 18.9-day periods and are targets of interest for atmospheric observations of young planets.
Here we study AU Mic's unocculted starspots using ground-based photometry and spectra in order to complement current and future transmission spectroscopy of its planets.
We gathered multi-color LCO 0.4m SBIG photometry to study the star’s rotational modulations and LCO NRES high-resolution spectra to measure the different spectral components within the integrated spectrum of the star, parameterized by 3 spectral components and their coverage fractions.
We find AU Mic's surface has at least 2 spectral components, a \Tamb$=$\ensembleTamb\ K ambient photosphere with cool spots that have a temperature of \Tspot$=$\ensembleTspot\ K covering a globally-averaged area of $39\pm4\%$ which increases and decreases by $5.1\pm0.3\%$ from the average throughout a rotation.
We also detect a third flux component with a filling factor less than $0.5\%$ and a largely uncertain temperature between 8500-10000K that we attribute to flare flux not entirely omitted when time-averaging the spectra.
We include measurements of spot characteristics using a 2-temperature model, which we find agree strongly with the 3-temperature results.
Our expanded use of various techniques to study starspots will help us better understand this system and may have applications for interpreting the transmission spectra for exoplanets transiting stars of a wide range of activity levels.
\end{abstract}

\section{Introduction} \label{intro}

Observations from JWST are now revealing exoplanet atmospheres in more detail than ever before \citep{Ahrer2023,Alderson2023,Feinstein2023,Rustamkulov2023,Fu2022} using an observational technique called transmission spectroscopy \citep{SeagerSasselov2000ApJ...537..916S,Brown2001,Pont2007,Berta2012,Sing2011}.
Transmission spectroscopy is done by measuring the transit depth (which is a proxy for the planet's radius) of an exoplanet as a function of wavelength and inferring atmospheric absorption \citep[e.g., ][]{SeagerSasselov2000ApJ...537..916S} and/or scattering \citep[e.g., ][]{Robinson2014,Sing2016} at wavelengths where the planet’s transit is deeper.
The stellar photons that are absorbed by the planet and its atmosphere originate specifically from the transit chord, the swathe of stellar surface occulted by the planet which in general is indistinguishable from the surrounding photosphere except in compact systems which exhibit transits at multiple latitudes.
Stellar surfaces can be homogeneous (i.e., spatially ``smooth" aside from granulation and limb-darkening effects), in which case the chord spectrum is the same as the disk-integrated spectrum, or they can be heterogeneous (containing active regions), in which case the transit chord is not necessarily representative of the disk-integrated stellar spectrum.

A homogeneous stellar background surface has typically been adopted in transmission studies.
While this assumption holds true in some cases, most stars do not have smooth, single-temperature surfaces but are instead spotted with activity-induced heterogeneities.
Spots are created where magnetic field lines pass through the photosphere and the magnetic pressure overwhelms the local gas pressure, suspending convection and causing the region within the intersecting field to cool.
Faculae arise from weaker concentrations of field lines where this pressure isn't enough to suspend convection but is enough to reduce the local opacity and increase the flux emanating from deeper in the photosphere, creating a brightening effect \citep{Basri2021}.

A non-axisymmetric distribution of spots and faculae creates time- and wavelength-dependent changes in the stellar surface flux, which has been observed in high-resolution stellar spectra \citep{Wing1967,AframBerdyugina2015}, color-magnitude relations \citep{Vogt1979, Olah1997}, stellar rotational modulation \citep{Vogt1979,Pass2023}, and more recently in exoplanet transits \citep{Brown2001,Pont2008,Sing2011,Sanchis-Ojeda2011}. 
When spots or faculae lie on the transit chord and are occulted by transiting planets, they create bumps or dips in the transit light curve that can bias the exoplanet radius measurement.
Occulted active regions, provided  they occupy discrete regions of the transit chord and their flux contrast appears above the noise, show up directly in transit light curves and can be identified and removed from the measured transmission spectrum.
Unocculted active regions, however, alter the disk-averaged spectrum such that it is no longer representative of the true source spectrum of photons entering the planetary atmosphere.
This in turn creates spurious $\lambda$-dependent changes to the exoplanet transit depth in what is now known as the transit light source effect \citep[TLSE, ][]{Rackham2018,Rackham2019}.

On cool stellar surfaces (below about 4000 K), molecular absorption lines \citep[like H$_2$O, VO, and TiO, ][]{Jones1995, Allard2012} begin to appear in the stellar spectrum and become entangled with molecular absorption signals in planetary atmospheres.
Cool unocculted spots with different or deeper molecular absorption lines than the surrounding photosphere will appear to add molecular absorption at those wavelengths in planetary transmission spectra and lead to mischaracterization of exoplanets and their atmospheres.
Additionally, unocculted spots give rise to an increasing transit depth toward bluer wavelengths as their contrast against the surrounding photosphere increases, which can be mistakenly identified as Rayleigh scattering in a transiting exoplanet’s atmosphere \citep[e.g., ][]{Robinson2014}.
Until we can precisely and reliably determine spot characteristics on our host stars, the signature of exoplanet atmospheres will be very challenging or impossible to disentangle from spot contamination for nearly all transmission observations of exoplanets M and K dwarfs.
This degeneracy, exemplified in recent transmission observations of sub-Neptune exoplanets TOI 270d \citep{Mikal-Evans2023}, Gl 486b \citep{Moran2023}, K2-33b \citep{Thao2023}, L 98-59 c \citep{Barclay2023}, and temperate terrestrial exoplanet TRAPPIST-1b \citep{Lim2023} is what we aim to mitigate for transmission observations of AU Mic b by precisely measuring spot characteristics for its host star in this work.

%%%%%%%%%%%%%%%%%%%%%%%%%%%%%%%%%%%%%%%%%%
% SYSTEM PARAMETERS TABLE - INTRO
%%%%%%%%%%%%%%%%%%%%%%%%%%%%%%%%%%%%%%%%%%
\begin{table}[htb]
    \centering
    \begin{tabular}{cc}
    \textbf{Quantity} & \textbf{Value}\\
    \textbf{AU Mic} & \\
    D [pc] & $9.714\pm0.002^{\rm a}$ \\
    T$_{\rm eff}$ [K] & $3665\pm31^{\rm b}$ \\
    M$_*$ [M$_{\Sun}$] & $0.60\pm0.03^{\rm c}$ \\
    R$_*$ [R$_{\Sun}$] & $0.82\pm0.05^{\rm c}$ \\
    P$_{\rm{rot}}$ [days] & $4.86\pm0.005^{\rm c,d}$ \\
    Metallicity [dex] & $0.12\pm0.10^{\rm c}$ \\
    log$g$ [log$_{10}$(cm/s$^2$)] & $4.52\pm0.05^{\rm c}$ \\
    \textbf{AU Mic b} & \\
    P [days] & 8.4631427$^{\rm e}$ \\
    R$_{p}$/R$_{*}$ & 0.0433$\pm$0.0017$^{\rm e}$ \\
    \end{tabular}
    \caption{System parameters relevant to this study.
    a: \citet{GaiaCollaboration2023},
    b: \citet{Cristofari2023},
    c: \citet{Donati2023},
    d: \citet{Martioli2021},
    e: \citet{Szabo2022}.}
    \label{tab:System_Parameters}
\end{table}

\paragraph{AU Microscopii}
AU Mic \citep{Torres1973} is a nearby \citep[9.7 pc;][]{GaiaCollaboration2023}, young \citep[$24\pm3$ Myr;][]{Mamajek2014}, rapidly rotating \citep[P$_{\rm{rot}}=4.86$ d;][]{Plavchan2020,Donati2023} pre-main sequence M dwarf with a debris disk \citep{Kalas2004, Chen2005b,MacGregor2013}.
This star has an inflated radius as it settles onto the main sequence, with M$=0.60$ M$_{\odot}$ and R$=0.82 $ R$_{\odot}$ \citep{Donati2023} and an effective temperature of 3600-3700K \citep[e.g., ][]{AframBerdyugina2019,Plavchan2020,Cristofari2023}
There are two transiting warm Neptunes on 8.46 and 18.86-day periods \citep{Hirano2020, Plavchan2020,Martioli2021,Gilbert2022,Zicher2022} and 2 unconfirmed candidate non-transiting planets on 12.74 and 33.39-day periods recently discovered through transit timing variations and radial velocity analysis \citep{Wittrock2022,Wittrock2023,Donati2023}.
The existence of an observable debris disk with interior transiting planets is a rare and exciting architecture that holds vast scientific potential.

Furthermore, this system is one of the best cases we have for studying star-planet interactions and the effects of young M dwarf activity on planetary atmospheres, an issue of great interest and concern in the search for terrestrial atmospheres and potentially habitable planets orbiting M dwarf stars \citep[e.g., ][]{Shields2016,Louca2023}.
This star's frequent high-energy flaring that may eventually lead to photo-evaporation of the atmospheres of AU Mic b and c \citep{Feinstein2022a} and even in the case where atmospheres are preserved, the long term implications of young M dwarf activity on planetary habitability are ominous.
By continuing to study AU Mic and its planets, we can build an internally consistent understanding of a nearby multi-planet system in the early stages of formation with a stellar surface evolving on months-long timescales \citep[e.g., ][]{Donati2023}.
System parameters are summarized in Table \ref{tab:System_Parameters} and a thorough review of AU Mic's stellar and planetary characteristics can be found in \citet{Donati2023}.

\paragraph{Transmission Observations of AU Mic b}
The observations we present in this work are part of a companion study alongside HST/WFC3 transmission spectra of AU Mic b; the first on 2021 Aug 30 (BJD 2459455.98) and the second on 2022 Apr 14 (BJD 2459684.40).
Because AU Mic's spots evolve noticeably over time \citep[e.g., ][]{Szabo2021,Szabo2022,Gilbert2022}, we have collected photometric and spectroscopic observations contemporaneous with the WFC3 observations to provide constraints on spot contamination at the time of both transmission visits.
This transmission spectrum will be analyzed in the context of our results and presented in a subsequent paper.

\paragraph{Objectives and Layout}
In order to characterize AU Mic’s spots and forward-model the spot contamination level in AU Mic b’s transmission spectrum, we assembled a self-consistent statistical framework that combines multi-color time-series photometry with contemporaneous high-resolution spectroscopy.
This method allows us to break observational degeneracies between spot coverage and spot contrast and better understand the \textit{bulk} characteristics of AU Mic's spots.
We use \textit{bulk} to mean the spatially-unresolved flux-weighted characteristics based on the treatment of spots as discrete regions without complex temperature profiles, ignoring for example the distinction between umbra and penumbra.
These are heavy assumptions, but we argue that our models are appropriate for the quality of our data and the information we hope to obtain.

This paper is laid out as follows: in Section \ref{observations_and_data} we describe the types of observations used in spot analysis and the specific observations we acquired, along with the data reduction and processing.
In Section \ref{methods} we describe our methods of analyzing AU Mic's rotational modulations (\ref{methods:rotation}), assembling the self-consistent statistical framework for modeling spot filling factor and temperature (\ref{methods:spotmodel}), and forward-modeling the spot contamination in AU Mic b's transmission spectrum (\ref{methods:contamination}).
In Section \ref{results} we report the results of our spot model and in Section \ref{discussion} we discuss the physical implications of our results, how they compare to previous studies, and the limitations of our approach.

\begin{figure*}[ht]
    \includegraphics[width=1.0\linewidth]{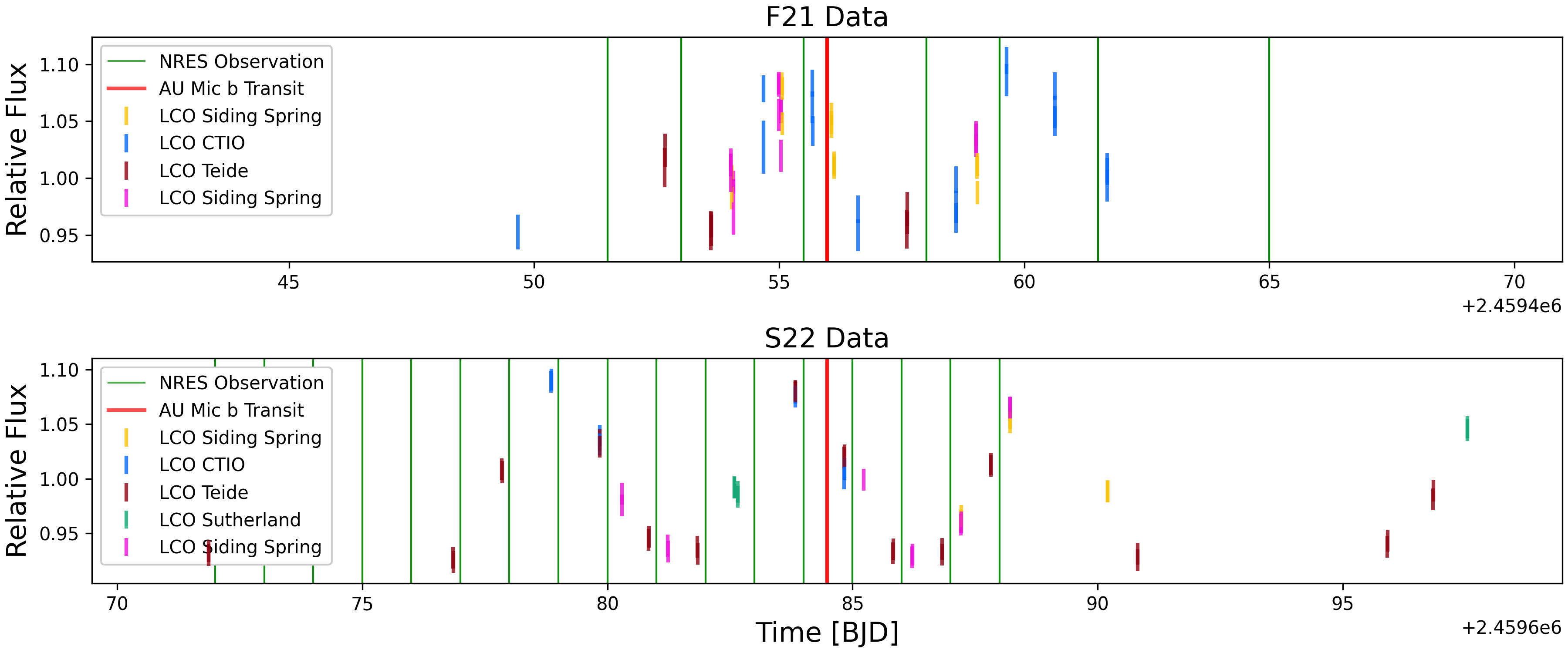}
\caption{Processed and baseline-corrected photometry data for all filters in visits F21 (top) and S22 (bottom) with vertical markers showing the temporal distribution of WFC3 (red) and NRES (green) observations. The color of each photometric data point is based on the LCO site where that data was observed. The time between transit observations is 229 days.}
\label{f:all_data}
\end{figure*}

%%%%%%%%%%%%%%%%%%%%%%%%%%%%%%%%%%%%%%%%%%%%%%%%%
% BEGIN OBSERVATIONS AND DATA REDUCTION SECTION
%%%%%%%%%%%%%%%%%%%%%%%%%%%%%%%%%%%%%%%%%%%%%%%%%
\section{Observations and Data Reduction} \label{observations_and_data}

\paragraph{Observing Spots}
Starspot characteristics are difficult to disentangle in practice as there is a degeneracy between spot coverage and temperature contrast that creates similar observational effects within a single waveband.
In addition, the physics and structure of stellar surfaces is poorly understood for all but the most heavily studied stars.
Stars of different type, rotation rate, and magnetic field strength exhibit differing forms of surface phenomena which are or will eventually be relevant to understand for the future of exoplanet discovery and characterization. 

Breaking spot degeneracies can be done by combining observations across optical and infrared wavelengths.
Short-wavelength broadband photometry allows us to probe starspots where they stand out the most against the stellar background (higher flux contrast toward the Wien limit), while long-wavelength observations are useful for identifying molecular characteristics of starspots where they overlap with planetary atmospheric absorption.
Broadband photometric variability measurements help us probe different temperature components on rotating stars, but generally only provides a lower limit on total spot coverage due to unknown axisymmetries in spot distribution \citep{Apai2018}.
Photometric variability amplitudes decrease with wavelength as the two flux components approach the Rayleigh-Jeans limit, so measuring rotational variability across the optical to the infrared provides strong constraints on the spot-to-photosphere temperature contrast \citep[e.g., ][]{StrassmeierOlah1992}.

High-resolution, time-series spectra have been used to study the relationships between stellar activity tracers and spot coverage \citep[e.g.,][]{Schofer2019,Medina2022}.
For spectroscopic studies, stellar spectra are modeled as a combination of two or more temperature components \citep[often referred to as \textit{spectral decomposition}, e.g., ][]{Gully-Santiago2017,Zhang2018,Wakeford2019}.
Specific molecular lines are often used as spot tracers including TiO \citep{Wing1967,Vogt1979}, CaH, MgH, FeH, and CrH \citep[e.g.,][]{Neff1995,AframBerdyugina2019}.
We recommend \citet{Berdyugina2005, Apai2018, Rackham2018} for a more thorough review of starspots and the techniques used to study them.

\subsection{Data} \label{data}
We acquired Las Cumbres Observatory (LCO) 0.4m $g'$-, $r'$-, and $i'$-band\footnote{\url{https://lco.global/observatory/instruments/filters/}} SBIG photometry and Network of Robotic Echelle Spectrographs (NRES) high-resolution echelle spectra on two separate visits spanning 2-3 weeks around their respective transmission observations (occurring in Fall 2021 and Spring 2022, hereafter F21 and S22).
These data were acquired contemporaneously with observations of AU Mic b's HST/WFC3 transmission spectrum (see Figure \ref{f:all_data}) with the hope of precisely constraining the magnitude of spot contamination at the appropriate stellar epoch and phase.

%%%%%%%%%%%%%%%%%%%%%%%%%%%%%%%%%%%%%%%%
% NRES SPECTRAL ORDERS TABLE
%%%%%%%%%%%%%%%%%%%%%%%%%%%%%%%%%%%%%%%%
\begin{table*}
    \centering
    \begin{tabular}{ccc}
    \textbf{Order} & $\lambda$ ($\mu$m) & \textbf{Note} \\
    53$^*$ & 0.873-0.888 & TiO line [8860 \AA]\\
    54 & 0.858-0.872  & \\
    55 & 0.842-0.857  & Excluded - Telluric Contamination \\
    56 & 0.826-0.841  & Excluded - Telluric Contamination \\
    57 & 0.810-0.826  & Excluded - Telluric Contamination \\
    58 & 0.798-0.812  & Excluded - Telluric Contamination \\
    59$^*$ & 0.784-0.798  & \\
    60 & 0.771-0.784  & Excluded - Poor Fit \\
    61$^*$ & 0.759-0.772  & TiO line [7600 \AA] \\
    62 & 0.746-0.760  & Excluded - Poor Fit \\
    63 & 0.734-0.748  & Excluded - Telluric Contamination \\
    64 & 0.723-0.736  & Excluded - Telluric Contamination \\
    65 & 0.712-0.724  & Excluded - Telluric Contamination | TiO line [7150 \AA]\\
    66 & 0.701-0.713  & Excluded - Telluric Contamination | TiO line [7050 \AA]\\
    67 & 0.690-0.702  & Excluded - Telluric Contamination\\
    68 & 0.680-0.692  & Excluded - Telluric Contamination\\
    69$^*$ & 0.671-0.682  & \\
    70 & 0.661-0.672  & Excluded - Poor Fit\\
    71$^*$ & 0.652-0.663  &  H-$\alpha$ band\\
    72$^*$ & 0.642-0.653  &  \\
    73$^*$ & 0.634-0.645  & Excluded - Poor Fit \\
    74 & 0.625-0.636  & Excluded - Telluric Contamination\\
    75$^*$ & 0.617-0.627  & \\
    76$^*$ & 0.609-0.619  & \\
    77 & 0.601-0.611  & Excluded - Poor Fit\\
    78 & 0.593-0.603  & Excluded - Telluric Contamination \\
    79 & 0.586-0.596  & Excluded - Telluric Contamination \\
    80 & 0.578-0.588  & Excluded - Poor Fit \\
    81$^*$ & 0.571-0.581  &  \\
    82$^*$ & 0.564-0.574  &  \\
    83$^*$ & 0.558-0.567  & \\
    \end{tabular}
    \caption{Details on the NRES Echelle spectra acquired for this study. The full spectrum spans 0.39-0.91 $\mu$m (orders 119-52) but we truncate the table and the analysis at orders 53 and 83 to focus on orders which are not dominated by noise. Orders that we omit from the final analysis are noted with a brief explanation, and further discussion of modeling specific orders is provided in the Appendix. Most omitted orders were heavily contaminated by telluric absorption, whereas the orders labeled ``Poor Fit" typically exhibit extremely cold spots, at the limit of the spectral library. Note that wavelength decreases with order.}
    \label{tab:NRES_spectrum_info}
\end{table*}

\subsubsection{SBIG Imaging}
Photometry was acquired between 2021 Aug 12-2021 Sept 03 (F21) and 2022 Apr 01-2022 Apr 27 (S22) using 5 separate telescopes automatically scheduled depending on weather, telescope availability, and target observability.
Typical exposure times were 20s in $g'$, 10s in $r'$, and 6s in $i'$.
Photometric data were automatically reduced into calibrated images by the BANZAI pipeline \citep{BANZAI2022} and downloaded from the LCO Science Archive\footnote{\url{https://archive.lco.global/}}.
We performed aperture photometry with AstroImageJ's multi-aperture photometry tool \citep{Collins2017}.
For each filter, the same 3 comparison stars are used to account for local atmospheric effects throughout the night and measure the target star's relative flux.
The fact that AU Mic is far brighter than its nearby comparison stars \citep[g', r', and i' magnitudes of 9.579$\pm$0.05,  8.636$\pm$0.09, and  7.355$\pm$0.14, respectively][]{2012yCat.1322....0Z} means that the photon noise in the photometry is set by the comparison star brightness, rather than by AU Mic itself.
Scintillation also contributes significantly to the photometric noise budget, particularly in the $i'$ photometry where exposure times are very short to avoid AU Mic saturating.
After performing aperture photometry, S/N per exposure was 300-600 in F21 $g'$ and $r'$, 200-300 in F21 $i'$, and 500-1000 in S22 $g'$ and $r'$.
For the first visit (F21), we have 450 exposures in SDSS $g'$, 1036 exposures in $r'$, and 1175 in $i'$.
For the second visit (S22) we have 328 in $g'$ and 330 in $r'$.

For each night of observations, we stitch together all data in a single filter and take the median in order to minimize the effect of flares.
We use \texttt{scipy.optimize} to fit an initial sinusoid model for sigma-clipping.
We clip $10-\sigma$ outliers from the initial model to account for flux variations outside of the rotational modulation (i.e., flares), generate an optimized fit with \texttt{scipy.optimize.minimize}, clip $5-\sigma$ outliers from that model, re-optimize a final time, and normalize uncertainties based on the reduced $\chi^2$ statistic from this final optimized fit.
The first cutoff is set at $10-\sigma$ because after median binning, the uncertainties were underestimated and a slightly wrong initial model could easily exclude otherwise useful data.
The second cutoff is set at $5-\sigma$ to account for any extreme outliers still remaining without being too restrictive, accounting for known uncertainties in the chosen model. 
After processing, the typical per-night S/N was 70-100 for $g'$ and $r'$ and 30-50 in $i'$, with a total of 19 data points for F21 $g'$, 18 for $r'$, 17 for F21 $i'$, 28 data points for S22 $g'$, and 30 for S22 $r'$. 
The per-night S/N only reaches to 100 because we set a minimum uncertainty of 1\% on the post-processing photometry based on the per-night spread in flux.

\begin{figure*}[ht]
    \includegraphics[width=1\linewidth]{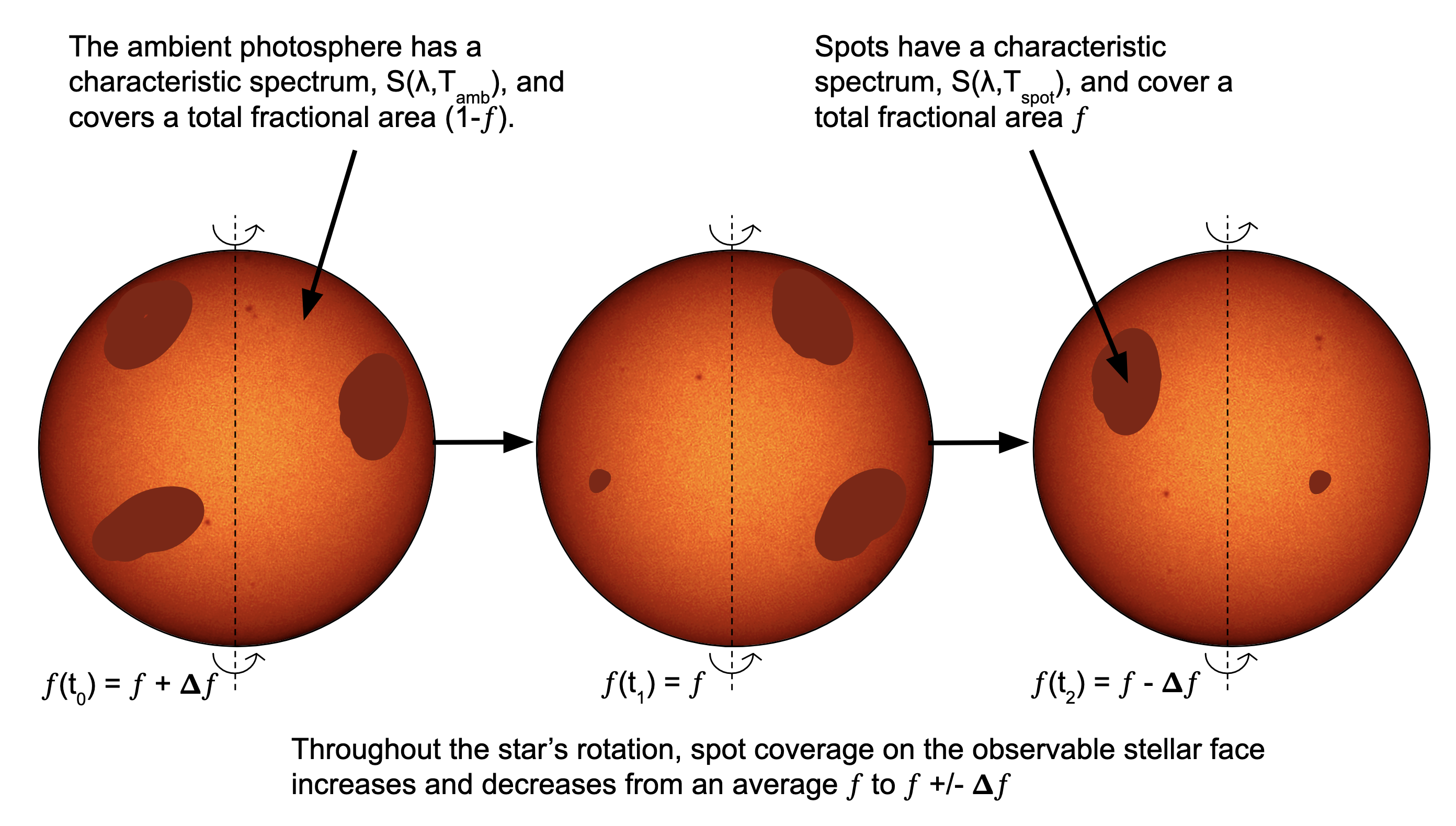}
\caption{Cartoon of a spotted star showing the parameters used in this study.}
\label{f:cartoon_parameters}
\end{figure*}

\subsubsection{NRES Spectra} \label{data:nres}
AU Mic's spectrum was observed in the $0.39-0.91\mu$m NRES bandpass with 600s exposure times resulting in a total of 38 observations from 2021 August 24-2021 September 09 (F21) and 48 observations from 2022 April 02-2022 April 18 (S22).
The R=53,000 NRES spectra were reduced by the BANZAI-NRES pipeline \citep{BANZAI2022} and downloaded from the LCO Science Archive.
Spectra from individual observations show typical peak S/N of 35 where the star is brightest (around order 60) and drops off to below 10 in order 52 and beyond order 83, so we omit orders outside of this range from our analysis. 
We processed the spectra with the \texttt{chromatic}\footnote{\url{https://github.com/zkbt/chromatic}} tool, first correcting for the velocity shift in each order's spectra from the movement of the Earth in different positions of its orbit.
This was -9.5km/s in F21 and 33.0 km/s in S22 derived from a $\chi^2$ grid search using a single-temperature optimized model spectrum and \texttt{scipy.optimize.minimize}.
We median-combine spectra in time to one averaged spectrum per night resulting in 7 spectra in F21 and 17 spectra in S22.
We also bin each spectrum to 0.05 nm (roughly R=11000-18000); this is greater than the Doppler broadening width of 0.015-0.024 nm that we estimate for the NRES bandpass based on AU Mic's v$sin$i of $8-9$ km/s \citep[e.g., ][]{Donati2023}.
Zeeman broadening is an additional effect which alters line profiles in magnetically active stars \citep[e.g., ][]{Gray1984}, but we calculate the Zeeman broadening \citep{Reiners2013} to be less than rotational broadening at these wavelengths and accounted for within our chosen bin size.
Many of our spectral orders overlap with absorption bands in Earth's atmosphere so we trim out any wavelengths where the molecular line transmission fraction of the atmosphere is $<0.995$ (i.e., any wavelength at which $\geq$0.5\% of the photons are absorbed) based on time-averaged telluric data from Skycalc \citep{Noll2012,Jones2013}.
We run a $\sigma$-clipping routine that first calculates an optimized single-temperature PHOENIX model \citep{Husseretal2013} for each spectrum, and second clips emission lines, defined as points $>3-\sigma$ \textit{above} the optimized model.
Uncertainties on the time-averaged spectra are inflated to give a reduced $\chi^2$ of 1 when fit against a 3650K template.
This is an increase in uncertainty of 7-55x depending on the order, resulting in typical per-order S/N of 10-40.

After processing, we omitted two-thirds of the spectral orders from the final analysis based on their level of telluric overlap or in some cases because the ambient or spotted component was very poorly constrained, possibly due to the spectral model fidelity problem \citep[][]{Iyer2020,Rackham2023}.
The orders we include in the final analysis are 53, 54, 59, 61, 69, 71, 72, 75, 76, 81, 82, and 83, with details in Table \ref{tab:NRES_spectrum_info}.
In F21, there are total of 1831 spectral data points, with 1828 in S22, for a total of 3659 spectral data points.
Discussion on the choice of orders to include and spectral model results for each individual order are in the Appendix.

%%%%%%%%%%%%%%%%%%%%%%%%%%%%%%%%%%%%%%%%
% BEGIN METHODS SECTION
%%%%%%%%%%%%%%%%%%%%%%%%%%%%%%%%%%%%%%%%
\section{Methods} \label{methods}

Our analysis is ordered in three steps:
\begin{enumerate} 
    \item Measuring the stellar rotation signal, where photometric data is modeled as a sine wave to infer the semi-amplitude of stellar variability.
    \item Modeling spot characteristics, where we infer spot characteristics based on AU Mic's \Teff, measured photometric variabilities, and the time-averaged spectra. This is the primary focus of our analysis.
    \item Forward modeling the TLSE, where we take the posterior samples from our modeling to calculate the range of spectral contamination we can expect in the HST/WFC3 transmission data for AU Mic b.
\end{enumerate}

Figure \ref{f:cartoon_parameters} shows the cartoon stellar surface we model as a combination of ambient photosphere with a characteristic spectrum S($\lambda$,\Tamb) and spots with characteristic spectra S($\lambda$,\Tspot) covering a globe-averaged \fspot\, which deviates from the average coverage by $\pm$\deltafspot\ throughout the stellar rotation.
We include a third flux component (not shown on the figure) and label this component ``hot" (with attributes \fhot\ and \Thot) as we are uncertain about the physical source of the measured hot component, if it exists.
The primary results we report in section \ref{results} come from this 3-temperature modeling, but we also test a 2-temperature model and discuss its results and implications in section \ref{discussion}.
The two visits are modeled with the same set of parameters (implying no change to the surface components between visits), an assumption we test and discuss later in the paper.

\subsection{Measuring Photometric Variability} \label{methods:rotation}

Once the photometry data were processed as described in Section \ref{observations_and_data}, we measure the rotational variability with a sinusoidal model with the following form:
\begin{equation} \label{eqn: sine_wave_1}
    F(t)={A~\sin}(2\pi t/P) + {B\cos}(2\pi t/P) + {C}_{k},
\end{equation}
where $t$ is the time of an individual data point, $P$ is the stellar rotation period which we keep fixed at 4.86 days, $A$ and $B$ are amplitude parameters, and $C_{\rm k}$ is the offset parameter unique to each camera in each visit.
We fit each camera's data separately as we expect different cameras to have slightly different responses and should be normalized to their separate average fluxes.
While AU Mic has a notably asymmetric light curve \citep[e.g., the TESS light curves shown in][]{Martioli2021} and \citet{Angus2018} caution against using simple sinusoidal models to fit stellar rotation curves, our goal here is not to infer a precise rotation curve morphology and spot distribution but to measure the relative amplitude of flux variability between separate bandpasses.

Using \texttt{emcee} \citep{Foreman-Mackey2012}, we ran a Markov Chain Monte Carlo (MCMC) with 100 walkers, 1000 steps, and 25\%~burn-in.
We used the auto-correlation time to judge when the sampler had converged for each parameter \citep[e.g., ][]{Hogg2010}.
Median-value parameters and their 1-$\sigma$ uncertainties calculated from the sample distributions are propagated through to the following reformulation of Equation \ref{eqn: sine_wave_1}:
\begin{equation} \label{eqn: sine_wave_2}
    F(t) = X\sin(2\pi t/P + \theta)+{C}_{k},
\end{equation}
where $X=\sqrt{A^2+B^2}$ and $\theta = \arccos(A/X)$.
$X$ is the photometric semi-amplitude of variability (or $\rm \frac{\Delta S}{S_{avg}}$) for a given photometric bandpass, which we use as inputs for the spot characteristics model described below.
We inflate the uncertainty on each variability measurement by 25$\%$ to account for the assumptions of a simple rotation curve and negligible facular contribution.

%%%%%%%%%%%%%%%%%%%%%%%%%%%%%%
% METHODS TABLE: DATA PRIORS 
%%%%%%%%%%%%%%%%%%%%%%%%%%%%%%
\begin{table}[ht]
    \centering
    \begin{tabular}{cc}
    \textbf{Quantity} &  \textbf{Prior} \\
    \Teff\ [K]	&  $3650\pm100$\\
    \Thot [K] &  $\mathcal{U}$[\Tamb, 12000] \\
    \Tspot\ [K] &  $\mathcal{U}$[2300, \Tamb] \\
    \Tamb\ [K] &  $\mathcal{U}$[\Tspot, \Thot]\\
    \fhot &  $\mathcal{U}$[0,0.5]\\
    \fspot\ &  $\mathcal{U}$[0,(1.0-\fhot)]\\
    \deltafspot\ &  $\mathcal{U}$[0, \fspot]\\
    \end{tabular}
    \caption{Priors placed on our model parameters in the spot characteristics Monte Carlo simulation.}
    \label{tab:Priors}
\end{table}

\subsection{Spot Characteristics Model} \label{methods:spotmodel}
To draw inferences about AU Mic's spot and facula characteristics, we assembled a model for 3 data components: The effective temperature, \Teff, the photometric semi-amplitude of variability, $\frac{\Delta S}{S_{\rm{avg}}}$, and the time-averaged stellar spectrum, S$_{\rm{avg}}$.
Each of these components is modeled as a function of some combination of \fhot, \fspot, \Thot, \Tspot, and \Tamb, with model priors described in Table \ref{tab:Priors}.

Modeling photometric variability requires one additional parameter, the $|$peak-average$|$ amplitude of the change in spot coverage throughout a rotation, \deltafspot.
This parameter represents a change in spot coverage relative to the average coverage in a way that is not relevant to our models of the time-averaged spectral data or \Teff.
It can range from 0, where the surface is homogeneous or the surface features are distributed symmetrically around the rotation axis, to \deltafspot$=$\fspot, where the total spot coverage is clustered on the surface such that it rotates entirely in and out of view.
While photometric variabilities only provide a lower limit on the \textit{average} spot coverage fraction, the magnitude of variability depends strongly on this change in spot coverage throughout a rotation and can be precisely constrained with sufficient evidence of the spectral contrast between the ambient and spotted photosphere.

\paragraph{Effective Temperature}
Similar to \citet{Libby-Roberts2022}, we treat \Teff\ in the following form:
\begin{equation} \label{eqn: Teff}
    \mathrm{T_{eff}}^4=f\mathrm{_{spot}T_{spot}}^4+f\mathrm{_{hot}T_{hot}}^4+f\mathrm{_{amb}T_{amb}}^4,
\end{equation}
where \fspot~is the globally-averaged spot coverage fraction with temperature \Tspot, \fhot is the average coverage of any potential third component (which may be faculae, flares, or something else) with temperature \Thot, and $f_{\rm amb}$ is the coverage of the ambient photosphere which has temperature \Tamb.
The ambient coverage is not a unique parameter in the model but is calculated as $f_{\rm amb}=1-(f_{\rm spot}+f_{\rm hot})$.
This constraint effectively ensures that whatever combination of spectral components is being modeled accurately reproduces the known surface-averaged bolometric flux emitted from the stellar surface. 

\paragraph{Photometric Variability}
Following the formalism in \citet{Libby-Roberts2022}, we can calculate the semi-amplitude of variability due to spots as the following: 
\begin{equation} \label{eqn: variability_calculation}
    \frac{\Delta S(\lambda)}{S_{\rm avg}(\lambda)} = -\Delta f_{\rm spot} \left(\frac{1-\frac{S(\lambda, \rm T_{spot})}{S(\lambda, \rm T_{amb})}}{1-f_{\rm spot}[1-\frac{S(\lambda, \rm T_{spot})}{S(\lambda, \rm T_{amb})}]}\right).
\end{equation}
The expression above, the only calculation in our model which depends on \deltafspot, is integrated across the filter bandpasses to generate a single variability datum for each filter.
We account for the filter response curves by normalizing our variability integral by the filter response function:
\begin{equation} \label{eqn: delta_s_over_s}
    \frac{\Delta S}{S_{\rm avg}} = \frac{\int^{\lambda_2}_{\lambda_1}\frac{\Delta S(\lambda)}{S_{\rm{avg}(\lambda)}}W(\lambda)S(\lambda, \rm T_{eff})d\lambda}{\int^{\lambda_2}_{\lambda_1}W(\lambda)S(\lambda, \rm T_{eff})d\lambda},
\end{equation}
where the SDSS filter response functions (W$_{\lambda}$) are acquired through \texttt{Speclite}\footnote{\url{https://speclite.readthedocs.io/en/latest/filters.html}}.
The stellar spectrum term ($S(\lambda, \rm T_{eff})$, calculated at T$_{\rm{eff}}=3650K$) accounts for the non-uniform distribution of stellar flux emitted across bandpasses.
These bandpass-integrated model variabilities are then fit to the broadband variability measurements extracted from the stellar rotation curve models (section \ref{methods:rotation}).

We ignore a facular contribution to the rotational variability because magnetically active stars are expected have photometric variabilities dominated by spots \citep{Shapiro2016}, an assumption we argue is valid considering the low filling factor of this component we measure, discussed in the results.
It does factor into the calculation of \Teff\ however, and therefore indirectly affects the model results when we examine the variability separate from the spectra.
The variability light curve is also highly under-sampled, so adding another component for this stage would be over-fitting the very few (5) photometric variability data points we have.

\paragraph{Average Spectrum}
From the NRES Echelle spectra, we calculate a time-averaged spectrum which we model as a combination of a spotted spectrum and an ambient spectrum weighted by their globally-averaged coverage:
\begin{equation}
    \mathrm{S_{avg}}=f_{\rm spot} S(\lambda, \mathrm{T_{spot}})+f_{\rm hot} S(\lambda, \mathrm{T_{hot}})+f_{\rm amb} S(\lambda, \mathrm{T_{amb}}).
\end{equation}

Older studies of starspots have been limited in this approach due to the computation time required to model thousands of spectral lines and as a result they typically probed specific regions and rotational or vibrational temperatures which may not be indicative of the bulk spot properties.
Here we modeled as many possible regions of the spectrum as possible, including orders which have weak or non-existent spot signatures as well as those with strong signatures indicative of very cool regions, to understand the most complete picture of the star provided by the spectral data.

\subsection{Spot Contamination Model} \label{methods:contamination}

Atmospheric absorption will induce a wavelength-dependent change in the transit depth ($\Delta\rm{D(\lambda)}$) of the planet, expressed as
\begin{equation} \label{eqn: delta_D_lambda}
    \Delta\rm{D(\lambda)} = \left(\frac{R_p}{R_*}\right)^2 + \Delta\rm{D(\lambda)_{atm}} + \Delta\rm{D(\lambda)_{spot}},
\end{equation}
with $\Delta\rm{D(\lambda)_{atm}}$ is defined as:
\begin{equation} \label{eqn:atmospheric absorption}
    \Delta\rm{D(\lambda)_{atm}} = \frac{2R_p}{R^2_*}H\times n(\lambda),
\end{equation}
where H is the scale height and $n(\lambda)$ is the number of opaque scale heights at each wavelength, which typically varies between 0-5 for cloud-free atmospheres \citep{Seager2000ApJ...540..504S}.

Following the derivations in \citet{Rackham2018, Zhang2018, Libby-Roberts2022},  we can express $\Delta\rm{D(\lambda)_{spot}}$ as:
\begin{equation} \label{eqn: delta_D_spot}
    \mathrm{\Delta D(\lambda)_{spot} = \left(\frac{R_p}{R_*}\right)^2}\left[\frac{(1-f_{\rm spot,tra})+f_{\rm spot,tra}\mathrm{\frac{S(\lambda, T_{spot})}{S(\lambda, T_{amb})}}}{(1-f_{\rm spot})+f_{\rm spot}\mathrm{\frac{S(\lambda, T_{spot})}{S(\lambda, T_{amb})}}}-1\right].
\end{equation}

This expression can similarly be used to calculate facular depth contribution but we assume this contribution is negligible on AU Mic.
Equation \ref{eqn: delta_D_spot} does not explicitly rely on \deltafspot, but the value we derive for \deltafspot\ can be used to project the spot coverage at a given time or phase, which is needed to account for spot contamination at the time of transit. 
In this work we assume $f_{tra}$ to be zero for both spots and faculae, which implies that the contamination calculated for a given set of parameters represents an upper limit relative to a spotted transit chord.
Samples for \fspot, \fhot, \Tspot, \Thot, and \Tamb\, are injected into this model to generate a posterior distribution of $\Delta\rm{D(\lambda)_{spot}}$.

\subsection{Experimental Design}
We run several different iterations of the spot characteristics model in order to examine the information and constraints provided by each component of the data: we model the photometric measurements separate from the spectra, the spectra without photometric models, and the ensemble model which includes both data types.
The model is set up as a Monte-Carlo simulation using the \texttt{emcee} sampler \citep{Foreman-Mackey2012}, run with 100 walkers and 2000 steps with a 25$\%$ burn-in.
Models ran past convergence in accordance with the auto-correlation time of the sampler chains \citep[for a discussion of convergence and autocorrelation, see][]{Foreman-Mackey2012}.
This framework maintains a distinction between the temperatures of the different spectral surface components (\Tspot$\leq$\Tamb$\leq$\Thot) while allowing \fspot\ and \fhot\ to vary, enabling models to arrive at solutions where the the surface is $\geq$50\% covered in spots.
Technically \fhot\ is allowed to vary as high as 50\% but in practice the models almost never preferred values of \fhot\ greater than a few percent.
Time-domain analysis of the 7 NRES spectra from F21 and 17 spectra from S22 should contain information about the change in spot coverage with stellar phase but no periodic signal could be found so we do not include a time-domain spectral model in the analysis.
Independent modeling of the separate visits returned strongly consistent measurements for \fspot, \fhot, and \deltafspot\, which could be be a robust finding, considering the time between visits is $2\times$ the 120-150 day activity evolution timescale (which we can approximate to be a spot decay timescale) measured by \citet{Donati2023}.

%%%%%%%%%%%%%%%%%%%%%%%%%%%%%%%%%%%%%%%%
% BEGIN RESULTS SECTION
%%%%%%%%%%%%%%%%%%%%%%%%%%%%%%%%%%%%%%%%
\section{Results} \label{results}

\subsection{Variability Amplitude Results}
\begin{figure*}[ht!]
    \subfloat{\includegraphics[width=0.32\textwidth]{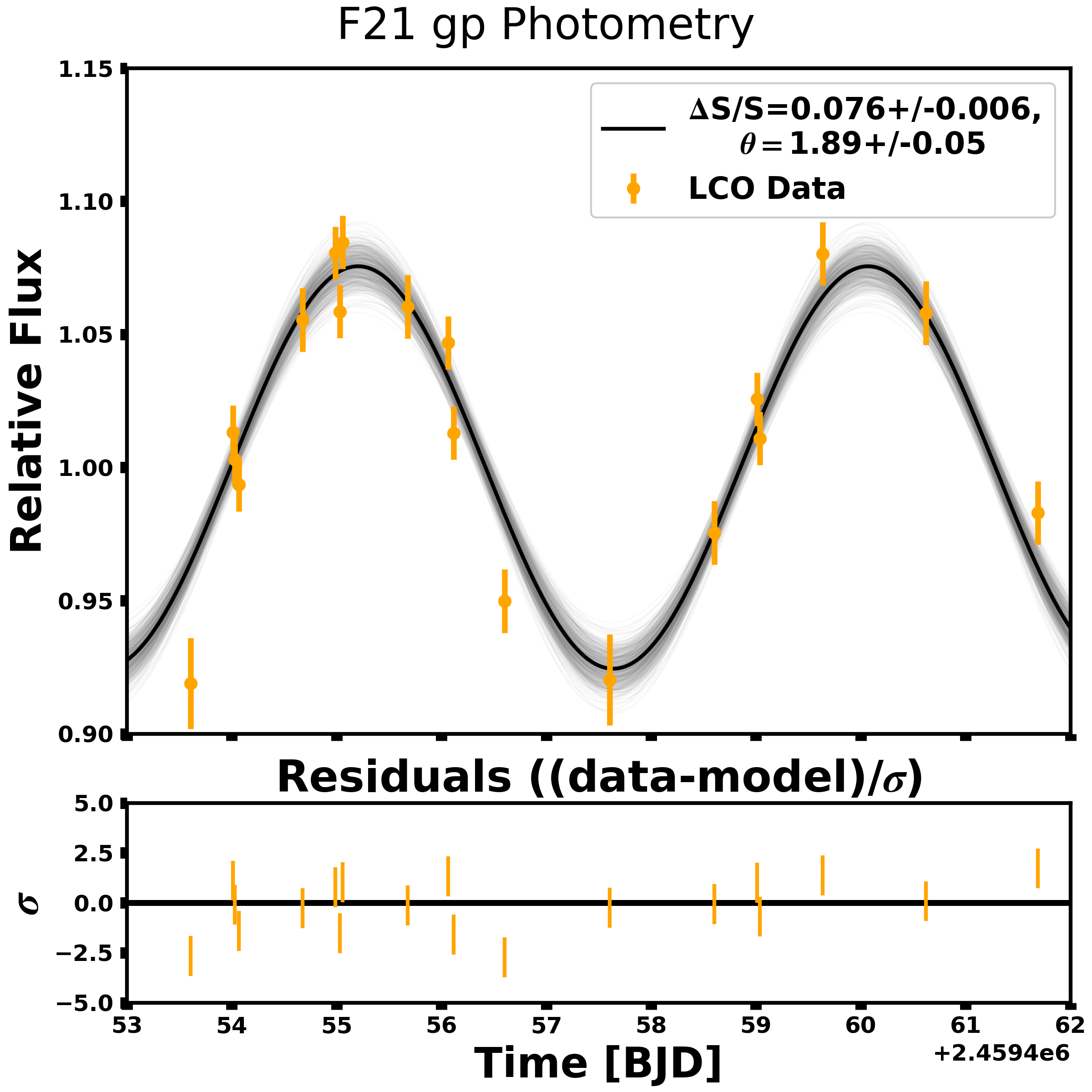}}
    \subfloat{\includegraphics[width=0.32\linewidth]{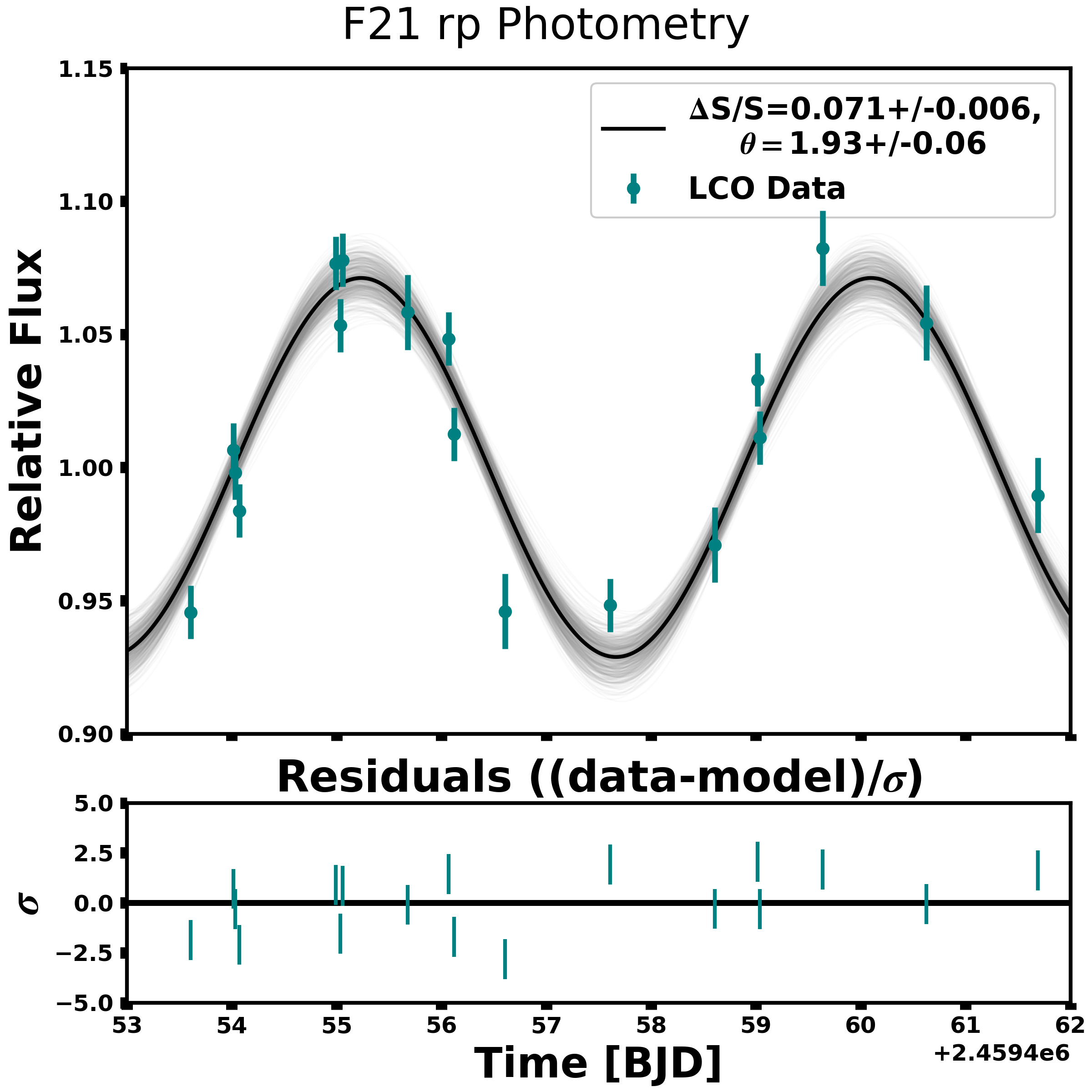}}
    \subfloat{\includegraphics[width=0.32\linewidth]{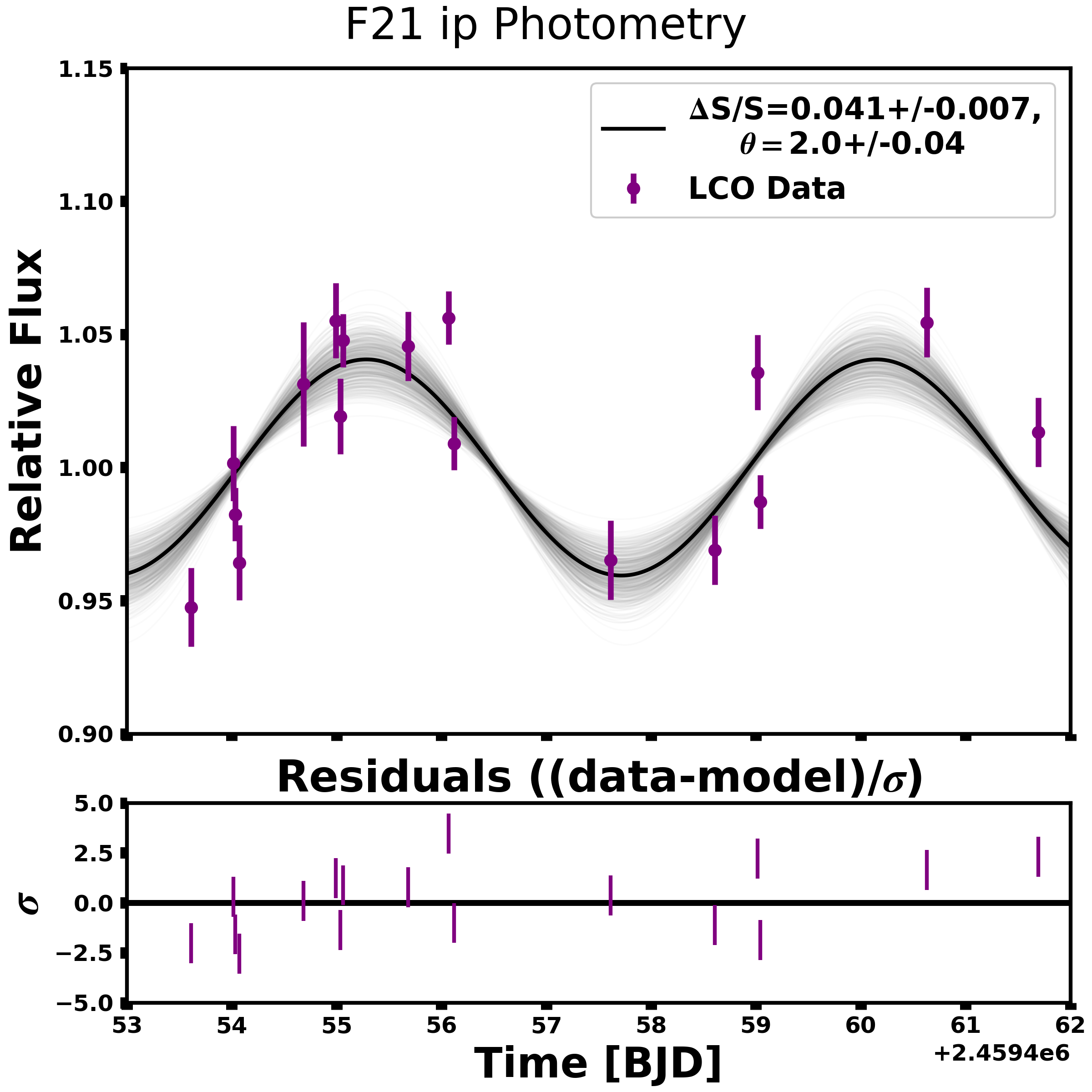}}\\
    \subfloat{\includegraphics[width=0.32\linewidth]{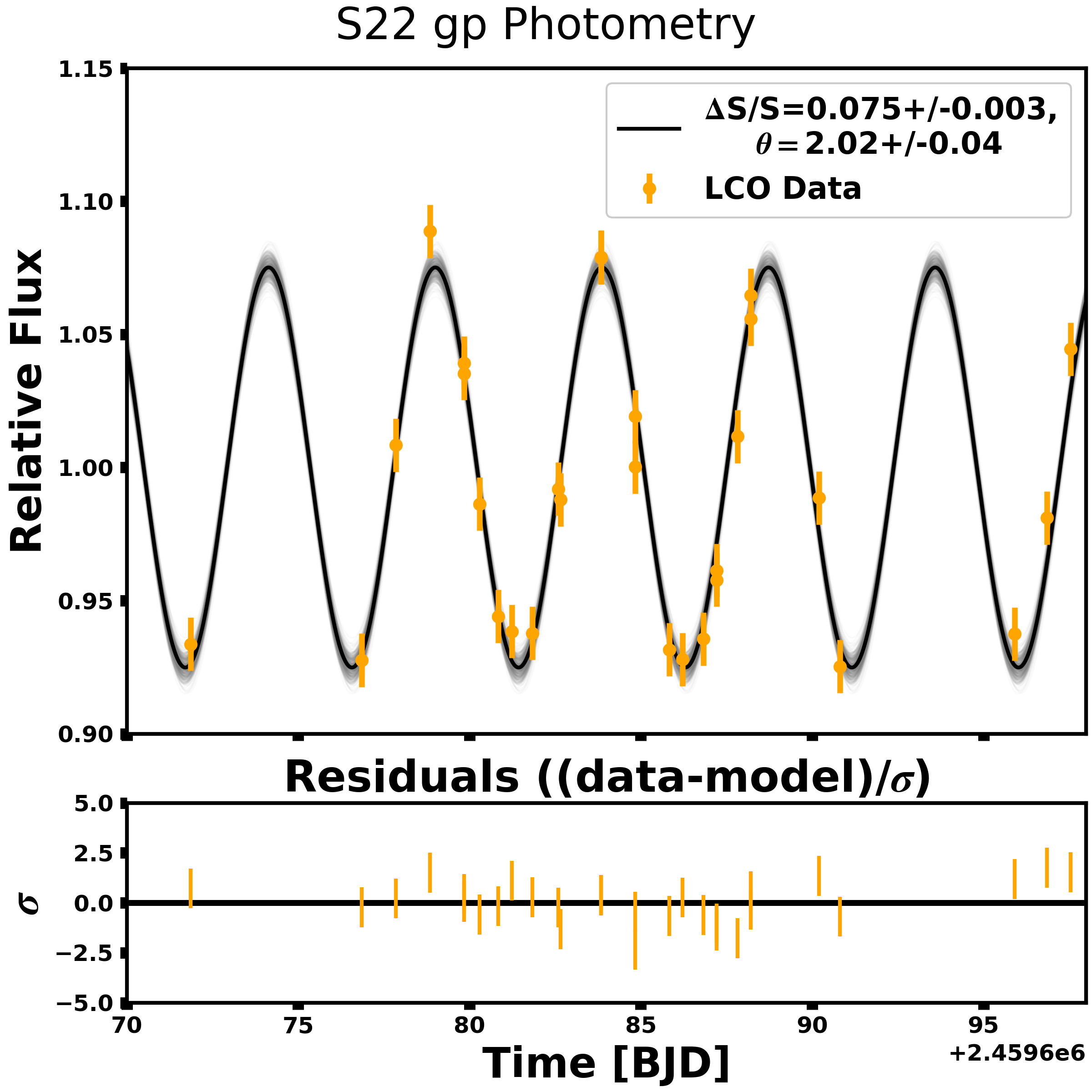}}
    \subfloat{\includegraphics[width=0.32\linewidth]{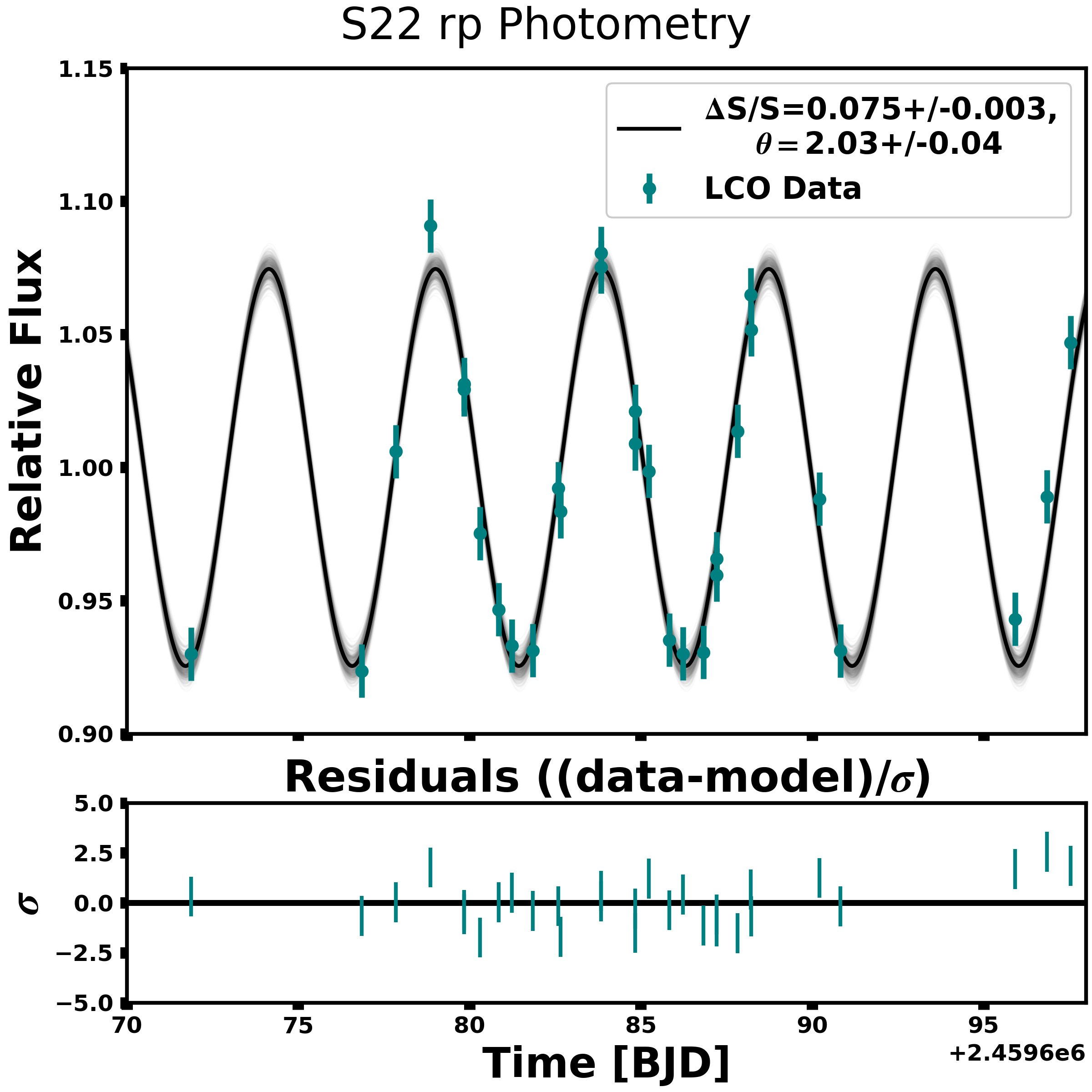}} 
\caption{LCO 0.4m photometry of AU Mic in the $g'$, $r'$, and $i'$ filters with the rotation model fits described in section \ref{observations_and_data}. Variability decreases with wavelength, exhibiting a significant decrease between $r'$ and $i'$ measurements. The shaded regions are randomly sampled models showing the distribution we quote as uncertainty on the amplitude, and the color of each data set corresponds to the filter response curves in Figure \ref{f:Teff_Phot_model_results}. Model results are provided in Table \ref{tab:rotational_variability_results}.}
\label{f:Rotation_Models}
\end{figure*}

For the F21 visit, we measure variability semi-amplitudes, $\Delta S/S$, in $g'$, $r'$, and $i'$ of \Fallgp, \Fallrp, and \Fallip, and for the S22 visit we measure $\Delta S/S$ of \Springgp\ in $g$' and \Springrp\ in $r$', shown in Figure \ref{f:Rotation_Models} and summarized in Table \ref{tab:rotational_variability_results}.

\begin{table}
    \centering
    \begin{tabular}{ccc}
    \textbf{Filter} & \textbf{Amplitude} & \textbf{phase}  \\
    First Visit (F21) \\
    $g$' & \Fallgp & $1.89\pm0.05$ \\
    $r$' & \Fallrp & $1.93\pm0.05$ \\
    $i$	'& \Fallip & $2.00\pm0.04$ \\
    Second Visit (S22) \\
    $g$' & \Springgp & $2.02\pm0.04$ \\
    $r$' & \Springrp & $2.03\pm0.04$ \\
    \end{tabular}
    \caption{Parameters from the MCMC fits of the photometry. Period was kept fixed at the literature period of 4.86d while the phase and amplitude were modeled as a combination of sine and cosine terms.}
    \label{tab:rotational_variability_results}
\end{table}

Measuring the signal for $i$' was slightly challenging because, as the reddest bandpass, this filter showed the weakest variability signal and therefore smallest signal-to-noise.
Moreover, AU Mic is much brighter than the nearby comparison stars and exposure times are short, scintillation noise is prevalent.
To account for poorly constrained measurements in $i'$ and improve the internal consistency of our multi-color variability measurements, we impose a prior on the phase ($\theta$) for $i$'.
The independently-modeled $g'$ and $r'$ rotation curves agree within their 1-$\sigma$ uncertainties in phase, so we use the average of their phases (1.91+/-0.04) as a prior when modeling $i$', resulting in a modeled phase in agreement with $g$' and $r$'.
This is a reasonable approach to improving our $i'$ results because we would expect the phase of 3 separate but contemporaneous data sets to be equal, so forcing the $i'$ phase to be consistent with $g'$ and $r'$ increases the consistency between the three measurements and lends confidence to the relative amplitudes our signal measurements.
The phase is not used further in this work but will be useful for the analysis of AU Mic b's atmospheric transmission spectrum when we need to estimate the spot coverage at the time of transit.

\subsection{Spot Characteristics Model}
We examine the spot characteristics results when we model only the stellar effective temperature and photometric variability data (excluding NRES spectral models) and similarly when we model only the NRES spectra with stellar effective temperature (excluding the multi-color photometric variability data).
Finally, we examine the results of modeling the stellar \Teff, photometric variabilities, and NRES spectra together in an ``ensemble" model which models the 12 spectral orders and multi-color variabilities for both visits, from which we report the final results.

\subsubsection{Photometric Variability}
\begin{figure}[ht]
    \includegraphics[width=\columnwidth]{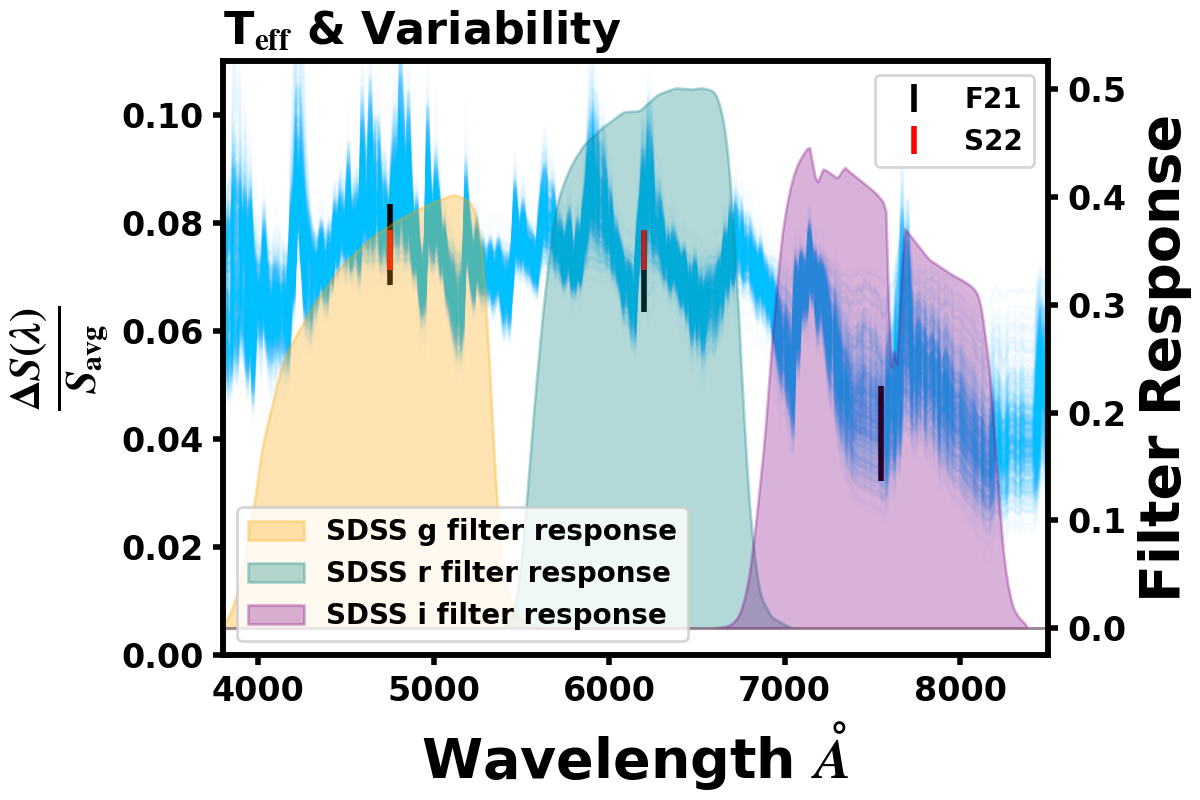}
\caption{All 5 measured rotational variabilities with random models (blue) drawn from the variability model samples. Shaded regions in the background are the filter response curves for our observations. We fit both visits together because the photometric and spectroscopic data are consistent despite the 6 months between visits \citep[e.g., ][]{Robertson2020}. Variability decreases with wavelength as the spot-to-photosphere flux contrast decreases, and when modeling the visits separately we find that our solutions were very sensitive to the magnitude and uncertainty of the $i$' measurement. We have photometry for all 3 filters in F21 but only $g'$ and $r'$ in S22.}
\label{f:variability_models}
\end{figure}

Figure \ref{f:variability_models} shows our measured photometric variability semi-amplitudes with random samples drawn from the photometry-only posterior distributions, with statistical results in Figure \ref{f:samples}.
The photometric variabilities, along with AU Mic' \Teff, show evidence for essentially any spot coverage between 10-90$\%$ with temperature \PhotTspot\ K, changing throughout an orbit by $6\pm3\%$.
Solutions for ambient temperature are \PhotTamb\ K, with an upper limit of 40$\%$ coverage of a hot component with temperature \PhotThot K.
The hot component is only relevant to the effective temperature calculation in the variability-only modeling, so it is acting more as an extra free parameter to improve the fit than it is related to anything physical.
The measurement of \Tspot\ is surprisingly accurate compared to the other measurements in this paper and others.  

\textbf{2-Temperature Models} of the variabilities return tighter constraints on the \textit{ratio} of \Tspot\ to \Tamb\ and their individual measurements, but the measurement of \Tspot\ is significantly warmer (\Tspot=\PhotTspottwoT\ K) than virtually every other measurement of \Tspot\ we present in this paper.
Measurements of \fspot\ and \deltafspot\ are consistent with the 3-temperature fits.
Results for two-temperature modeling are in Table \ref{tab:twoT_final_results}.

\subsubsection{Spectral Decomposition Results}
\begin{figure*}
    \subfloat{\includegraphics[width=\linewidth]{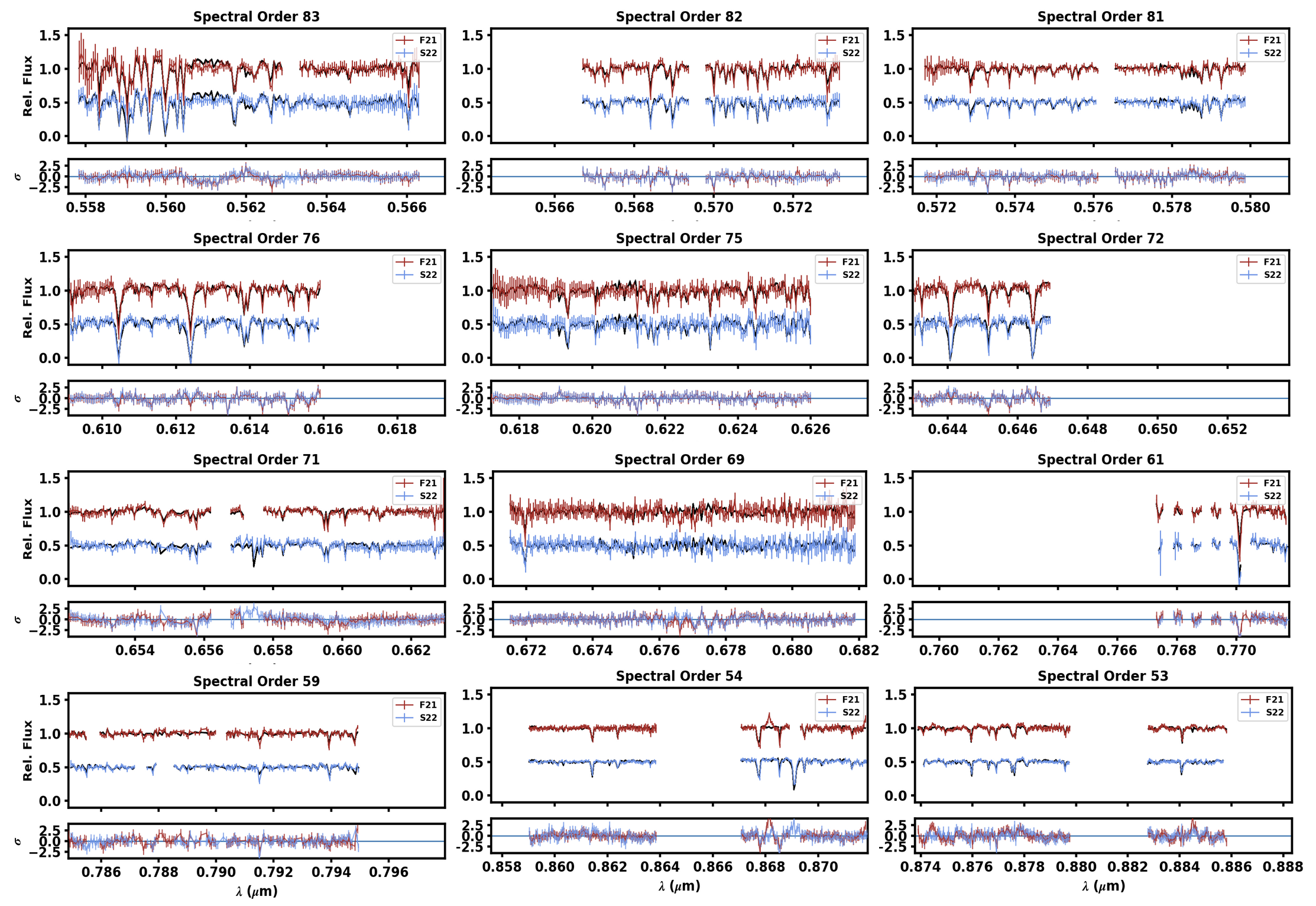}}
\caption{Spectra used in this analysis. The red and blue data points (F21 and S22, respectively) are median-averaged in time and have had their uncertainties normalized so a 3650 K template model has a reduced $\chi^2$ of 1. 100 randomly sampled models are plotted in black, which in most cases is a tight spread and difficult to notice in these plots. Despite some poorly fit line depths, there is a tight constraint on spot temperatures and filling factor from the spectra with or without photometric variabilities.}
\label{f:spectral_collage}
\end{figure*}

When modeling the 12 spectral orders (Figure \ref{f:spectral_collage}) simultaneously and without photometric variability, we find more precisely constrained temperatures and coverages of all three components.
The change in spot coverage, \deltafspot\, is unconstrained by these models because we are modeling time-averaged spectra.
The stellar spectra indicate a large fraction (\fspot=\Specfspot) of cool spots with temperature \SpecTspot\ K, a dominant "ambient" (warmer) photosphere with temperature \SpecTamb\ K, and a very tenuous detection of a hot component with temperature \SpecThot\ K covering less than 0.5$\%$ of the surface. 

\textbf{2-Temperature Models} of the spectra result in fully consistent measurements of each parameter.
The spot component measured to cover $41\pm3\%$ of AU Mic with temperature \SpecTspottwoT\ K, and the ambient photosphere temperature is measured to be \SpecTambtwoT\ K.

\subsubsection {Ensemble Model}
\begin{figure*}[ht]
    \subfloat{\includegraphics[width=\textwidth]{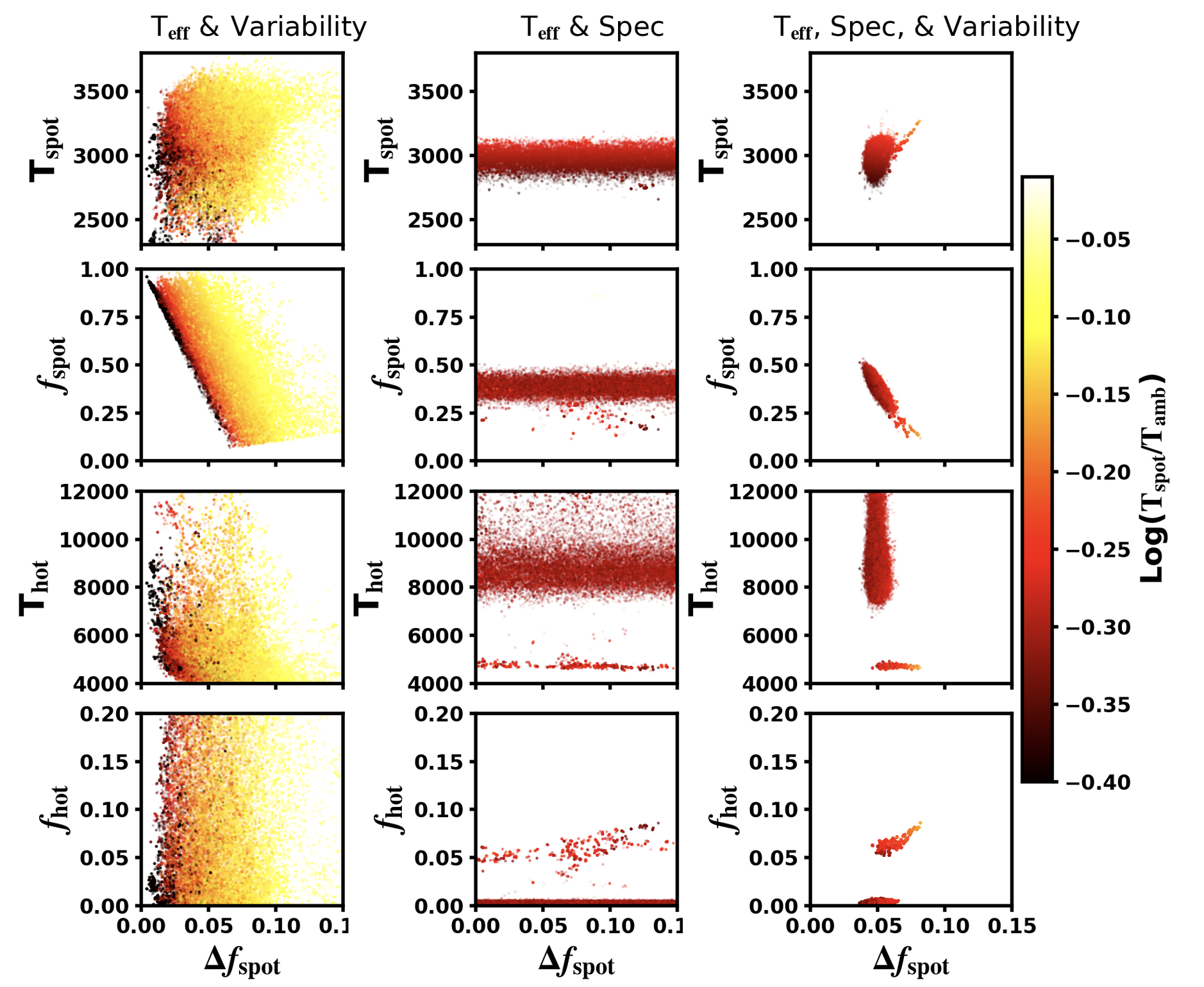}}
\caption{Posterior samples for the variability model (left), spectral model (middle) and the ensemble model (right), which retains characteristics of the separate model results. The spectra constrain how hot spot temperatures can be while the photometry constrains how cool they can be. Limits on spot coverage fraction are mostly provided by the spectral modeling, with the variability providing constraints on \deltafspot. The color of each point corresponds to the log of the temperature ratio (\Tspot/\Tamb), with redder points being spots that are further from the ambient temperature.}
\label{f:samples}
\end{figure*}

\begin{figure*}
    \subfloat{\includegraphics[width=0.48\textwidth]{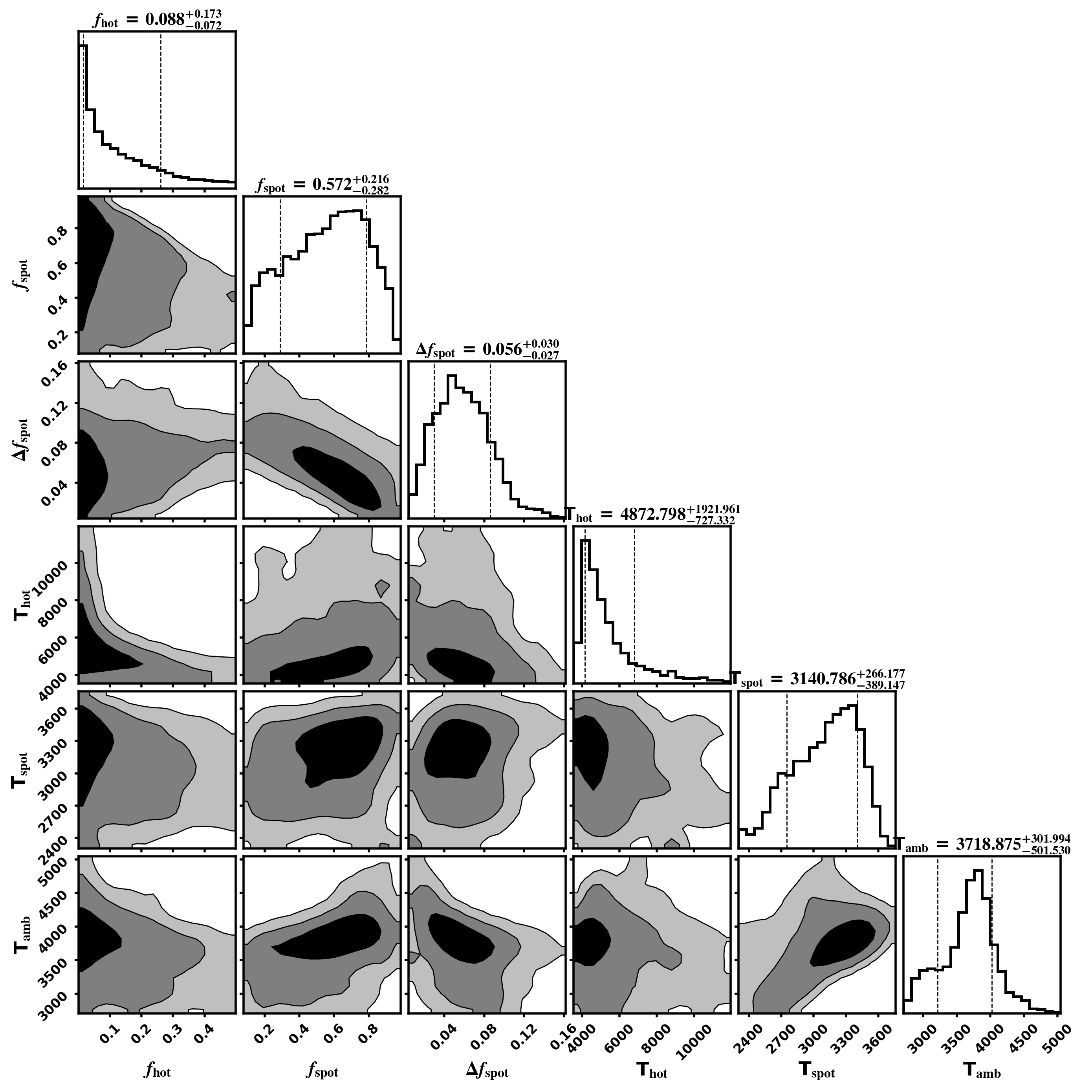}}
    \subfloat{\includegraphics[width=0.48\textwidth]{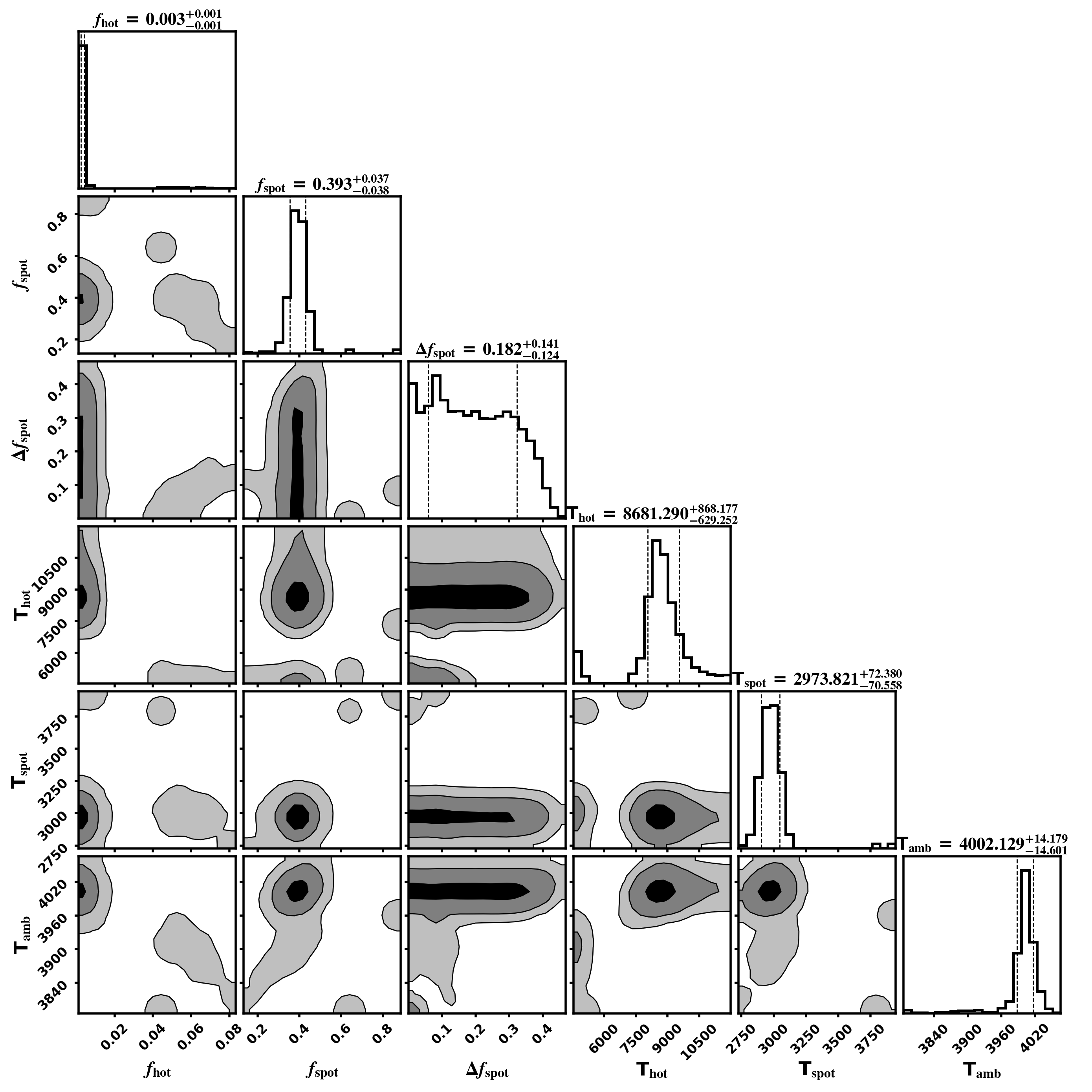}} \\
    \centering
    \subfloat{\includegraphics[width=0.68\textwidth]{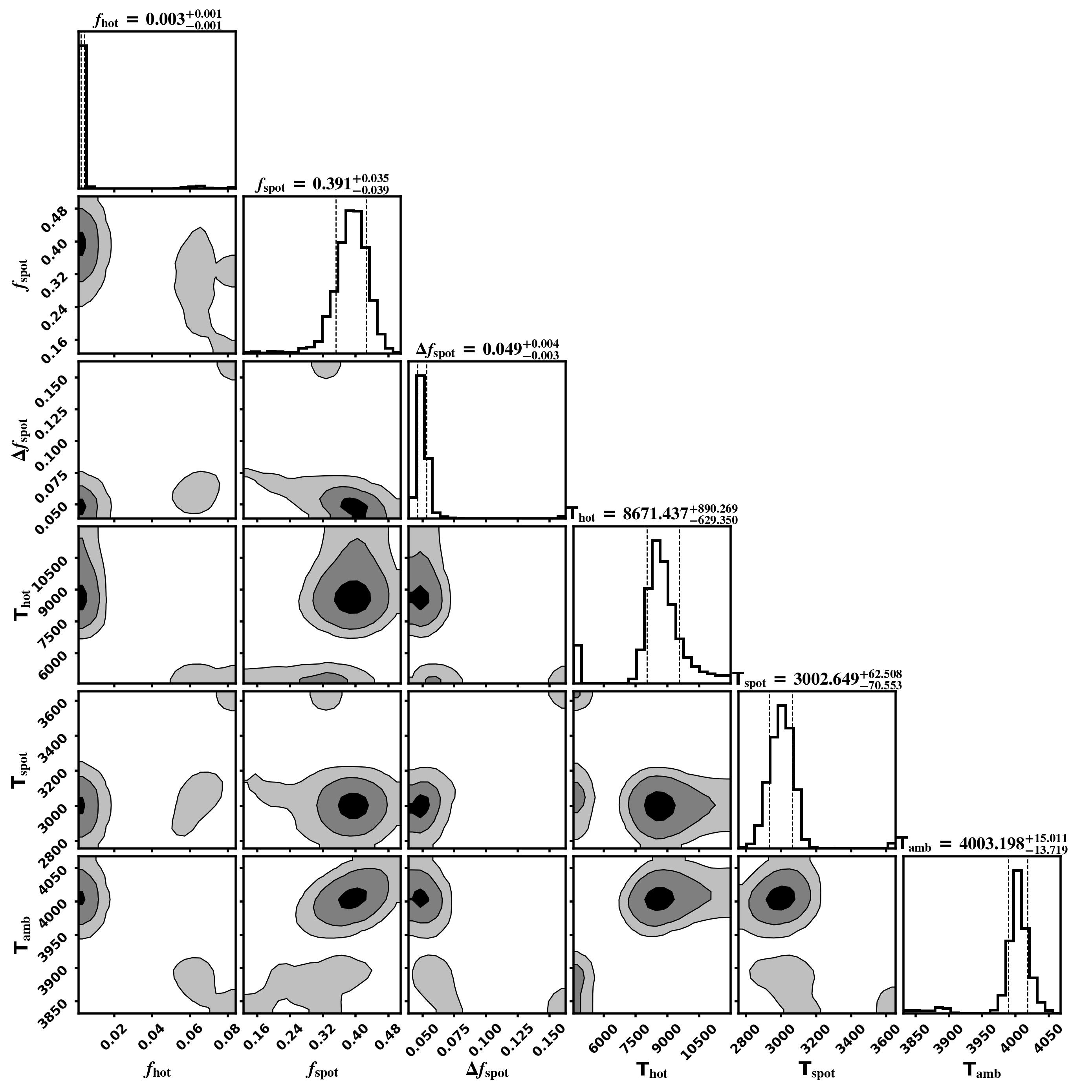}}
\caption{Posterior histograms for the variability (top left) and spectral models (top right), with the ensemble results (bottom) exhibiting what looks like the separate-model posteriors multiplied together.}
\label{f:Corner_Plots}
\end{figure*}

\begin{figure}
    \includegraphics[width=\columnwidth]{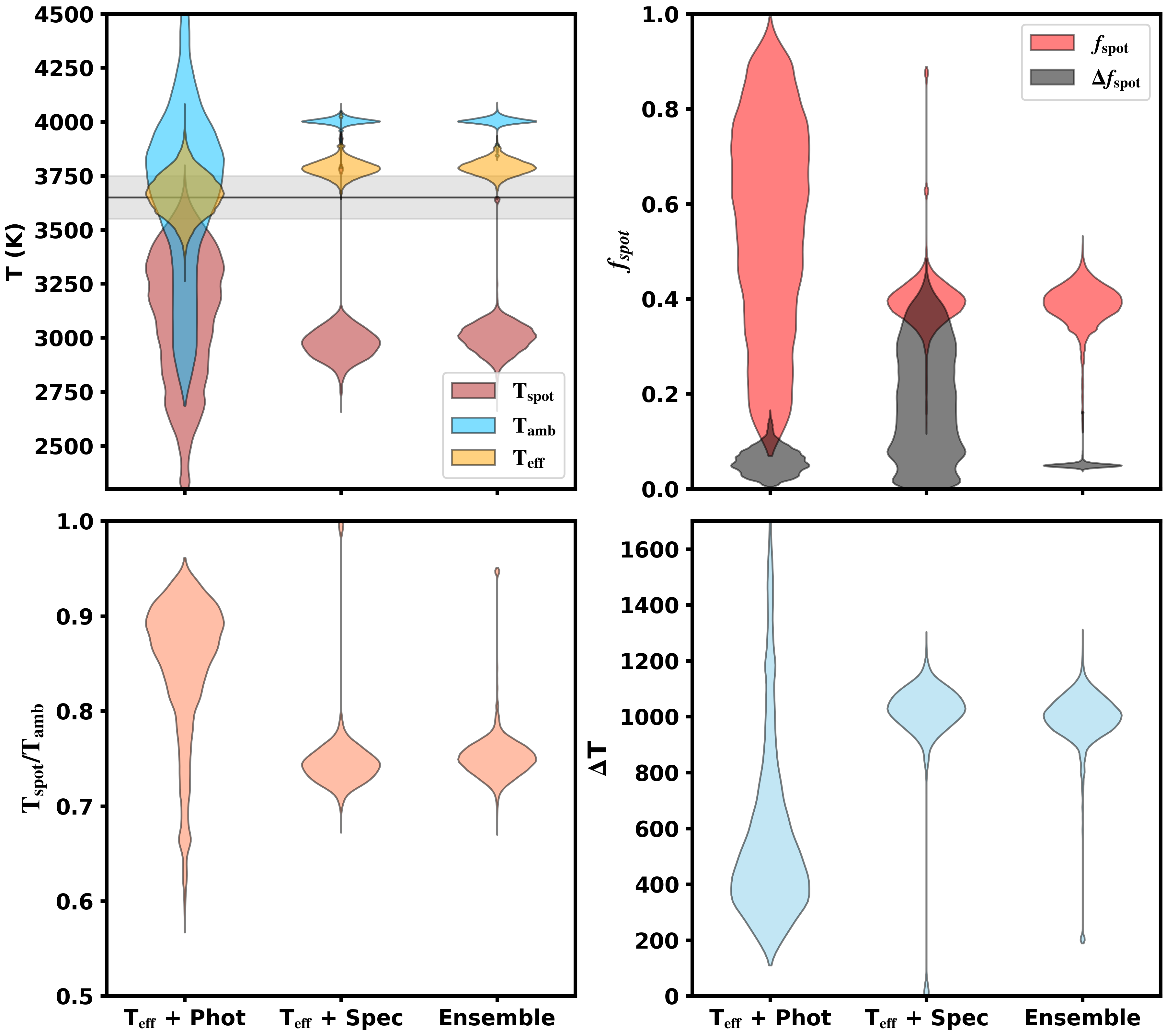}
\caption{Violin plot posteriors for the variability modeling (left distribution on each panel), spectral modeling (middle distribution), and the ensemble model (right distribution). Top left: temperature posteriors showing \Tspot, \Tamb, and \Teff. Top right: \fspot and \deltafspot\ posteriors. Bottom left: temperature ratio, \Tspot/\Tamb, and bottom right: $\Delta$T, the difference between the spotted and ambient temperatures. The ensemble results exhibit a narrower parameter space for spot characteristics which is effectively the product of the variability and spectral posteriors.}
\label{f:ensemble_violin}
\end{figure}

Results from our ensemble model, where all 5 photometric variability measurements are jointly modeled with the spectra of both visits, are shown in Figures \ref{f:samples}, \ref{f:Corner_Plots}, and \ref{f:ensemble_violin}, along with Table \ref{tab:final_results}. 
This model finds well-constrained spot characteristics of \fspot=\ensemblefspot, \deltafspot=\ensembledeltafspot, and \Tspot=\ensembleTspot\ K, with \Tamb=\ensembleTamb\ K.

Some degenerate solutions can be seen which prefer what looks like a fourth flux component between 3600-3900~K with a coverage fraction of 6-10$\%$.
This could be very weak evidence of either the spot penumbra or faculae, but attempts to extract that component were unsuccessful and can be pursued more in future work.

We can see that this ensemble model exhibits characteristics of both the photometry-only and spectra-only models.
The photometric variabilities strongly constrain \deltafspot\ and the temperature ratio but provide poor constraints on the quantity of spots.
Variability measurements in multiple wavebands across the optical-NIR constrain temperature contrast because the relative change in variability with wavelength is set by the ratio of spectral temperature components. Further into the red, the spectral components are more similar and variability decreases. At bluer wavelengths, variability reaches a maximum as the difference in spectra is greatest.

The spectral decomposition returns a precise estimate of \fspot, which is further constrained with the inclusion of photometric modeling.
The cold spot preference of the spectral models is balanced by the photometric limits on how cool the spots can be, given the contrast at longer wavelengths.
The spot coverage fraction is unconstrained when modeling the photometry due to degenerate observational effects between \fspot\ and \Tspot\, but is precisely constrained when spectral modeling is included.
The ensemble model results are primarily driven by the spectra, with the photometry being most important for the measurement of \deltafspot.

We argue that the consistency between measurements of the separate components based on different data-model combinations indicates that the results we report from the ensemble model are physically realistic.

\textbf{2-Temperature Models} agree with the 3-temperature results remarkably well, showing spot coverage of $41\pm3\%$ that changes throughout a rotation by $5.1\pm0.3\%$, with \Tspot=\ensembleTspottwoT\ and \Tamb=\ensembleTambtwoT.

\begin{table*}[t]
    \centering
    \begin{tabular}{cccc}
    \textbf{Parameter} & Photometry Model & Spectral Model & \textbf{Ensemble Model}\\
    \Tspot\ (K) & \PhotTspot  & \SpecTspot & \ensembleTspot \\
    \fspot	& \Photfspot & \Specfspot & \ensemblefspot \\
    \deltafspot & \Photdeltafspot & \Specdeltafspot  & \ensembledeltafspot \\
    \fhot	& \Photfhot & \Specfhot & \ensemblefhot \\
    \Thot\ (K) & \PhotThot & \SpecThot  & \ensembleThot \\
    \Tamb\ (K) & \PhotTamb & \SpecTamb  & \ensembleTamb \\
    \Teff\ (K) & \PhotTeff & \SpecTeff  & \ensembleTeff \\
    $\chi^2$ & \Photchisq & \Specchisq & \ensemblechisq \\
    \end{tabular}
    \caption{Parameters from fitting the different data combinations with a 3-temperature model. The photometry model uses the 5 photometric variability measurements, the spectral model uses 12 spectral orders from both visits. The ensemble model is applied to all the photometric and spectroscopic data. Results are broadly consistent with the 2-T model, most importantly recovering an approximately 3000K spot in either case. The results we recommend citing for AU Mic's surface components are the Ensemble Model results in the rightmost column of this table.}
    \label{tab:final_results}
\end{table*}

\subsection{Spot Contamination} \label{results: contamination}
\begin{figure*}[ht]
    \includegraphics[width=\textwidth]{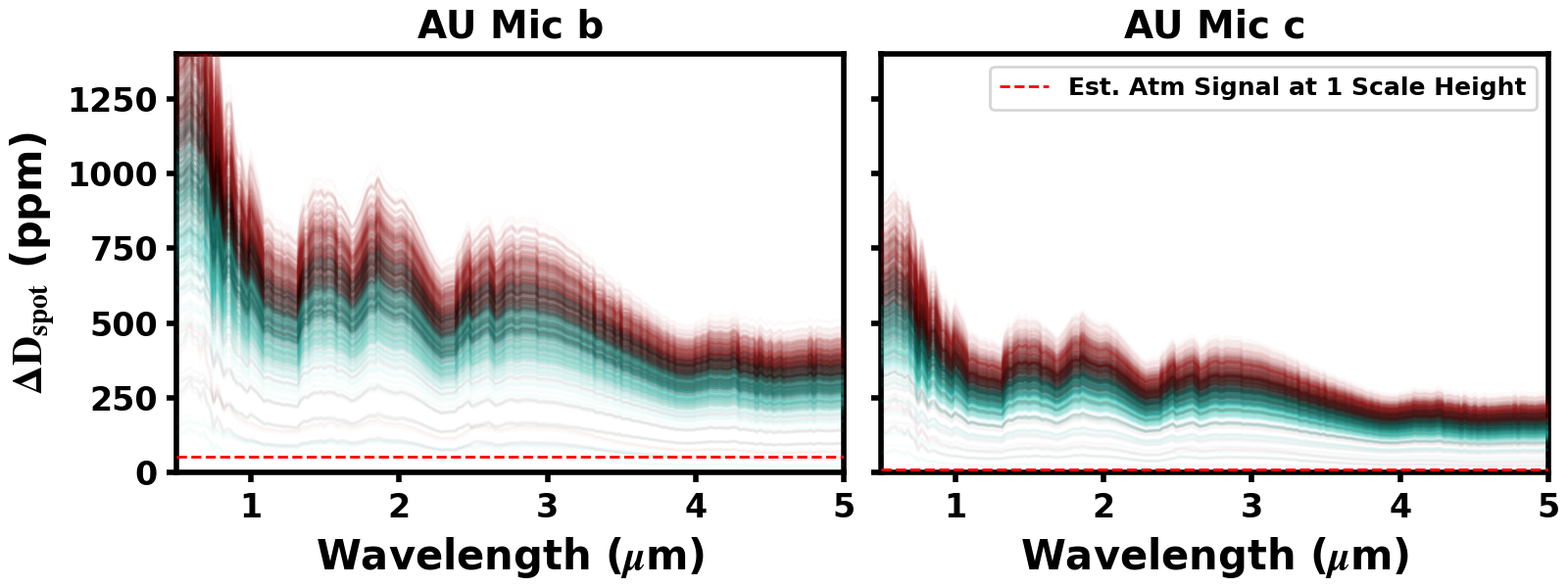}
\caption{Forward-modeled spot contamination in the transmission spectrum of AU Mic b (left) and AU Mic c (right) with models calculated using sampled parameters from the ensemble model posterior. Red corresponds to models with $f=$\fspot$+$\deltafspot\ (The point of maximum spot coverage and minimum flux throughout AU Mic's rotation), black to $f=$\fspot\, and turquoise to models with $f=$\fspot$-$\deltafspot.
Atmospheric depth estimates come from Equation \ref{eqn:atmospheric absorption} and are calculated to be 52 ppm for AU Mic b and 10ppm for AU Mic c at one scale height. For an optimistic case where we can measure 5 scale heights of a cloud-free atmosphere, AU Mic b's atmospheric features will be comparable to lower estimates of spot contamination, while AU Mic c's atmospheric features will still be a factor of a few below the lowest contamination scenarios.}
\label{f:final_contamination}
\end{figure*}

We forward-modeled spot contamination (Equation \ref{eqn: delta_D_spot}) under the assumption of $0\%$ spot coverage on the transit chord using ensemble model posteriors generated in the previous step, shown in Figure \ref{f:final_contamination}.
At short wavelengths like the TESS bandpass (0.6-1$\mu$m), contamination ranges from 550-1400 ppm.
In the WFC3 bandpass, contamination is between 500-750ppm, and decreases to 250-500ppm at longer wavelengths.
The transit depth of AU Mic b reported in \citet{Szabo2022} calculated from TESS and CHEOPS (0.33-1.1$\mu$m) transits translates to about 1875 ppm, so the spot contamination is 25-75$\%$ of that signal if $f_{\rm tra}=0$.
This means that without accounting for spot contamination, AU Mic b's true radius (3.46 R$_{\rm E}$ if the \citet{Szabo2022} value is uncontaminated) may be overestimated by as much as 0.5-1.7 Earth radii.
For AU Mic c, the \citet{Szabo2022} depth is about 980 ppm, with contamination calculated to be between 400-750 ppm.
This translates to an over-estimated planetary radius of between 0.6-1.3 Earth radii.

The magnitude of this shorter-wavelength contamination may be much different at different points in AU Mic's activity cycle and long term magnetic evolution, so these are rough estimates and contemporaneous measurements of transit depths across the optical to infrared would shed more light on the true radius of AU Mic b.

%%%%%%%%%%%%%%%%%%%%%%%%%%%%%%%%%%%%%%%%
% BEGIN DISCUSSION SECTION
%%%%%%%%%%%%%%%%%%%%%%%%%%%%%%%%%%%%%%%%
\section{Discussion} \label{discussion}

In this work, we present an analysis of broadband photometry and high-resolution spectroscopy to constrain the spot coverage fraction and temperature contrast on AU Mic.
This is crucial for addressing the spot contamination in the transmission spectrum of AU Mic b and therefore understanding its atmosphere.
With observational data of stellar effective temperature, multi-color time-series photometry, and high resolution stellar spectra, we parameterized the spectroscopic and photometric effects of starspots into spot coverage and temperature, with a well-constrained ambient temperature and tentative measurement of a hot component.

\begin{table*}[t]
    \centering
    \begin{tabular}{cccc}
    \textbf{Parameter} & Photometry Model & Spectral Model & Ensemble Model\\
    \Tspot\ (K) & \PhotTspottwoT  & \SpecTspottwoT & \ensembleTspottwoT \\
    \fspot	& \PhotfspottwoT & \SpecfspottwoT & \ensemblefspottwoT \\
    \deltafspot & \PhotdeltafspottwoT & \SpecdeltafspottwoT  & \ensembledeltafspottwoT \\
    \Tamb\ (K) & \PhotTambtwoT & \SpecTambtwoT  & \ensembleTambtwoT \\
    \Teff\ (K) & \PhotTefftwoT & \SpecTefftwoT  & \ensembleTefftwoT \\
    $\chi^2$ & \PhotchisqtwoT & \SpecchisqtwoT & \ensemblechisqtwoT \\
    \end{tabular}
    \caption{Parameters from fitting the different data combinations with a 2-temperature model. The photometry model uses the 5 photometric variability measurements, the spectral model uses 12 spectral orders from both visits. The ensemble model is applied to all the photometric and spectroscopic data. The primary difference compared to 3-T results can be seen in the temperatures measured for \Tspot\ and \Tamb.}
    \label{tab:twoT_final_results}
\end{table*}

\subsection{The Photometric Variability Spectrum}
Photometric variability has wavelength-dependent characteristic which can be used to constrain the primary flux components.
Amplitudes increase to some maximum at blue wavelengths as the flux contrast increases to 1, and decreases at redder wavelengths toward the Rayleigh-Jeans limit.
The shape and extent of the amplitude of rotational variability as a function of wavelength provides an important constraint on spot temperatures.
Because our measured variability amplitude is significantly less in $i'$ than $g'$ or $r'$, the spot temperature contrast is very sensitive to the magnitude and uncertainty of $i'$.
The $i'$ variability measurement thus carries more weight than any other single data point in this study, as it strongly constrains spot contrast and limits how cool the spots can be.
If the $i'$ measurement showed greater variability, this would allow spot temperature solutions to be cooler, as the contrast would be greater into the red.
The variability must instead decrease in this bandpass, which forces the range of solutions to be narrower and the spot contrast to be within a certain range.
The importance of this $i'$ measurement indicates that further photometric studies of spotted stars must be sure to include multiple bandpasses in the optical-to-infrared in order to precisely determine the wavelength regime where variability decrease or \textit{``turn-off"} happens.
The wavelengths where this turn-off is observed are highly descriptive of the stars spot spectra.

\subsection{Two Flux Components or Three?}
The primary results in this work are reported from the 3-temperature modeling, but we  examine how those results changed when modeled with only two temperatures, finding it returns tightly constrained and consistent measurements for \fspot, \deltafspot, \Tspot, and \Tamb.
When we allow a third component, it finds a poorly constrained temperature between 7000-10000 K, with a small number of solutions closer to the ambient photosphere, around 4000-5000 K. 
If the 7000-10000 K component is physical, it may be due to flares which are persistent enough to leave their flux in the median-averaged spectra.

Faculae on M dwarfs may be up to a few hundred K above the ambient photosphere \citep[e.g., ][]{Norris2023} and a small cluster of solutions in this temperature range stand apart from the primary results (seen in Figures \ref{f:samples} and \ref{f:Corner_Plots}).
This could be indicative of faculae but we did not recover that component in concerted attempts and its filling factor must be even less than the filling factor measured for the hotter temperature, detected at less than 0.5$\%$.

There is one key difference in outcome between 2- and 3-temperature models.
When modeling only \Teff\ and photometric variabilities, the 2-temperature model returns a significantly warmer \Tspot\ and a tighter ratio of \Tspot\ to \Tamb\ compared with the 3-temperature model. 
A third temperature, which only factors into the calculation of \Teff, brings spot temperature solutions into agreement with the other measurements we report, with larger uncertainties and a weaker (though still noticeable) relationship between spotted and unspotted temperatures.

Further investigation into multi-component rotation models and independent measurements of spot temperatures will be needed to help clarify when 2- or 3-temperature rotation models are most appropriate, and what cautions to impose on interpreting spot characteristics from photometric variabilities alone.

\subsection{Physical Interpretation of AU Mic's Spot Characteristics}
Much work has been done to measure and theoretically determine spot temperatures, distributions, and filling factors as a function of stellar type.
Still, observational evidence of consistent spot temperatures and precise filling factors for AU Mic is tenuous.
Spot characteristics are notoriously difficult to determine and different approaches can lead to inconsistent results.
Here we will discuss the results, limitations of this approach, and physical interpretation of AU Mic's starspots.

\paragraph{Spot Temperature}
We report a characteristic spot temperature for AU Mic of \Tspot$=$\ensembleTspot\ K.
A study of diatomic molecular lines in AU Mic's spot spectra found in \citet{Berdyugina2011} implies a $\Delta$T (\Teff-\Tspot) of 500-700 K, which would translate to spot temperatures of 3000-3200 K for a star with \Teff\ of 3700 K.
More recently, \citet{Ikuta2023} calculate \Tspot$=3140\pm64$K for AU Mic based on Equation 4 of \citet{Herbst2021}, which is based on the work done by \citet{Berdyugina2005}.
\citet{Rackham2018} suggest that for cool stars, \Tspot~can be estimated as 0.86$\times$T$_{\rm phot}$ (which we have labeled \Tamb\, in this work).
Given our modeled \Tamb\ of \ensembleTamb, the estimated spot temperature would be roughly 3300-3400K, which is over 200K warmer than our spectral and ensemble results but in agreement with our photometry-only model results. 
Recent work by \citet{Flagg2022} found evidence of a cold H$_2$ layer in AU Mic's photosphere with $1000\rm K<$\Tspot$<2400$K.
\citet{AframBerdyugina2019} use molecular lines to measure spot temperatures and find that M0 stars may have spot temperatures $\sim$2200K less than the stellar effective temperature, implying spots on AU Mic may have T$_{\rm{spot}}~\lesssim~1500$~K and T$_{\rm{spot}}$-to-T$_{\rm{eff}}$ ratios $\sim\leq0.4$.

Placed in context, the spot temperatures we measure in this work are much cooler than the $\Delta T=400$K or \Tspot=0.86\Tamb\ estimates for low mass stars and slightly cooler than (but consistent with) the \citet{Ikuta2023} and \citet{Berdyugina2011} estimates for AU Mic, warmer than some of the molecular and vibrational estimates for spots, but consistent with older measurements of spot temperatures on young active stars similar to AU Mic \citep[e.g., ][]{RamseyNations1980, Vogt1981ApJ...250..327V}.
The range of temperatures measured using different methods may be indicative that our assumption of spots as a feature with a single descriptive bulk temperature is breaking down, and we need to further develop our theoretical spot models to account for a temperature gradient.
If the spots on AU Mic have complex temperature profiles and extensive penumbrae, then the \fspot recovered in these models may be an overestimate of the true spot umbra and an under-estimate of the spot umbra + penumbra.

\paragraph{Spot Filling Factor}
Our models indicate that the surface of AU Mic is \fspot$ = $ \ensemblefspot\ covered in cool spots, with a change in spot coverage throughout a stellar rotation of \deltafspot$ = $ \ensembledeltafspot.
This estimate is within the very broad and uncertain range of possible spot coverage fractions measured for sun-like and cooler stars and indicates a heavily spotted stellar surface.
\citet{Yamashita2022} estimate \fspot\ of between 1-21\% for Zero Age Main Sequence stars using a fixed starspot temperature variability model, but such models are typically underestimates on \fspot\ \citep{Rackham2018,Apai2018}.
\citet{Cao2022} use APOGEE H-band spectra to measure average spot filling factors to be 0.248$\pm$0.005 for active stars in the Pleiades cluster and 0.03$\pm$0.008 for main sequence G and K stars in M67.
Other estimates of \fspot\ for low mass stars range from $<1\%$ to $50\%$ \citep[see Table 3 in ][]{Rackham2018}.
Our results are consistent with the range of expected coverage fractions, and our derived \deltafspot\ is consistent with values modeled by \citet{Libby-Roberts2022} of \deltafspot$\leq0.1$, though this is a much older and slowly rotating star.

\subsection{Caveats}
Here we will briefly describe some of the limitations of this work and possible directions for future spot studies.
\begin{itemize}
\item We use a simple sinusoidal rotation model, which is not necessarily the best choice for complex rotation curves like AU Mic's, but the photometry is not densely sampled and individual measurements have large uncertainties, so for the primary purpose of measuring the \textit{relative} amplitude of variability between filters we argue this is an appropriate model. 
\item AU Mic is a bright star, so ground-based differential photometry is difficult with LCO's field of view. Without similarly bright field stars, our photometry is very noisy, limiting the precision with which we can measure variability signals, especially in redder bandpasses.
\item Our spot contamination forward-models assume that $f_{\rm{tra}} = 0$, which means our contamination results are upper-limit cases for different sets of spot parameters. In reality, there may be some non-zero fraction of spot coverage on the transit chord, which will lower the contamination level in the transmission spectrum.
\item It is unclear what the source of the hot component is or whether it is truly a real signal, and the choice to include a third component affects the measurement of spot temperature from photometric variability models.
\item Our understanding of M star photosphere spectra is limited with the current generation of high-resolution synthetic spectral libraries like the \citet{Husseretal2013} PHOENIX models, described in detail by \citet{Iyer2020} and \citet{Rackham2023}. Choosing the best spectral template is a problem for cool stars generally, but describing AU Mic with these models is further complicated by its pre-main sequence age, a specific environment for which no truly appropriate model spectra yet exist. Alternative spectral models for M dwarfs exist, such as BT-SETTL \citep{Allard2011, Allard2014} or SPHINX \citep{Iyer2023}, but those are low- to mid- resolution libraries whereas the \citet{Husseretal2013} library has a resolution of R=100000-500000 and is more appropriate for the R=53000 echelle spectra we acquired. 

\end{itemize}

%%%%%%%%%%%%%%%%%%%%%%%%%%%%%%%%%%%%%%%%
% BEGIN CONCLUSIONS SECTION
%%%%%%%%%%%%%%%%%%%%%%%%%%%%%%%%%%%%%%%%
\section{Conclusions} \label{conclusions}

In this work, we have demonstrated the use in acquiring broadband photometry and high-resolution spectroscopy of only the star (without data acquired during transit) for the purpose of studying starspots and estimating their effect on transmission spectra. The results of our modeling lead us to the following conclusions:

\begin{itemize}

    \item At the time of these observations, AU Mic may have been \fspot=39$\pm4$\% covered in spots with a change of \deltafspot=6$\pm$1\% throughout a rotation. The spots have a bulk (flux-weighted, not distinguishing umbra from penumbra) temperature of \Tspot=\ensembleTspot\ K surrounded by a \Tamb=\ensembleTamb\ K photosphere. We found very weak evidence for facular coverage, and tentatively detect evidence of flux from flares with characteristic temperature \ensembleThot\ K. The fractional uncertainties measured on our final results for \Tamb\ and \Tspot\ come out to $0.25\%$ and $2\%$, respectively. \citet{Berardo2023} report theoretical precision limits on stellar spectral models with current instruments and spectral libraries, finding we can constrain photospheric spectra to $\geq0.2\%$ and spot spectra down to $1-5\%$, which translate to temperature uncertainties of $\leq0.05\%$ on \Tamb\ and $0.25-1.25\%$ on \Tspot\. If our results are physically accurate, we have measured the temperature of AU Mic's spots with a precision near but slightly worse than the \citet{Berardo2023} theoretical precision limit, and AU Mic's ambient photosphere temperature down to near its limit.

    \item \textbf{Spot contamination in transmission spectra for AU Mic b will be significant, adding between $250-1200$ppm contamination across the 0.5-5$\mu$m range}, overlapping with wavelengths where we measure planetary absorption features. This contamination is based on zero spot coverage on the transit chord, which could likely be an incorrect assumption if the large spot coverage fraction we measure is accurate.
    Nevertheless, spot contamination may be causing us to significantly overestimate the radii of AU Mic b and c. Determining the true planetary radii may require further acquisition and study of contemporaneous multi-color transit observations.

    \item From these measurements we calculate \Tamb-\Tspot$=$ 1000 K and \Tspot/\Tamb$=$ 0.75. Spots this cool should show up noticeably if occulted in transit if our measurements and interpretation of AU Mic's photosphere are correct, but none have yet been confirmed. With AU Mic's short period and large radius, it is possible that spots exist primarily or only at high latitudes, in which case spot crossings would be rarely or never observed.

    \item For stellar surfaces dominated by a single heterogeneity like spots (\textit{or} faculae, but the picture becomes more complicated with significant filling factors of both), \textbf{multi-color photometric variabilities provide significant constraints on the ratio of \Tspot/\Tamb (or T$_{\rm fac}$/\Tamb) and potentially accurate measurements of \Tspot}.
    Sampling the sensitive regions of a star's  \textit{variability spectrum}, even with broad photometric bandpasses, is useful for roughly estimating the temperatures of different spectral components.

    \item Modeling the data with either 2-T or 3-T spectral decomposition \textbf{did not significantly change measurements of \Tspot, \Tamb, or \deltafspot,} which indicates that these measurements are robust. As we are uncertain of the nature and significance of the hot component we detect, further investigation of the phase-resolved stellar spectrum may help distinguish the nature of the different flux components.
    
    \item Modeling only the variabilities provides a less accurate result for the coverage fraction or when only 2 temperatures are included, even though this only affects the calculation of \Teff.
    The 3-T \Teff\ and variability model returns an accurate (but highly uncertain) spot temperature while the 2-T \Teff\ and variability model prefers spots of nearly half the temperature.
    We recommend further investigation into the accuracy of measuring spot and/or facula temperatures with multi-color rotation modulations using both 2- and 3-temperature modeling.

    \item Improving measurements of spot characteristics on exoplanet host stars and understanding spot contamination is challenging but tractable with a multi-modal approach that covers a broad range of the visible-IR electromagnetic spectrum. As spot models and synthetic spectra improve, spot characteristics on our host stars will become much clearer.
\end{itemize}

%%%%%%%%%%%%%%%%%%%%%%%%%%%%%%%%%%%%%%%%
% Software and Acknowledgements
%%%%%%%%%%%%%%%%%%%%%%%%%%%%%%%%%%%%%%%%
\paragraph{Software} Python code used in this paper is available on the author's GitHub (\url{https://github.com/will-waalkes/AUMicTLSE}). This project made use of many publicly available tools and packages for which the authors are immensely grateful. In addition to the software cited throughout the paper, we also used \texttt{Astropy} \citep{2013A&A...558A..33A, 2018AJ....156..123A, 2022ApJ...935..167A}, \texttt{NumPy} \citep{numpy}, \texttt{Matplotlib} \citep{matplotlib}, \texttt{Pandas} \citep{pandas}, and Anaconda's \texttt{JupyterLab}.

\paragraph{Acknowledgements}

The authors thank the anonymous referee for their timely, thorough, and encouraging review which led to significant improvements to the manuscript.
This material is based upon work supported by the National Science Foundation Graduate Research Fellowship Program under Grant No. DGE-1650115.
Any opinions, findings, and conclusions or recommendations expressed in this material are those of the authors and do not necessarily reflect the views of the National Science Foundation.
This work makes use of observations from the LCOGT network.
WW thanks Jessica Libby-Roberts, Hannalore Gerling-Dunsmore, Dennis Tilipman, Girish Duvuuri, Ward Howard, John Monnier, Steve Cranmer, Gibor Basri, and Steve Vogt for conversations that helped aspects of this project.
WW and ZKBT were supported through STScI grant HST-GO-15788 and the NSF CAREER program (AST-1945633).
AWM was supported by grants from the NSF CAREER program (AST-2143763) and NASA's exoplanet research program (XRP 80NSSC21K0393).
ERN was supported by HST-GO-15836.

\bibliography{main.bib}

\begin{thebibliography}{}
\expandafter\ifx\csname natexlab\endcsname\relax\def\natexlab#1{#1}\fi
\providecommand{\url}[1]{\href{#1}{#1}}
\providecommand{\dodoi}[1]{doi:~\href{http://doi.org/#1}{\nolinkurl{#1}}}
\providecommand{\doeprint}[1]{\href{http://ascl.net/#1}{\nolinkurl{http://ascl.net/#1}}}
\providecommand{\doarXiv}[1]{\href{https://arxiv.org/abs/#1}{\nolinkurl{https://arxiv.org/abs/#1}}}

\bibitem[{{Afram} \& {Berdyugina}(2015)}]{AframBerdyugina2015}
{Afram}, N., \& {Berdyugina}, S.~V. 2015, \aap, 576, A34, \dodoi{10.1051/0004-6361/201425314}

\bibitem[{{Afram} \& {Berdyugina}(2019)}]{AframBerdyugina2019}
---. 2019, \aap, 629, A83, \dodoi{10.1051/0004-6361/201935793}

\bibitem[{{Ahrer} {et~al.}(2023){Ahrer}, {Stevenson}, {Mansfield}, {Moran}, {Brande}, {Morello}, {Murray}, {Nikolov}, {Petit dit de la Roche}, {Schlawin}, {Wheatley}, {Zieba}, {Batalha}, {Damiano}, {Goyal}, {Lendl}, {Lothringer}, {Mukherjee}, {Ohno}, {Batalha}, {Battley}, {Bean}, {Beatty}, {Benneke}, {Berta-Thompson}, {Carter}, {Cubillos}, {Daylan}, {Espinoza}, {Gao}, {Gibson}, {Gill}, {Harrington}, {Hu}, {Kreidberg}, {Lewis}, {Line}, {L{\'o}pez-Morales}, {Parmentier}, {Powell}, {Sing}, {Tsai}, {Wakeford}, {Welbanks}, {Alam}, {Alderson}, {Allen}, {Anderson}, {Barstow}, {Bayliss}, {Bell}, {Blecic}, {Bryant}, {Burleigh}, {Carone}, {Casewell}, {Changeat}, {Chubb}, {Crossfield}, {Crouzet}, {Decin}, {D{\'e}sert}, {Feinstein}, {Flagg}, {Fortney}, {Gizis}, {Heng}, {Iro}, {Kempton}, {Kendrew}, {Kirk}, {Knutson}, {Komacek}, {Lagage}, {Leconte}, {Lustig-Yaeger}, {MacDonald}, {Mancini}, {May}, {Mayne}, {Miguel}, {Mikal-Evans}, {Molaverdikhani}, {Palle}, {Piaulet}, {Rackham}, {Redfield}, {Rogers}, {Roy}, {Rustamkulov},
  {Shkolnik}, {Sotzen}, {Taylor}, {Tremblin}, {Tucker}, {Turner}, {de Val-Borro}, {Venot}, \& {Zhang}}]{Ahrer2023}
{Ahrer}, E.-M., {Stevenson}, K.~B., {Mansfield}, M., {et~al.} 2023, \nat, 614, 653, \dodoi{10.1038/s41586-022-05590-4}

\bibitem[{{Alderson} {et~al.}(2023){Alderson}, {Wakeford}, {Alam}, {Batalha}, {Lothringer}, {Adams Redai}, {Barat}, {Brande}, {Damiano}, {Daylan}, {Espinoza}, {Flagg}, {Goyal}, {Grant}, {Hu}, {Inglis}, {Lee}, {Mikal-Evans}, {Ramos-Rosado}, {Roy}, {Wallack}, {Batalha}, {Bean}, {Benneke}, {Berta-Thompson}, {Carter}, {Changeat}, {Col{\'o}n}, {Crossfield}, {D{\'e}sert}, {Foreman-Mackey}, {Gibson}, {Kreidberg}, {Line}, {L{\'o}pez-Morales}, {Molaverdikhani}, {Moran}, {Morello}, {Moses}, {Mukherjee}, {Schlawin}, {Sing}, {Stevenson}, {Taylor}, {Aggarwal}, {Ahrer}, {Allen}, {Barstow}, {Bell}, {Blecic}, {Casewell}, {Chubb}, {Crouzet}, {Cubillos}, {Decin}, {Feinstein}, {Fortney}, {Harrington}, {Heng}, {Iro}, {Kempton}, {Kirk}, {Knutson}, {Krick}, {Leconte}, {Lendl}, {MacDonald}, {Mancini}, {Mansfield}, {May}, {Mayne}, {Miguel}, {Nikolov}, {Ohno}, {Palle}, {Parmentier}, {Petit dit de la Roche}, {Piaulet}, {Powell}, {Rackham}, {Redfield}, {Rogers}, {Rustamkulov}, {Tan}, {Tremblin}, {Tsai}, {Turner}, {de Val-Borro},
  {Venot}, {Welbanks}, {Wheatley}, \& {Zhang}}]{Alderson2023}
{Alderson}, L., {Wakeford}, H.~R., {Alam}, M.~K., {et~al.} 2023, \nat, 614, 664, \dodoi{10.1038/s41586-022-05591-3}

\bibitem[{{Allard}(2014)}]{Allard2014}
{Allard}, F. 2014, in Exploring the Formation and Evolution of Planetary Systems, ed. M.~{Booth}, B.~C. {Matthews}, \& J.~R. {Graham}, Vol. 299, 271--272, \dodoi{10.1017/S1743921313008545}

\bibitem[{{Allard} {et~al.}(2011){Allard}, {Homeier}, \& {Freytag}}]{Allard2011}
{Allard}, F., {Homeier}, D., \& {Freytag}, B. 2011, in Astronomical Society of the Pacific Conference Series, Vol. 448, 16th Cambridge Workshop on Cool Stars, Stellar Systems, and the Sun, ed. C.~{Johns-Krull}, M.~K. {Browning}, \& A.~A. {West}, 91, \dodoi{10.48550/arXiv.1011.5405}

\bibitem[{{Allard} {et~al.}(2012){Allard}, {Homeier}, {Freytag}, \& {Sharp}}]{Allard2012}
{Allard}, F., {Homeier}, D., {Freytag}, B., \& {Sharp}, C.~M. 2012, in EAS Publications Series, Vol.~57, EAS Publications Series, ed. C.~{Reyl{\'e}}, C.~{Charbonnel}, \& M.~{Schultheis}, 3--43, \dodoi{10.1051/eas/1257001}

\bibitem[{{Angus} {et~al.}(2018){Angus}, {Morton}, {Aigrain}, {Foreman-Mackey}, \& {Rajpaul}}]{Angus2018}
{Angus}, R., {Morton}, T., {Aigrain}, S., {Foreman-Mackey}, D., \& {Rajpaul}, V. 2018, \mnras, 474, 2094, \dodoi{10.1093/mnras/stx2109}

\bibitem[{{Apai} {et~al.}(2018){Apai}, {Rackham}, {Giampapa}, {Angerhausen}, {Teske}, {Barstow}, {Carone}, {Cegla}, {Domagal-Goldman}, {Espinoza}, {Giles}, {Gully-Santiago}, {Haywood}, {Hu}, {Jordan}, {Kreidberg}, {Line}, {Llama}, {L{\'o}pez-Morales}, {Marley}, \& {de Wit}}]{Apai2018}
{Apai}, D., {Rackham}, B.~V., {Giampapa}, M.~S., {et~al.} 2018, arXiv e-prints, arXiv:1803.08708.
\newblock \doarXiv{1803.08708}

\bibitem[{{Astropy Collaboration} {et~al.}(2013){Astropy Collaboration}, {Robitaille}, {Tollerud}, {Greenfield}, {Droettboom}, {Bray}, {Aldcroft}, {Davis}, {Ginsburg}, {Price-Whelan}, {Kerzendorf}, {Conley}, {Crighton}, {Barbary}, {Muna}, {Ferguson}, {Grollier}, {Parikh}, {Nair}, {Unther}, {Deil}, {Woillez}, {Conseil}, {Kramer}, {Turner}, {Singer}, {Fox}, {Weaver}, {Zabalza}, {Edwards}, {Azalee Bostroem}, {Burke}, {Casey}, {Crawford}, {Dencheva}, {Ely}, {Jenness}, {Labrie}, {Lim}, {Pierfederici}, {Pontzen}, {Ptak}, {Refsdal}, {Servillat}, \& {Streicher}}]{2013A&A...558A..33A}
{Astropy Collaboration}, {Robitaille}, T.~P., {Tollerud}, E.~J., {et~al.} 2013, \aap, 558, A33, \dodoi{10.1051/0004-6361/201322068}

\bibitem[{{Astropy Collaboration} {et~al.}(2018){Astropy Collaboration}, {Price-Whelan}, {Sip{\H{o}}cz}, {G{\"u}nther}, {Lim}, {Crawford}, {Conseil}, {Shupe}, {Craig}, {Dencheva}, {Ginsburg}, {VanderPlas}, {Bradley}, {P{\'e}rez-Su{\'a}rez}, {de Val-Borro}, {Aldcroft}, {Cruz}, {Robitaille}, {Tollerud}, {Ardelean}, {Babej}, {Bach}, {Bachetti}, {Bakanov}, {Bamford}, {Barentsen}, {Barmby}, {Baumbach}, {Berry}, {Biscani}, {Boquien}, {Bostroem}, {Bouma}, {Brammer}, {Bray}, {Breytenbach}, {Buddelmeijer}, {Burke}, {Calderone}, {Cano Rodr{\'\i}guez}, {Cara}, {Cardoso}, {Cheedella}, {Copin}, {Corrales}, {Crichton}, {D'Avella}, {Deil}, {Depagne}, {Dietrich}, {Donath}, {Droettboom}, {Earl}, {Erben}, {Fabbro}, {Ferreira}, {Finethy}, {Fox}, {Garrison}, {Gibbons}, {Goldstein}, {Gommers}, {Greco}, {Greenfield}, {Groener}, {Grollier}, {Hagen}, {Hirst}, {Homeier}, {Horton}, {Hosseinzadeh}, {Hu}, {Hunkeler}, {Ivezi{\'c}}, {Jain}, {Jenness}, {Kanarek}, {Kendrew}, {Kern}, {Kerzendorf}, {Khvalko}, {King}, {Kirkby}, {Kulkarni},
  {Kumar}, {Lee}, {Lenz}, {Littlefair}, {Ma}, {Macleod}, {Mastropietro}, {McCully}, {Montagnac}, {Morris}, {Mueller}, {Mumford}, {Muna}, {Murphy}, {Nelson}, {Nguyen}, {Ninan}, {N{\"o}the}, {Ogaz}, {Oh}, {Parejko}, {Parley}, {Pascual}, {Patil}, {Patil}, {Plunkett}, {Prochaska}, {Rastogi}, {Reddy Janga}, {Sabater}, {Sakurikar}, {Seifert}, {Sherbert}, {Sherwood-Taylor}, {Shih}, {Sick}, {Silbiger}, {Singanamalla}, {Singer}, {Sladen}, {Sooley}, {Sornarajah}, {Streicher}, {Teuben}, {Thomas}, {Tremblay}, {Turner}, {Terr{\'o}n}, {van Kerkwijk}, {de la Vega}, {Watkins}, {Weaver}, {Whitmore}, {Woillez}, {Zabalza}, \& {Astropy Contributors}}]{2018AJ....156..123A}
{Astropy Collaboration}, {Price-Whelan}, A.~M., {Sip{\H{o}}cz}, B.~M., {et~al.} 2018, \aj, 156, 123, \dodoi{10.3847/1538-3881/aabc4f}

\bibitem[{{Astropy Collaboration} {et~al.}(2022){Astropy Collaboration}, {Price-Whelan}, {Lim}, {Earl}, {Starkman}, {Bradley}, {Shupe}, {Patil}, {Corrales}, {Brasseur}, {N{\"o}the}, {Donath}, {Tollerud}, {Morris}, {Ginsburg}, {Vaher}, {Weaver}, {Tocknell}, {Jamieson}, {van Kerkwijk}, {Robitaille}, {Merry}, {Bachetti}, {G{\"u}nther}, {Aldcroft}, {Alvarado-Montes}, {Archibald}, {B{\'o}di}, {Bapat}, {Barentsen}, {Baz{\'a}n}, {Biswas}, {Boquien}, {Burke}, {Cara}, {Cara}, {Conroy}, {Conseil}, {Craig}, {Cross}, {Cruz}, {D'Eugenio}, {Dencheva}, {Devillepoix}, {Dietrich}, {Eigenbrot}, {Erben}, {Ferreira}, {Foreman-Mackey}, {Fox}, {Freij}, {Garg}, {Geda}, {Glattly}, {Gondhalekar}, {Gordon}, {Grant}, {Greenfield}, {Groener}, {Guest}, {Gurovich}, {Handberg}, {Hart}, {Hatfield-Dodds}, {Homeier}, {Hosseinzadeh}, {Jenness}, {Jones}, {Joseph}, {Kalmbach}, {Karamehmetoglu}, {Ka{\l}uszy{\'n}ski}, {Kelley}, {Kern}, {Kerzendorf}, {Koch}, {Kulumani}, {Lee}, {Ly}, {Ma}, {MacBride}, {Maljaars}, {Muna}, {Murphy}, {Norman},
  {O'Steen}, {Oman}, {Pacifici}, {Pascual}, {Pascual-Granado}, {Patil}, {Perren}, {Pickering}, {Rastogi}, {Roulston}, {Ryan}, {Rykoff}, {Sabater}, {Sakurikar}, {Salgado}, {Sanghi}, {Saunders}, {Savchenko}, {Schwardt}, {Seifert-Eckert}, {Shih}, {Jain}, {Shukla}, {Sick}, {Simpson}, {Singanamalla}, {Singer}, {Singhal}, {Sinha}, {Sip{\H{o}}cz}, {Spitler}, {Stansby}, {Streicher}, {{\v{S}}umak}, {Swinbank}, {Taranu}, {Tewary}, {Tremblay}, {de Val-Borro}, {Van Kooten}, {Vasovi{\'c}}, {Verma}, {de Miranda Cardoso}, {Williams}, {Wilson}, {Winkel}, {Wood-Vasey}, {Xue}, {Yoachim}, {Zhang}, {Zonca}, \& {Astropy Project Contributors}}]{2022ApJ...935..167A}
{Astropy Collaboration}, {Price-Whelan}, A.~M., {Lim}, P.~L., {et~al.} 2022, \apj, 935, 167, \dodoi{10.3847/1538-4357/ac7c74}

\bibitem[{{Barclay} {et~al.}(2023){Barclay}, {Sheppard}, {Latouf}, {Mandell}, {Quintana}, {Gilbert}, {Liuzzi}, {Villanueva}, {Arney}, {Brande}, {Col{\'o}n}, {Covone}, {Crossfield}, {Damiano}, {Domagal-Goldman}, {Fauchez}, {Fiscale}, {Gallo}, {Hedges}, {Hu}, {Kite}, {Koll}, {Kopparapu}, {Kostov}, {Kreidberg}, {Lopez}, {Mang}, {Morley}, {Mullally}, {Mullally}, {Pidhorodetska}, {Schlieder}, {Vega}, {Youngblood}, \& {Zieba}}]{Barclay2023}
{Barclay}, T., {Sheppard}, K.~B., {Latouf}, N., {et~al.} 2023, arXiv e-prints, arXiv:2301.10866, \dodoi{10.48550/arXiv.2301.10866}

\bibitem[{Basri(2021)}]{Basri2021}
Basri, G. 2021, An Introduction to Stellar Magnetic Activity, 2514-3433 (IOP Publishing), \dodoi{10.1088/2514-3433/ac2956}

\bibitem[{{Berardo} {et~al.}(2023){Berardo}, {de Wit}, \& {Rackham}}]{Berardo2023}
{Berardo}, D., {de Wit}, J., \& {Rackham}, B.~V. 2023, arXiv e-prints, arXiv:2307.04785, \dodoi{10.48550/arXiv.2307.04785}

\bibitem[{{Berdyugina}(2005)}]{Berdyugina2005}
{Berdyugina}, S.~V. 2005, Living Reviews in Solar Physics, 2, 8, \dodoi{10.12942/lrsp-2005-8}

\bibitem[{{Berdyugina} {et~al.}(2011){Berdyugina}, {Berdyugin}, {Fluri}, \& {Piirola}}]{Berdyugina2011}
{Berdyugina}, S.~V., {Berdyugin}, A.~V., {Fluri}, D.~M., \& {Piirola}, V. 2011, \apjl, 728, L6, \dodoi{10.1088/2041-8205/728/1/L6}

\bibitem[{{Berta} {et~al.}(2012){Berta}, {Charbonneau}, {D{\'e}sert}, {Miller-Ricci Kempton}, {McCullough}, {Burke}, {Fortney}, {Irwin}, {Nutzman}, \& {Homeier}}]{Berta2012}
{Berta}, Z.~K., {Charbonneau}, D., {D{\'e}sert}, J.-M., {et~al.} 2012, \apj, 747, 35, \dodoi{10.1088/0004-637X/747/1/35}

\bibitem[{{Brown} {et~al.}(2001){Brown}, {Charbonneau}, {Gilliland}, {Noyes}, \& {Burrows}}]{Brown2001}
{Brown}, T.~M., {Charbonneau}, D., {Gilliland}, R.~L., {Noyes}, R.~W., \& {Burrows}, A. 2001, \apj, 552, 699, \dodoi{10.1086/320580}

\bibitem[{{Cao} {et~al.}(2022){Cao}, {Pinsonneault}, {Hillenbrand}, \& {Kuhn}}]{Cao2022}
{Cao}, L., {Pinsonneault}, M.~H., {Hillenbrand}, L.~A., \& {Kuhn}, M.~A. 2022, \apj, 924, 84, \dodoi{10.3847/1538-4357/ac307f}

\bibitem[{{Chen} {et~al.}(2005){Chen}, {Patten}, {Werner}, {Dowell}, {Stapelfeldt}, {Song}, {Stauffer}, {Blaylock}, {Gordon}, \& {Krause}}]{Chen2005b}
{Chen}, C.~H., {Patten}, B.~M., {Werner}, M.~W., {et~al.} 2005, \apj, 634, 1372, \dodoi{10.1086/497124}

\bibitem[{{Collins} {et~al.}(2017){Collins}, {Kielkopf}, {Stassun}, \& {Hessman}}]{Collins2017}
{Collins}, K.~A., {Kielkopf}, J.~F., {Stassun}, K.~G., \& {Hessman}, F.~V. 2017, \aj, 153, 77, \dodoi{10.3847/1538-3881/153/2/77}

\bibitem[{{Cristofari} {et~al.}(2023){Cristofari}, {Donati}, {Folsom}, {Masseron}, {Fouqu{\'e}}, {Moutou}, {Artigau}, {Carmona}, {Petit}, {Delfosse}, {Martioli}, \& {the SLS consortium}}]{Cristofari2023}
{Cristofari}, P.~I., {Donati}, J.~F., {Folsom}, C.~P., {et~al.} 2023, \mnras, 522, 1342, \dodoi{10.1093/mnras/stad865}

\bibitem[{{Donati} {et~al.}(2023){Donati}, {Cristofari}, {Finociety}, {Klein}, {Moutou}, {Gaidos}, {Cadieux}, {Artigau}, {Correia}, {Bou{\'e}}, {Cook}, {Carmona}, {Lehmann}, {Bouvier}, {Martioli}, {Morin}, {Fouqu{\'e}}, {Delfosse}, {Doyon}, {H{\'e}brard}, {Alencar}, {Laskar}, {Arnold}, {Petit}, {K{\'o}sp{\'a}l}, {Vidotto}, {Folsom}, \& {collaboration}}]{Donati2023}
{Donati}, J.~F., {Cristofari}, P.~I., {Finociety}, B., {et~al.} 2023, \mnras, 525, 455, \dodoi{10.1093/mnras/stad1193}

\bibitem[{{Feinstein} {et~al.}(2022){Feinstein}, {France}, {Youngblood}, {Duvvuri}, {Teal}, {Cauley}, {Seligman}, {Gaidos}, {Kempton}, {Bean}, {Diamond-Lowe}, {Newton}, {Ginzburg}, {Plavchan}, {Gao}, \& {Schlichting}}]{Feinstein2022a}
{Feinstein}, A.~D., {France}, K., {Youngblood}, A., {et~al.} 2022, \aj, 164, 110, \dodoi{10.3847/1538-3881/ac8107}

\bibitem[{{Feinstein} {et~al.}(2023){Feinstein}, {Radica}, {Welbanks}, {Murray}, {Ohno}, {Coulombe}, {Espinoza}, {Bean}, {Teske}, {Benneke}, {Line}, {Rustamkulov}, {Saba}, {Tsiaras}, {Barstow}, {Fortney}, {Gao}, {Knutson}, {MacDonald}, {Mikal-Evans}, {Rackham}, {Taylor}, {Parmentier}, {Batalha}, {Berta-Thompson}, {Carter}, {Changeat}, {dos Santos}, {Gibson}, {Goyal}, {Kreidberg}, {L{\'o}pez-Morales}, {Lothringer}, {Miguel}, {Molaverdikhani}, {Moran}, {Morello}, {Mukherjee}, {Sing}, {Stevenson}, {Wakeford}, {Ahrer}, {Alam}, {Alderson}, {Allen}, {Batalha}, {Bell}, {Blecic}, {Brande}, {Caceres}, {Casewell}, {Chubb}, {Crossfield}, {Crouzet}, {Cubillos}, {Decin}, {D{\'e}sert}, {Harrington}, {Heng}, {Henning}, {Iro}, {Kempton}, {Kendrew}, {Kirk}, {Krick}, {Lagage}, {Lendl}, {Mancini}, {Mansfield}, {May}, {Mayne}, {Nikolov}, {Palle}, {Petit dit de la Roche}, {Piaulet}, {Powell}, {Redfield}, {Rogers}, {Roman}, {Roy}, {Nixon}, {Schlawin}, {Tan}, {Tremblin}, {Turner}, {Venot}, {Waalkes}, {Wheatley}, \&
  {Zhang}}]{Feinstein2023}
{Feinstein}, A.~D., {Radica}, M., {Welbanks}, L., {et~al.} 2023, \nat, 614, 670, \dodoi{10.1038/s41586-022-05674-1}

\bibitem[{{Flagg} {et~al.}(2022){Flagg}, {Johns-Krull}, {France}, {Herczeg}, {Najita}, {Youngblood}, {Carvalho}, {Carptenter}, {Kenyon}, {Newton}, \& {Rockcliffe}}]{Flagg2022}
{Flagg}, L., {Johns-Krull}, C.~M., {France}, K., {et~al.} 2022, \apj, 934, 8, \dodoi{10.3847/1538-4357/ac7643}

\bibitem[{{Foreman-Mackey} {et~al.}(2013){Foreman-Mackey}, {Hogg}, {Lang}, \& {Goodman}}]{Foreman-Mackey2012}
{Foreman-Mackey}, D., {Hogg}, D.~W., {Lang}, D., \& {Goodman}, J. 2013, \pasp, 125, 306, \dodoi{10.1086/670067}

\bibitem[{{Fu} {et~al.}(2022){Fu}, {Espinoza}, {Sing}, {Lothringer}, {Dos Santos}, {Rustamkulov}, {Deming}, {Kempton}, {Komacek}, {Knutson}, {Albert}, {Pontoppidan}, {Volk}, \& {Filippazzo}}]{Fu2022}
{Fu}, G., {Espinoza}, N., {Sing}, D.~K., {et~al.} 2022, \apjl, 940, L35, \dodoi{10.3847/2041-8213/ac9977}

\bibitem[{{Gaia Collaboration} {et~al.}(2023){Gaia Collaboration}, {Vallenari}, {Brown}, {Prusti}, {de Bruijne}, {Arenou}, {Babusiaux}, {Biermann}, {Creevey}, {Ducourant}, {Evans}, {Eyer}, {Guerra}, {Hutton}, {Jordi}, {Klioner}, {Lammers}, {Lindegren}, {Luri}, {Mignard}, {Panem}, {Pourbaix}, {Randich}, {Sartoretti}, {Soubiran}, {Tanga}, {Walton}, {Bailer-Jones}, {Bastian}, {Drimmel}, {Jansen}, {Katz}, {Lattanzi}, {van Leeuwen}, {Bakker}, {Cacciari}, {Casta{\~n}eda}, {De Angeli}, {Fabricius}, {Fouesneau}, {Fr{\'e}mat}, {Galluccio}, {Guerrier}, {Heiter}, {Masana}, {Messineo}, {Mowlavi}, {Nicolas}, {Nienartowicz}, {Pailler}, {Panuzzo}, {Riclet}, {Roux}, {Seabroke}, {Sordo}, {Th{\'e}venin}, {Gracia-Abril}, {Portell}, {Teyssier}, {Altmann}, {Andrae}, {Audard}, {Bellas-Velidis}, {Benson}, {Berthier}, {Blomme}, {Burgess}, {Busonero}, {Busso}, {C{\'a}novas}, {Carry}, {Cellino}, {Cheek}, {Clementini}, {Damerdji}, {Davidson}, {de Teodoro}, {Nu{\~n}ez Campos}, {Delchambre}, {Dell'Oro}, {Esquej},
  {Fern{\'a}ndez-Hern{\'a}ndez}, {Fraile}, {Garabato}, {Garc{\'\i}a-Lario}, {Gosset}, {Haigron}, {Halbwachs}, {Hambly}, {Harrison}, {Hern{\'a}ndez}, {Hestroffer}, {Hodgkin}, {Holl}, {Jan{\ss}en}, {Jevardat de Fombelle}, {Jordan}, {Krone-Martins}, {Lanzafame}, {L{\"o}ffler}, {Marchal}, {Marrese}, {Moitinho}, {Muinonen}, {Osborne}, {Pancino}, {Pauwels}, {Recio-Blanco}, {Reyl{\'e}}, {Riello}, {Rimoldini}, {Roegiers}, {Rybizki}, {Sarro}, {Siopis}, {Smith}, {Sozzetti}, {Utrilla}, {van Leeuwen}, {Abbas}, {{\'A}brah{\'a}m}, {Abreu Aramburu}, {Aerts}, {Aguado}, {Ajaj}, {Aldea-Montero}, {Altavilla}, {{\'A}lvarez}, {Alves}, {Anders}, {Anderson}, {Anglada Varela}, {Antoja}, {Baines}, {Baker}, {Balaguer-N{\'u}{\~n}ez}, {Balbinot}, {Balog}, {Barache}, {Barbato}, {Barros}, {Barstow}, {Bartolom{\'e}}, {Bassilana}, {Bauchet}, {Becciani}, {Bellazzini}, {Berihuete}, {Bernet}, {Bertone}, {Bianchi}, {Binnenfeld}, {Blanco-Cuaresma}, {Blazere}, {Boch}, {Bombrun}, {Bossini}, {Bouquillon}, {Bragaglia}, {Bramante}, {Breedt},
  {Bressan}, {Brouillet}, {Brugaletta}, {Bucciarelli}, {Burlacu}, {Butkevich}, {Buzzi}, {Caffau}, {Cancelliere}, {Cantat-Gaudin}, {Carballo}, {Carlucci}, {Carnerero}, {Carrasco}, {Casamiquela}, {Castellani}, {Castro-Ginard}, {Chaoul}, {Charlot}, {Chemin}, {Chiaramida}, {Chiavassa}, {Chornay}, {Comoretto}, {Contursi}, {Cooper}, {Cornez}, {Cowell}, {Crifo}, {Cropper}, {Crosta}, {Crowley}, {Dafonte}, {Dapergolas}, {David}, {David}, {de Laverny}, {De Luise}, {De March}, {De Ridder}, {de Souza}, {de Torres}, {del Peloso}, {del Pozo}, {Delbo}, {Delgado}, {Delisle}, {Demouchy}, {Dharmawardena}, {Di Matteo}, {Diakite}, {Diener}, {Distefano}, {Dolding}, {Edvardsson}, {Enke}, {Fabre}, {Fabrizio}, {Faigler}, {Fedorets}, {Fernique}, {Fienga}, {Figueras}, {Fournier}, {Fouron}, {Fragkoudi}, {Gai}, {Garcia-Gutierrez}, {Garcia-Reinaldos}, {Garc{\'\i}a-Torres}, {Garofalo}, {Gavel}, {Gavras}, {Gerlach}, {Geyer}, {Giacobbe}, {Gilmore}, {Girona}, {Giuffrida}, {Gomel}, {Gomez}, {Gonz{\'a}lez-N{\'u}{\~n}ez},
  {Gonz{\'a}lez-Santamar{\'\i}a}, {Gonz{\'a}lez-Vidal}, {Granvik}, {Guillout}, {Guiraud}, {Guti{\'e}rrez-S{\'a}nchez}, {Guy}, {Hatzidimitriou}, {Hauser}, {Haywood}, {Helmer}, {Helmi}, {Sarmiento}, {Hidalgo}, {Hilger}, {H{\l}adczuk}, {Hobbs}, {Holland}, {Huckle}, {Jardine}, {Jasniewicz}, {Jean-Antoine Piccolo}, {Jim{\'e}nez-Arranz}, {Jorissen}, {Juaristi Campillo}, {Julbe}, {Karbevska}, {Kervella}, {Khanna}, {Kontizas}, {Kordopatis}, {Korn}, {K{\'o}sp{\'a}l}, {Kostrzewa-Rutkowska}, {Kruszy{\'n}ska}, {Kun}, {Laizeau}, {Lambert}, {Lanza}, {Lasne}, {Le Campion}, {Lebreton}, {Lebzelter}, {Leccia}, {Leclerc}, {Lecoeur-Taibi}, {Liao}, {Licata}, {Lindstr{\o}m}, {Lister}, {Livanou}, {Lobel}, {Lorca}, {Loup}, {Madrero Pardo}, {Magdaleno Romeo}, {Managau}, {Mann}, {Manteiga}, {Marchant}, {Marconi}, {Marcos}, {Marcos Santos}, {Mar{\'\i}n Pina}, {Marinoni}, {Marocco}, {Marshall}, {Martin Polo}, {Mart{\'\i}n-Fleitas}, {Marton}, {Mary}, {Masip}, {Massari}, {Mastrobuono-Battisti}, {Mazeh}, {McMillan}, {Messina}, {Michalik},
  {Millar}, {Mints}, {Molina}, {Molinaro}, {Moln{\'a}r}, {Monari}, {Mongui{\'o}}, {Montegriffo}, {Montero}, {Mor}, {Mora}, {Morbidelli}, {Morel}, {Morris}, {Muraveva}, {Murphy}, {Musella}, {Nagy}, {Noval}, {Oca{\~n}a}, {Ogden}, {Ordenovic}, {Osinde}, {Pagani}, {Pagano}, {Palaversa}, {Palicio}, {Pallas-Quintela}, {Panahi}, {Payne-Wardenaar}, {Pe{\~n}alosa Esteller}, {Penttil{\"a}}, {Pichon}, {Piersimoni}, {Pineau}, {Plachy}, {Plum}, {Poggio}, {Pr{\v{s}}a}, {Pulone}, {Racero}, {Ragaini}, {Rainer}, {Raiteri}, {Rambaux}, {Ramos}, {Ramos-Lerate}, {Re Fiorentin}, {Regibo}, {Richards}, {Rios Diaz}, {Ripepi}, {Riva}, {Rix}, {Rixon}, {Robichon}, {Robin}, {Robin}, {Roelens}, {Rogues}, {Rohrbasser}, {Romero-G{\'o}mez}, {Rowell}, {Royer}, {Ruz Mieres}, {Rybicki}, {Sadowski}, {S{\'a}ez N{\'u}{\~n}ez}, {Sagrist{\`a} Sell{\'e}s}, {Sahlmann}, {Salguero}, {Samaras}, {Sanchez Gimenez}, {Sanna}, {Santove{\~n}a}, {Sarasso}, {Schultheis}, {Sciacca}, {Segol}, {Segovia}, {S{\'e}gransan}, {Semeux}, {Shahaf}, {Siddiqui}, {Siebert},
  {Siltala}, {Silvelo}, {Slezak}, {Slezak}, {Smart}, {Snaith}, {Solano}, {Solitro}, {Souami}, {Souchay}, {Spagna}, {Spina}, {Spoto}, {Steele}, {Steidelm{\"u}ller}, {Stephenson}, {S{\"u}veges}, {Surdej}, {Szabados}, {Szegedi-Elek}, {Taris}, {Taylor}, {Teixeira}, {Tolomei}, {Tonello}, {Torra}, {Torra}, {Torralba Elipe}, {Trabucchi}, {Tsounis}, {Turon}, {Ulla}, {Unger}, {Vaillant}, {van Dillen}, {van Reeven}, {Vanel}, {Vecchiato}, {Viala}, {Vicente}, {Voutsinas}, {Weiler}, {Wevers}, {Wyrzykowski}, {Yoldas}, {Yvard}, {Zhao}, {Zorec}, {Zucker}, \& {Zwitter}}]{GaiaCollaboration2023}
{Gaia Collaboration}, {Vallenari}, A., {Brown}, A.~G.~A., {et~al.} 2023, \aap, 674, A1, \dodoi{10.1051/0004-6361/202243940}

\bibitem[{{Gilbert} {et~al.}(2022){Gilbert}, {Barclay}, {Quintana}, {Walkowicz}, {Vega}, {Schlieder}, {Monsue}, {Cale}, {Collins}, {Gaidos}, {El Mufti}, {Reefe}, {Plavchan}, {Tanner}, {Wittenmyer}, {Wittrock}, {Jenkins}, {Latham}, {Ricker}, {Rose}, {Seager}, {Vanderspek}, \& {Winn}}]{Gilbert2022}
{Gilbert}, E.~A., {Barclay}, T., {Quintana}, E.~V., {et~al.} 2022, \aj, 163, 147, \dodoi{10.3847/1538-3881/ac23ca}

\bibitem[{{Gray}(1984)}]{Gray1984}
{Gray}, D.~F. 1984, \apj, 277, 640, \dodoi{10.1086/161735}

\bibitem[{{Gully-Santiago} {et~al.}(2017){Gully-Santiago}, {Herczeg}, {Czekala}, {Somers}, {Grankin}, {Covey}, {Donati}, {Alencar}, {Hussain}, {Shappee}, {Mace}, {Lee}, {Holoien}, {Jose}, \& {Liu}}]{Gully-Santiago2017}
{Gully-Santiago}, M.~A., {Herczeg}, G.~J., {Czekala}, I., {et~al.} 2017, \apj, 836, 200, \dodoi{10.3847/1538-4357/836/2/200}

\bibitem[{{Herbst} {et~al.}(2021){Herbst}, {Papaioannou}, {Airapetian}, \& {Atri}}]{Herbst2021}
{Herbst}, K., {Papaioannou}, A., {Airapetian}, V.~S., \& {Atri}, D. 2021, \apj, 907, 89, \dodoi{10.3847/1538-4357/abcc04}

\bibitem[{{Hirano} {et~al.}(2020){Hirano}, {Krishnamurthy}, {Gaidos}, {Flewelling}, {Mann}, {Narita}, {Plavchan}, {Kotani}, {Tamura}, {Harakawa}, {Hodapp}, {Ishizuka}, {Jacobson}, {Konishi}, {Kudo}, {Kurokawa}, {Kuzuhara}, {Nishikawa}, {Omiya}, {Serizawa}, {Ueda}, \& {Vievard}}]{Hirano2020}
{Hirano}, T., {Krishnamurthy}, V., {Gaidos}, E., {et~al.} 2020, \apjl, 899, L13, \dodoi{10.3847/2041-8213/aba6eb}

\bibitem[{{Hogg} {et~al.}(2010){Hogg}, {Bovy}, \& {Lang}}]{Hogg2010}
{Hogg}, D.~W., {Bovy}, J., \& {Lang}, D. 2010, arXiv e-prints, arXiv:1008.4686, \dodoi{10.48550/arXiv.1008.4686}

\bibitem[{Hunter(2007)}]{matplotlib}
Hunter, J.~D. 2007, Computing in Science \& Engineering, 9, 90, \dodoi{10.1109/MCSE.2007.55}

\bibitem[{{Husser} {et~al.}(2013){Husser}, {Wende-von Berg}, {Dreizler}, {Homeier}, {Reiners}, {Barman}, \& {Hauschildt}}]{Husseretal2013}
{Husser}, T.-O., {Wende-von Berg}, S., {Dreizler}, S., {et~al.} 2013, \aap, 553, A6, \dodoi{10.1051/0004-6361/201219058}

\bibitem[{{Ikuta} {et~al.}(2023){Ikuta}, {Namekata}, {Notsu}, {Maehara}, {Okamoto}, {Honda}, {Nogami}, \& {Shibata}}]{Ikuta2023}
{Ikuta}, K., {Namekata}, K., {Notsu}, Y., {et~al.} 2023, arXiv e-prints, arXiv:2302.09249, \dodoi{10.48550/arXiv.2302.09249}

\bibitem[{{Iyer} \& {Line}(2020)}]{Iyer2020}
{Iyer}, A.~R., \& {Line}, M.~R. 2020, \apj, 889, 78, \dodoi{10.3847/1538-4357/ab612e}

\bibitem[{{Iyer} {et~al.}(2023){Iyer}, {Line}, {Muirhead}, {Fortney}, \& {Gharib-Nezhad}}]{Iyer2023}
{Iyer}, A.~R., {Line}, M.~R., {Muirhead}, P.~S., {Fortney}, J.~J., \& {Gharib-Nezhad}, E. 2023, \apj, 944, 41, \dodoi{10.3847/1538-4357/acabc2}

\bibitem[{{Jones} {et~al.}(2013){Jones}, {Noll}, {Kausch}, {Szyszka}, \& {Kimeswenger}}]{Jones2013}
{Jones}, A., {Noll}, S., {Kausch}, W., {Szyszka}, C., \& {Kimeswenger}, S. 2013, \aap, 560, A91, \dodoi{10.1051/0004-6361/201322433}

\bibitem[{{Jones} {et~al.}(1995){Jones}, {Longmore}, {Allard}, {Hauschildt}, {Miller}, \& {Tennyson}}]{Jones1995}
{Jones}, H. R.~A., {Longmore}, A.~J., {Allard}, F., {et~al.} 1995, \mnras, 277, 767, \dodoi{10.1093/mnras/277.3.767}

\bibitem[{{Kalas} {et~al.}(2004){Kalas}, {Liu}, \& {Matthews}}]{Kalas2004}
{Kalas}, P., {Liu}, M.~C., \& {Matthews}, B.~C. 2004, Science, 303, 1990, \dodoi{10.1126/science.1093420}

\bibitem[{{Libby-Roberts} {et~al.}(2022){Libby-Roberts}, {Berta-Thompson}, {Diamond-Lowe}, {Gully-Santiago}, {Irwin}, {Kempton}, {Rackham}, {Charbonneau}, {D{\'e}sert}, {Dittmann}, {Hofmann}, {Morley}, \& {Newton}}]{Libby-Roberts2022}
{Libby-Roberts}, J.~E., {Berta-Thompson}, Z.~K., {Diamond-Lowe}, H., {et~al.} 2022, \aj, 164, 59, \dodoi{10.3847/1538-3881/ac75de}

\bibitem[{{Lim} {et~al.}(2023){Lim}, {Benneke}, {Doyon}, {MacDonald}, {Piaulet}, {Artigau}, {Coulombe}, {Radica}, {L'Heureux}, {Albert}, {Rackham}, {de Wit}, {Salhi}, {Roy}, {Flagg}, {Fournier-Tondreau}, {Taylor}, {Cook}, {Lafreni{\`e}re}, {Cowan}, {Kaltenegger}, {Rowe}, {Espinoza}, {Dang}, \& {Darveau-Bernier}}]{Lim2023}
{Lim}, O., {Benneke}, B., {Doyon}, R., {et~al.} 2023, arXiv e-prints, arXiv:2309.07047, \dodoi{10.48550/arXiv.2309.07047}

\bibitem[{{Louca} {et~al.}(2023){Louca}, {Miguel}, {Tsai}, {Froning}, {Loyd}, \& {France}}]{Louca2023}
{Louca}, A.~J., {Miguel}, Y., {Tsai}, S.-M., {et~al.} 2023, \mnras, 521, 3333, \dodoi{10.1093/mnras/stac1220}

\bibitem[{{MacGregor} {et~al.}(2013){MacGregor}, {Wilner}, {Rosenfeld}, {Andrews}, {Matthews}, {Hughes}, {Booth}, {Chiang}, {Graham}, {Kalas}, {Kennedy}, \& {Sibthorpe}}]{MacGregor2013}
{MacGregor}, M.~A., {Wilner}, D.~J., {Rosenfeld}, K.~A., {et~al.} 2013, \apjl, 762, L21, \dodoi{10.1088/2041-8205/762/2/L21}

\bibitem[{{Mamajek} \& {Bell}(2014)}]{Mamajek2014}
{Mamajek}, E.~E., \& {Bell}, C. P.~M. 2014, \mnras, 445, 2169, \dodoi{10.1093/mnras/stu1894}

\bibitem[{{Martioli} {et~al.}(2021){Martioli}, {H{\'e}brard}, {Correia}, {Laskar}, \& {Lecavelier des Etangs}}]{Martioli2021}
{Martioli}, E., {H{\'e}brard}, G., {Correia}, A.~C.~M., {Laskar}, J., \& {Lecavelier des Etangs}, A. 2021, \aap, 649, A177, \dodoi{10.1051/0004-6361/202040235}

\bibitem[{{McCully} {et~al.}(2022){McCully}, {Turner}, {Collom}, \& {Daily}}]{BANZAI2022}
{McCully}, C., {Turner}, M., {Collom}, D., \& {Daily}, M. 2022, {BANZAI: Beautiful Algorithms to Normalize Zillions of Astronomical Images}, Astrophysics Source Code Library, record ascl:2207.031.
\newblock \doeprint{2207.031}

\bibitem[{McKinney(2011)}]{pandas}
McKinney, W. 2011, Python for High Performance and Scientific Computing, 14

\bibitem[{{Medina} {et~al.}(2022){Medina}, {Charbonneau}, {Winters}, {Irwin}, \& {Mink}}]{Medina2022}
{Medina}, A.~A., {Charbonneau}, D., {Winters}, J.~G., {Irwin}, J., \& {Mink}, J. 2022, \apj, 928, 185, \dodoi{10.3847/1538-4357/ac5738}

\bibitem[{{Mikal-Evans} {et~al.}(2023){Mikal-Evans}, {Madhusudhan}, {Dittmann}, {G{\"u}nther}, {Welbanks}, {Van Eylen}, {Crossfield}, {Daylan}, \& {Kreidberg}}]{Mikal-Evans2023}
{Mikal-Evans}, T., {Madhusudhan}, N., {Dittmann}, J., {et~al.} 2023, \aj, 165, 84, \dodoi{10.3847/1538-3881/aca90b}

\bibitem[{{Moran} {et~al.}(2023){Moran}, {Stevenson}, {Sing}, {MacDonald}, {Kirk}, {Lustig-Yaeger}, {Peacock}, {Mayorga}, {Bennett}, {L{\'o}pez-Morales}, {May}, {Rustamkulov}, {Valenti}, {Adams Redai}, {Alam}, {Batalha}, {Fu}, {Gonzalez-Quiles}, {Highland}, {Kruse}, {Lothringer}, {Ortiz Ceballos}, {Sotzen}, \& {Wakeford}}]{Moran2023}
{Moran}, S.~E., {Stevenson}, K.~B., {Sing}, D.~K., {et~al.} 2023, \apjl, 948, L11, \dodoi{10.3847/2041-8213/accb9c}

\bibitem[{{Neff} {et~al.}(1995){Neff}, {O'Neal}, \& {Saar}}]{Neff1995}
{Neff}, J.~E., {O'Neal}, D., \& {Saar}, S.~H. 1995, \apj, 452, 879, \dodoi{10.1086/176356}

\bibitem[{{Noll} {et~al.}(2012){Noll}, {Kausch}, {Barden}, {Jones}, {Szyszka}, {Kimeswenger}, \& {Vinther}}]{Noll2012}
{Noll}, S., {Kausch}, W., {Barden}, M., {et~al.} 2012, \aap, 543, A92, \dodoi{10.1051/0004-6361/201219040}

\bibitem[{{Norris} {et~al.}(2023){Norris}, {Unruh}, {Witzke}, {Solanki}, {Krivova}, {Shapiro}, {Yeo}, {Cameron}, \& {Beeck}}]{Norris2023}
{Norris}, C.~M., {Unruh}, Y.~C., {Witzke}, V., {et~al.} 2023, \mnras, 524, 1139, \dodoi{10.1093/mnras/stad1738}

\bibitem[{{Olah} {et~al.}(1997){Olah}, {K{\H{o}}v{\'a}ri}, {Bartus}, {Strassmeier}, {Hall}, \& {Henry}}]{Olah1997}
{Olah}, K., {K{\H{o}}v{\'a}ri}, Z., {Bartus}, J., {et~al.} 1997, \aap, 321, 811

\bibitem[{{Pass} {et~al.}(2023){Pass}, {Winters}, {Charbonneau}, {Irwin}, \& {Medina}}]{Pass2023}
{Pass}, E.~K., {Winters}, J.~G., {Charbonneau}, D., {Irwin}, J.~M., \& {Medina}, A.~A. 2023, \aj, 166, 16, \dodoi{10.3847/1538-3881/acd6a2}

\bibitem[{{Plavchan} {et~al.}(2020){Plavchan}, {Barclay}, {Gagn{\'e}}, {Gao}, {Cale}, {Matzko}, {Dragomir}, {Quinn}, {Feliz}, {Stassun}, {Crossfield}, {Berardo}, {Latham}, {Tieu}, {Anglada-Escud{\'e}}, {Ricker}, {Vanderspek}, {Seager}, {Winn}, {Jenkins}, {Rinehart}, {Krishnamurthy}, {Dynes}, {Doty}, {Adams}, {Afanasev}, {Beichman}, {Bottom}, {Bowler}, {Brinkworth}, {Brown}, {Cancino}, {Ciardi}, {Clampin}, {Clark}, {Collins}, {Davison}, {Foreman-Mackey}, {Furlan}, {Gaidos}, {Geneser}, {Giddens}, {Gilbert}, {Hall}, {Hellier}, {Henry}, {Horner}, {Howard}, {Huang}, {Huber}, {Kane}, {Kenworthy}, {Kielkopf}, {Kipping}, {Klenke}, {Kruse}, {Latouf}, {Lowrance}, {Mennesson}, {Mengel}, {Mills}, {Morton}, {Narita}, {Newton}, {Nishimoto}, {Okumura}, {Palle}, {Pepper}, {Quintana}, {Roberge}, {Roccatagliata}, {Schlieder}, {Tanner}, {Teske}, {Tinney}, {Vanderburg}, {von Braun}, {Walp}, {Wang}, {Wang}, {Weigand}, {White}, {Wittenmyer}, {Wright}, {Youngblood}, {Zhang}, \& {Zilberman}}]{Plavchan2020}
{Plavchan}, P., {Barclay}, T., {Gagn{\'e}}, J., {et~al.} 2020, \nat, 582, 497, \dodoi{10.1038/s41586-020-2400-z}

\bibitem[{{Pont} {et~al.}(2008){Pont}, {Knutson}, {Gilliland}, {Moutou}, \& {Charbonneau}}]{Pont2008}
{Pont}, F., {Knutson}, H., {Gilliland}, R.~L., {Moutou}, C., \& {Charbonneau}, D. 2008, \mnras, 385, 109, \dodoi{10.1111/j.1365-2966.2008.12852.x}

\bibitem[{{Pont} {et~al.}(2007){Pont}, {Gilliland}, {Moutou}, {Charbonneau}, {Bouchy}, {Brown}, {Mayor}, {Queloz}, {Santos}, \& {Udry}}]{Pont2007}
{Pont}, F., {Gilliland}, R.~L., {Moutou}, C., {et~al.} 2007, \aap, 476, 1347, \dodoi{10.1051/0004-6361:20078269}

\bibitem[{{Rackham} {et~al.}(2018){Rackham}, {Apai}, \& {Giampapa}}]{Rackham2018}
{Rackham}, B.~V., {Apai}, D., \& {Giampapa}, M.~S. 2018, \apj, 853, 122, \dodoi{10.3847/1538-4357/aaa08c}

\bibitem[{{Rackham} {et~al.}(2019){Rackham}, {Apai}, \& {Giampapa}}]{Rackham2019}
---. 2019, \aj, 157, 96, \dodoi{10.3847/1538-3881/aaf892}

\bibitem[{{Rackham} \& {de Wit}(2023)}]{Rackham2023}
{Rackham}, B.~V., \& {de Wit}, J. 2023, arXiv e-prints, arXiv:2303.15418, \dodoi{10.48550/arXiv.2303.15418}

\bibitem[{{Ramsey} \& {Nations}(1980)}]{RamseyNations1980}
{Ramsey}, L.~W., \& {Nations}, H.~L. 1980, \apjl, 239, L121, \dodoi{10.1086/183306}

\bibitem[{{Reiners} {et~al.}(2013){Reiners}, {Shulyak}, {Anglada-Escud{\'e}}, {Jeffers}, {Morin}, {Zechmeister}, {Kochukhov}, \& {Piskunov}}]{Reiners2013}
{Reiners}, A., {Shulyak}, D., {Anglada-Escud{\'e}}, G., {et~al.} 2013, \aap, 552, A103, \dodoi{10.1051/0004-6361/201220437}

\bibitem[{{Robertson} {et~al.}(2020){Robertson}, {Stefansson}, {Mahadevan}, {Endl}, {Cochran}, {Beard}, {Bender}, {Diddams}, {Duong}, {Ford}, {Fredrick}, {Halverson}, {Hearty}, {Holcomb}, {Juan}, {Kanodia}, {Lubin}, {Metcalf}, {Monson}, {Ninan}, {Palafoutas}, {Ramsey}, {Roy}, {Schwab}, {Terrien}, \& {Wright}}]{Robertson2020}
{Robertson}, P., {Stefansson}, G., {Mahadevan}, S., {et~al.} 2020, \apj, 897, 125, \dodoi{10.3847/1538-4357/ab989f}

\bibitem[{{Robinson} {et~al.}(2014){Robinson}, {Maltagliati}, {Marley}, \& {Fortney}}]{Robinson2014}
{Robinson}, T.~D., {Maltagliati}, L., {Marley}, M.~S., \& {Fortney}, J.~J. 2014, Proceedings of the National Academy of Science, 111, 9042, \dodoi{10.1073/pnas.1403473111}

\bibitem[{{Rustamkulov} {et~al.}(2023){Rustamkulov}, {Sing}, {Mukherjee}, {May}, {Kirk}, {Schlawin}, {Line}, {Piaulet}, {Carter}, {Batalha}, {Goyal}, {L{\'o}pez-Morales}, {Lothringer}, {MacDonald}, {Moran}, {Stevenson}, {Wakeford}, {Espinoza}, {Bean}, {Batalha}, {Benneke}, {Berta-Thompson}, {Crossfield}, {Gao}, {Kreidberg}, {Powell}, {Cubillos}, {Gibson}, {Leconte}, {Molaverdikhani}, {Nikolov}, {Parmentier}, {Roy}, {Taylor}, {Turner}, {Wheatley}, {Aggarwal}, {Ahrer}, {Alam}, {Alderson}, {Allen}, {Banerjee}, {Barat}, {Barrado}, {Barstow}, {Bell}, {Blecic}, {Brande}, {Casewell}, {Changeat}, {Chubb}, {Crouzet}, {Daylan}, {Decin}, {D{\'e}sert}, {Mikal-Evans}, {Feinstein}, {Flagg}, {Fortney}, {Harrington}, {Heng}, {Hong}, {Hu}, {Iro}, {Kataria}, {Kempton}, {Krick}, {Lendl}, {Lillo-Box}, {Louca}, {Lustig-Yaeger}, {Mancini}, {Mansfield}, {Mayne}, {Miguel}, {Morello}, {Ohno}, {Palle}, {Petit dit de la Roche}, {Rackham}, {Radica}, {Ramos-Rosado}, {Redfield}, {Rogers}, {Shkolnik}, {Southworth}, {Teske}, {Tremblin},
  {Tucker}, {Venot}, {Waalkes}, {Welbanks}, {Zhang}, \& {Zieba}}]{Rustamkulov2023}
{Rustamkulov}, Z., {Sing}, D.~K., {Mukherjee}, S., {et~al.} 2023, \nat, 614, 659, \dodoi{10.1038/s41586-022-05677-y}

\bibitem[{{Sanchis-Ojeda} \& {Winn}(2011)}]{Sanchis-Ojeda2011}
{Sanchis-Ojeda}, R., \& {Winn}, J.~N. 2011, \apj, 743, 61, \dodoi{10.1088/0004-637X/743/1/61}

\bibitem[{{Sch{\"o}fer} {et~al.}(2019){Sch{\"o}fer}, {Jeffers}, {Reiners}, {Shulyak}, {Fuhrmeister}, {Johnson}, {Zechmeister}, {Ribas}, {Quirrenbach}, {Amado}, {Caballero}, {Anglada-Escud{\'e}}, {Bauer}, {B{\'e}jar}, {Cort{\'e}s-Contreras}, {Dreizler}, {Guenther}, {Kaminski}, {K{\"u}rster}, {Lafarga}, {Montes}, {Morales}, {Pedraz}, \& {Tal-Or}}]{Schofer2019}
{Sch{\"o}fer}, P., {Jeffers}, S.~V., {Reiners}, A., {et~al.} 2019, \aap, 623, A44, \dodoi{10.1051/0004-6361/201834114}

\bibitem[{{Seager} \& {Sasselov}(2000)}]{SeagerSasselov2000ApJ...537..916S}
{Seager}, S., \& {Sasselov}, D.~D. 2000, \apj, 537, 916, \dodoi{10.1086/309088}

\bibitem[{{Seager} {et~al.}(2000){Seager}, {Whitney}, \& {Sasselov}}]{Seager2000ApJ...540..504S}
{Seager}, S., {Whitney}, B.~A., \& {Sasselov}, D.~D. 2000, \apj, 540, 504, \dodoi{10.1086/309292}

\bibitem[{{Shapiro} {et~al.}(2016){Shapiro}, {Solanki}, {Krivova}, {Yeo}, \& {Schmutz}}]{Shapiro2016}
{Shapiro}, A.~I., {Solanki}, S.~K., {Krivova}, N.~A., {Yeo}, K.~L., \& {Schmutz}, W.~K. 2016, \aap, 589, A46, \dodoi{10.1051/0004-6361/201527527}

\bibitem[{Shields {et~al.}(2016)Shields, Ballard, \& Johnson}]{Shields2016}
Shields, A.~L., Ballard, S., \& Johnson, J.~A. 2016, Physics Reports, 663, 1, \dodoi{10.1016/j.physrep.2016.10.003}

\bibitem[{{Sing} {et~al.}(2011){Sing}, {Pont}, {Aigrain}, {Charbonneau}, {D{\'e}sert}, {Gibson}, {Gilliland}, {Hayek}, {Henry}, {Knutson}, {Lecavelier Des Etangs}, {Mazeh}, \& {Shporer}}]{Sing2011}
{Sing}, D.~K., {Pont}, F., {Aigrain}, S., {et~al.} 2011, \mnras, 416, 1443, \dodoi{10.1111/j.1365-2966.2011.19142.x}

\bibitem[{{Sing} {et~al.}(2016){Sing}, {Fortney}, {Nikolov}, {Wakeford}, {Kataria}, {Evans}, {Aigrain}, {Ballester}, {Burrows}, {Deming}, {D{\'e}sert}, {Gibson}, {Henry}, {Huitson}, {Knutson}, {Lecavelier Des Etangs}, {Pont}, {Showman}, {Vidal-Madjar}, {Williamson}, \& {Wilson}}]{Sing2016}
{Sing}, D.~K., {Fortney}, J.~J., {Nikolov}, N., {et~al.} 2016, \nat, 529, 59, \dodoi{10.1038/nature16068}

\bibitem[{{Strassmeier} \& {Olah}(1992)}]{StrassmeierOlah1992}
{Strassmeier}, K.~G., \& {Olah}, K. 1992, \aap, 259, 595

\bibitem[{{Szab{\'o}} {et~al.}(2021){Szab{\'o}}, {Gandolfi}, {Brandeker}, {Csizmadia}, {Garai}, {Billot}, {Broeg}, {Ehrenreich}, {Fortier}, {Fossati}, {Hoyer}, {Kiss}, {Lecavelier des Etangs}, {Maxted}, {Ribas}, {Alibert}, {Alonso}, {Anglada Escud{\'e}}, {B{\'a}rczy}, {Barros}, {Barrado}, {Baumjohann}, {Beck}, {Beck}, {Bekkelien}, {Bonfils}, {Benz}, {Borsato}, {Busch}, {Cabrera}, {Charnoz}, {Collier Cameron}, {Van Damme}, {Davies}, {Delrez}, {Deleuil}, {Demangeon}, {Demory}, {Erikson}, {Fridlund}, {Futyan}, {Garc{\'\i}a Mu{\~n}oz}, {Gillon}, {Guedel}, {Guterman}, {Heng}, {Isaak}, {Lacedelli}, {Laskar}, {Lendl}, {Lovis}, {Luntzer}, {Magrin}, {Nascimbeni}, {Olofsson}, {Osborn}, {Ottensamer}, {Pagano}, {Pall{\'e}}, {Peter}, {Piazza}, {Piotto}, {Pollacco}, {Queloz}, {Ragazzoni}, {Rando}, {Rauer}, {Santos}, {Scandariato}, {S{\'e}gransan}, {Serrano}, {Sicilia}, {Simon}, {Smith}, {Sousa}, {Steller}, {Thomas}, {Udry}, {Van Grootel}, {Walton}, \& {Wilson}}]{Szabo2021}
{Szab{\'o}}, G.~M., {Gandolfi}, D., {Brandeker}, A., {et~al.} 2021, \aap, 654, A159, \dodoi{10.1051/0004-6361/202140345}

\bibitem[{{Szab{\'o}} {et~al.}(2022){Szab{\'o}}, {Garai}, {Brandeker}, {Gandolfi}, {Wilson}, {Deline}, {Olofsson}, {Fortier}, {Queloz}, {Borsato}, {Kiefer}, {Lecavelier des Etangs}, {Lendl}, {Serrano}, {Sulis}, {Ulmer Moll}, {Van Grootel}, {Alibert}, {Alonso}, {Anglada}, {B{\'a}rczy}, {Barrado y Navascues}, {Barros}, {Baumjohann}, {Beck}, {Beck}, {Benz}, {Billot}, {Bonfanti}, {Bonfils}, {Broeg}, {Cabrera}, {Charnoz}, {Collier Cameron}, {Csizmadia}, {Davies}, {Deleuil}, {Delrez}, {Demangeon}, {Demory}, {Ehrenreich}, {Erikson}, {Fossati}, {Fridlund}, {Gillon}, {G{\"u}del}, {Heng}, {Hoyer}, {Isaak}, {Kiss}, {Laskar}, {Lovis}, {Magrin}, {Maxted}, {Mecina}, {Nascimbeni}, {Ottensamer}, {Pagano}, {Pall{\'e}}, {Peter}, {Piotto}, {Pollacco}, {Ragazzoni}, {Rando}, {Rauer}, {Ribas}, {Santos}, {Sarajlic}, {Scandariato}, {S{\'e}gransan}, {Simon}, {Smith}, {Sousa}, {Steller}, {Thomas}, {Udry}, {Verrecchia}, {Walton}, \& {Wolter}}]{Szabo2022}
{Szab{\'o}}, G.~M., {Garai}, Z., {Brandeker}, A., {et~al.} 2022, \aap, 659, L7, \dodoi{10.1051/0004-6361/202243076}

\bibitem[{{Thao} {et~al.}(2023){Thao}, {Mann}, {Gao}, {Owens}, {Vanderburg}, {Newton}, {Tang}, {Fields}, {David}, {Irwin}, {Husser}, {Charbonneau}, \& {Ballard}}]{Thao2023}
{Thao}, P.~C., {Mann}, A.~W., {Gao}, P., {et~al.} 2023, \aj, 165, 23, \dodoi{10.3847/1538-3881/aca07a}

\bibitem[{{Torres} \& {Ferraz Mello}(1973)}]{Torres1973}
{Torres}, C.~A.~O., \& {Ferraz Mello}, S. 1973, \aap, 27, 231

\bibitem[{{van der Walt} {et~al.}(2011){van der Walt}, {Colbert}, \& {Varoquaux}}]{numpy}
{van der Walt}, S., {Colbert}, S.~C., \& {Varoquaux}, G. 2011, Computing in Science and Engineering, 13, 22, \dodoi{10.1109/MCSE.2011.37}

\bibitem[{{Vogt}(1979)}]{Vogt1979}
{Vogt}, S.~S. 1979, \pasp, 91, 616, \dodoi{10.1086/130549}

\bibitem[{{Vogt}(1981)}]{Vogt1981ApJ...250..327V}
---. 1981, \apj, 250, 327, \dodoi{10.1086/159379}

\bibitem[{{Wakeford} {et~al.}(2019){Wakeford}, {Lewis}, {Fowler}, {Bruno}, {Wilson}, {Moran}, {Valenti}, {Batalha}, {Filippazzo}, {Bourrier}, {H{\"o}rst}, {Lederer}, \& {de Wit}}]{Wakeford2019}
{Wakeford}, H.~R., {Lewis}, N.~K., {Fowler}, J., {et~al.} 2019, \aj, 157, 11, \dodoi{10.3847/1538-3881/aaf04d}

\bibitem[{{Wing} {et~al.}(1967){Wing}, {Peimbert}, \& {Spinrad}}]{Wing1967}
{Wing}, R.~F., {Peimbert}, M., \& {Spinrad}, H. 1967, \pasp, 79, 351, \dodoi{10.1086/128496}

\bibitem[{{Wittrock} {et~al.}(2022){Wittrock}, {Dreizler}, {Reefe}, {Morris}, {Plavchan}, {Lowrance}, {Demory}, {Ingalls}, {Gilbert}, {Barclay}, {Cale}, {Collins}, {Collins}, {Crossfield}, {Dragomir}, {Eastman}, {Mufti}, {Feliz}, {Gagn{\'e}}, {Gaidos}, {Gao}, {Geneser}, {Hebb}, {Henze}, {Horne}, {Jenkins}, {Jensen}, {Kane}, {Kaye}, {Martioli}, {Monsue}, {Pall{\'e}}, {Quintana}, {Radford}, {Roccatagliata}, {Schlieder}, {Schwarz}, {Shporer}, {Stassun}, {Stockdale}, {Tan}, {Tanner}, {Vanderburg}, {Vega}, \& {Wang}}]{Wittrock2022}
{Wittrock}, J.~M., {Dreizler}, S., {Reefe}, M.~A., {et~al.} 2022, \aj, 164, 27, \dodoi{10.3847/1538-3881/ac68e5}

\bibitem[{{Wittrock} {et~al.}(2023){Wittrock}, {Plavchan}, {Cale}, {Barclay}, {Gilbert}, {Ludwig}, {Schwarz}, {Mekarnia}, {Triaud}, {Abe}, {Suarez}, {Guillot}, {Conti}, {Collins}, {Waite}, {Kielkopf}, {Collins}, {Dreizler}, {El Mufti}, {Feliz}, {Gaidos}, {Geneser}, {Horne}, {Kane}, {Lowrance}, {Martioli}, {Radford}, {Reefe}, {Roccatagliata}, {Shporer}, {Stassun}, {Stockdale}, {Tan}, {Tanner}, \& {Vega}}]{Wittrock2023}
{Wittrock}, J.~M., {Plavchan}, P., {Cale}, B.~L., {et~al.} 2023, arXiv e-prints, arXiv:2302.04922, \dodoi{10.48550/arXiv.2302.04922}

\bibitem[{{Yamashita} {et~al.}(2022){Yamashita}, {Itoh}, \& {Oasa}}]{Yamashita2022}
{Yamashita}, M., {Itoh}, Y., \& {Oasa}, Y. 2022, \pasj, 74, 1295, \dodoi{10.1093/pasj/psac069}

\bibitem[{{Zacharias} {et~al.}(2012){Zacharias}, {Finch}, {Girard}, {Henden}, {Bartlett}, {Monet}, \& {Zacharias}}]{2012yCat.1322....0Z}
{Zacharias}, N., {Finch}, C.~T., {Girard}, T.~M., {et~al.} 2012, VizieR Online Data Catalog, I/322A

\bibitem[{{Zhang} {et~al.}(2018){Zhang}, {Zhou}, {Rackham}, \& {Apai}}]{Zhang2018}
{Zhang}, Z., {Zhou}, Y., {Rackham}, B.~V., \& {Apai}, D. 2018, \aj, 156, 178, \dodoi{10.3847/1538-3881/aade4f}

\bibitem[{{Zicher} {et~al.}(2022){Zicher}, {Barrag{\'a}n}, {Klein}, {Aigrain}, {Owen}, {Gandolfi}, {Lagrange}, {Serrano}, {Kaye}, {Nielsen}, {Rajpaul}, {Grandjean}, {Goffo}, \& {Nicholson}}]{Zicher2022}
{Zicher}, N., {Barrag{\'a}n}, O., {Klein}, B., {et~al.} 2022, \mnras, 512, 3060, \dodoi{10.1093/mnras/stac614}

\end{thebibliography}

\appendix \label{appendix}

\begin{figure}
\plotone{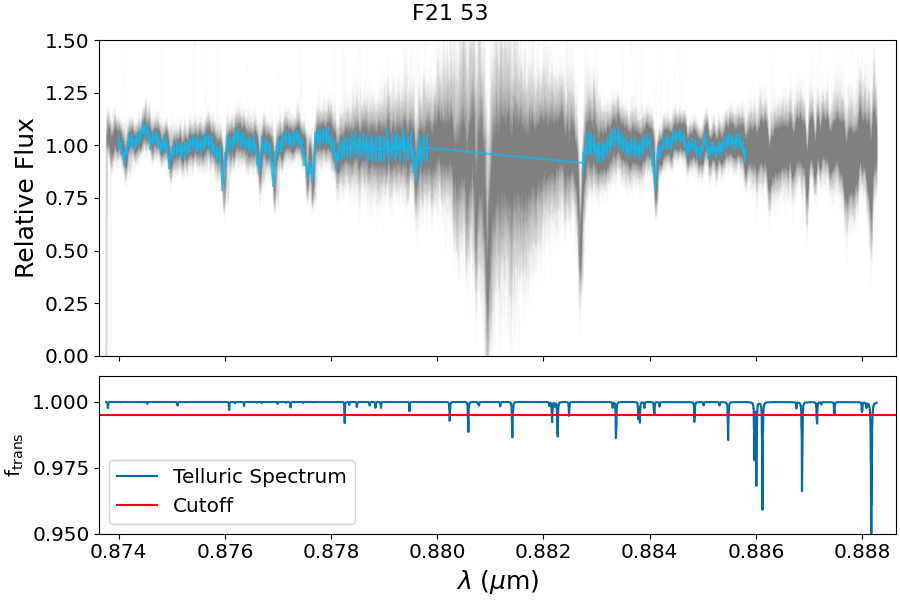}
\caption{Order 53 from visit F21 with spectral orders used in this analysis extending from order 53 (0.881 $\mu$m) to order 83 (0.562 $\mu$m). Stacked in gray are individual spectra collected from 2021 Aug 24 to 2021 Sep 9. In blue is a single spectrum derived from the time series observations which has been binned to 0.05 nm, telluric-corrected, median-averaged in time, and continuum normalized. The complete figure set (31 images) is available in the online journal.} \label{f:F21_figset_poster}
\end{figure}

\begin{figure}
\plotone{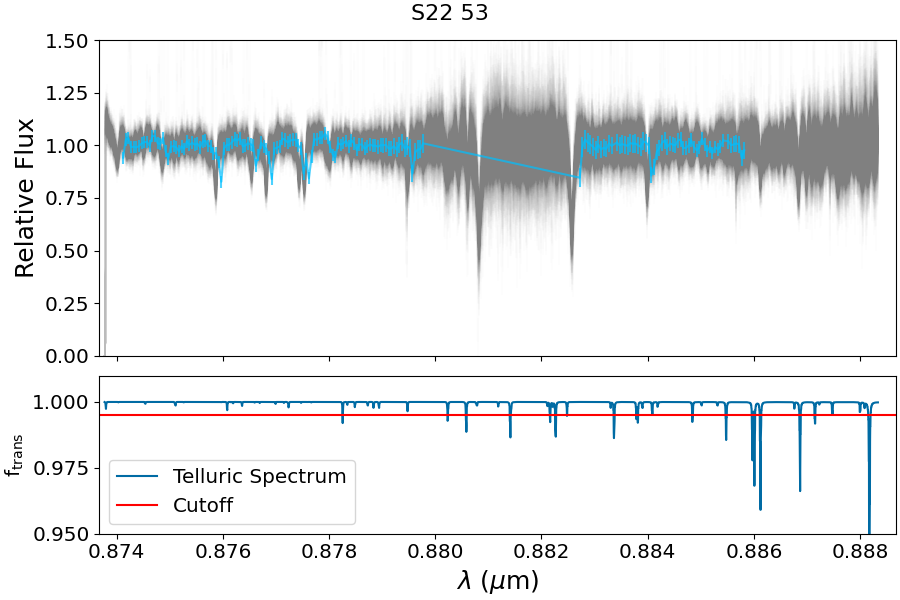}
\caption{Order 53 from visit S22 with spectral orders used in this analysis extending from order 53 (0.881 $\mu$m) to order 83 (0.562 $\mu$m). Stacked in gray are individual spectra collected from 2022 Apr 02 to 2022 Apr 17. In blue is a single spectrum derived from the time series observations which has been binned to 0.05 nm, telluric-corrected, median-averaged in time, and continuum normalized. The complete figure set (31 images) is available in the online journal.} \label{f:S22_figset_poster}
\end{figure}

We include this appendix to show some of the spectral data (Figures \ref{f:F21_figset_poster}, \ref{f:S22_figset_poster}), visit-specific models of the photometric variabilities (Figure \ref{f:Teff_Phot_model_results}), and order-specific modeling of the NRES spectra (Figure \ref{f:combined_Teff_Spec_violin}).
Comparing F21 and S22 variability models demonstrates how important the i' measurement is and the relative constraints provided with or without it. 
The order-specific spectral models show agreement in many ways, as well as some poorly constrained orders which are likely due to telluric contamination. 

Figures \ref{f:F21_figset_poster} and \ref{f:S22_figset_poster} show spectral order 53 from both visits as examples of how the data looked before and after being processed and the full figure sets can be found in the online journal, including orders that were cut from final analysis.

\figsetstart
\figsettitle{NRES Spectra - Visit F21}

\figsetgrpstart
\figsetnum{10}
\figsetgrpnum{1.1}
\figsetgrptitle{Order 53}
\figsetplot{F21_order53_nres_spectrum.png}
\figsetgrpnote{Order 53 from visit F21 with spectral orders used in this analysis extending from order 53 (0.881 $\mu$m) to order 83 (0.562 $\mu$m). Stacked in gray are individual spectra collected from 2021 Aug 24 to 2021 Sep 9. In blue is a single spectrum derived from the time series observations which has been binned to 0.05 nm, telluric-corrected, median-averaged in time, and continuum normalized. The complete figure set (31 images) is available in the online journal.}\figsetgrpend

\figsetend

\figsetstart
\figsettitle{NRES Spectra - Visit S22}

\figsetgrpstart
\figsetgrpnum{2.1}
\figsetgrptitle{Order 53}
\figsetplot{S22_order53_nres_spectrum.png}
\figsetgrpnote{Order 53 from visit S22 with spectral orders used in this analysis extending from order 53 (0.881 $\mu$m) to order 83 (0.562 $\mu$m). Stacked in gray are individual spectra collected from 2022 Apr 02 to 2022 Apr 17. In blue is a single spectrum derived from the time series observations which has been binned to 0.05 nm, telluric-corrected, median-averaged in time, and continuum normalized. The complete figure set (31 images) is available in the online journal.}
\figsetgrpend

\figsetend

\section{Visit-Specific Variability Measurements}
\begin{figure*}
    \subfloat{\includegraphics[width=0.5\columnwidth]{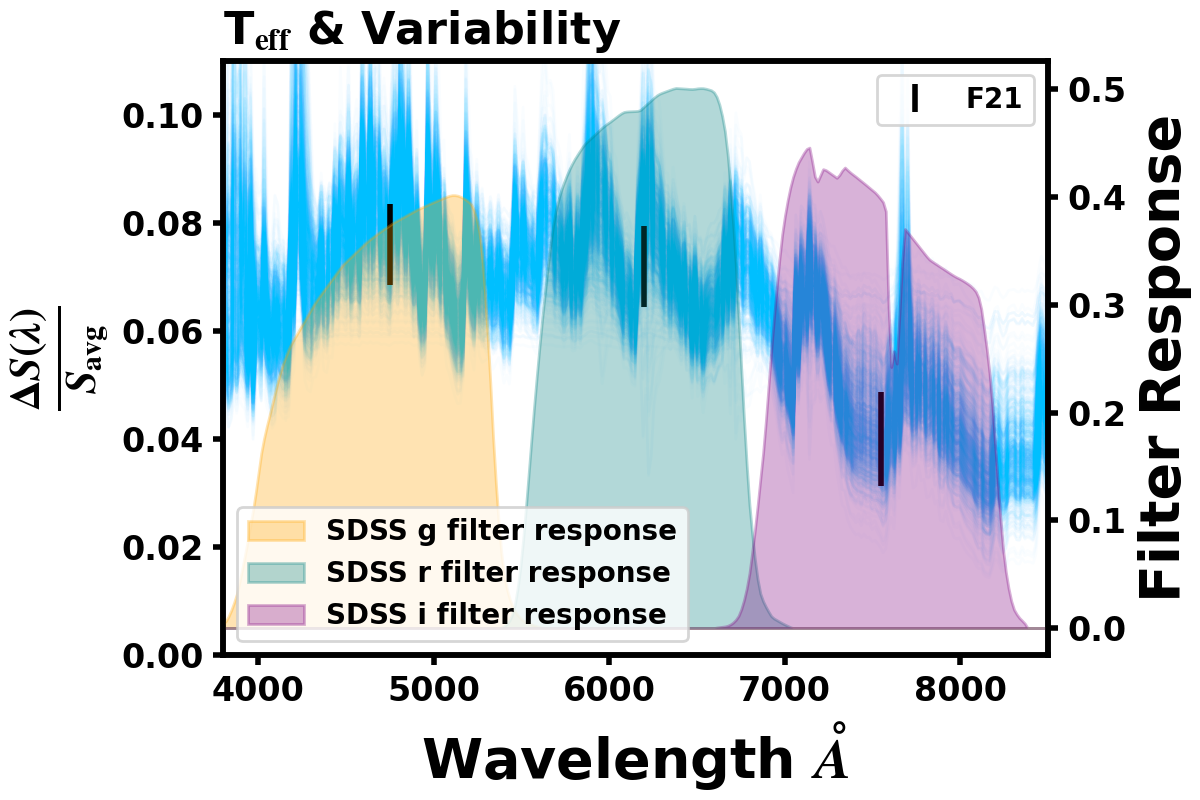}}
    \subfloat{\includegraphics[width=0.5\columnwidth]{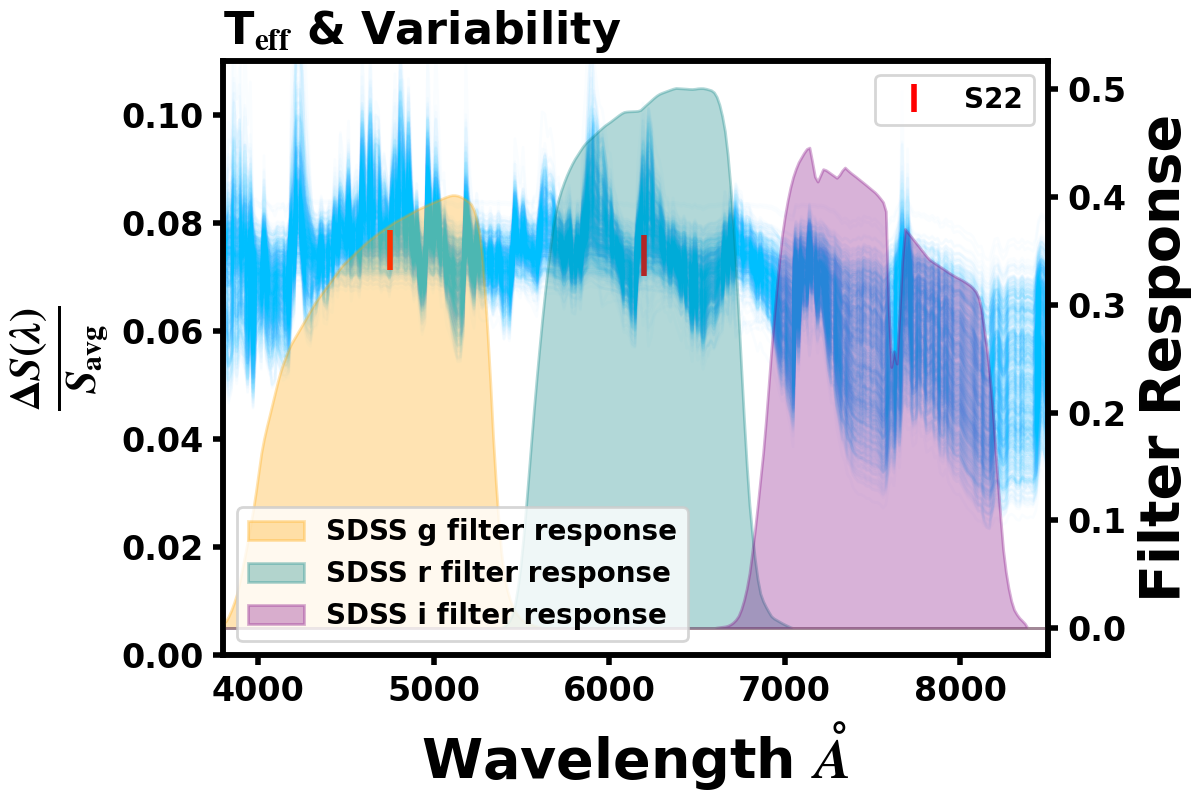}}
\caption{F21 (left) and S22 (right) sampled variability models (without spectral fits). The tighter model constraints imposed by an $i'$ measurement can be seen in the spread of model solutions in the red when comparing visits. The presence or absence of $i'$ is strongly constraining and further work should emphasize multi-band measurements to better constrain the spectral temperature contrast.}
\label{f:Teff_Phot_model_results}
\end{figure*}

While the magnitude of stellar variability in any given waveband is related to degeneracies between spot temperature and spot coverage, the relative variability between photometric bandpasses and the wavelengths where the variability begins to decrease contain a lot of information about the spectral temperatures on the surface of the star.
If the bulk spot temperatures are very cold relative to the photosphere, the variability will remain high at redder wavelengths.
Similarly, if spot temperatures are closer to the ambient photosphere temperature, then the variability will decrease within optical and bluer wavelengths.
The right panel in Figure \ref{f:Teff_Phot_model_results} shows the greater spread in models at redder wavelengths resulting from the absence of an $i'$ measurement, allowing the spot temperatures can be much cooler and much further from the ambient photosphere. 
The different model solutions possible are strongly constrained by this $i'$ variability, and we extend that observation to say that multi-color photometric variabilities covering the optical-NIR provide very strong constraints on the temperature contrast for spotted stars.

%%%%%%%%%%%%%%%%%%%%%%%%%%%%%%%%%%%%%%%%%%%%%%%%%%%%%%
% Photometry only plot
%%%%%%%%%%%%%%%%%%%%%%%%%%%%%%%%%%%%%%%%%%%%%%%%%%%%%%

\section{Spectral Decomposition by Order}
\begin{figure*}
    \subfloat{\includegraphics[width=0.95\textwidth]{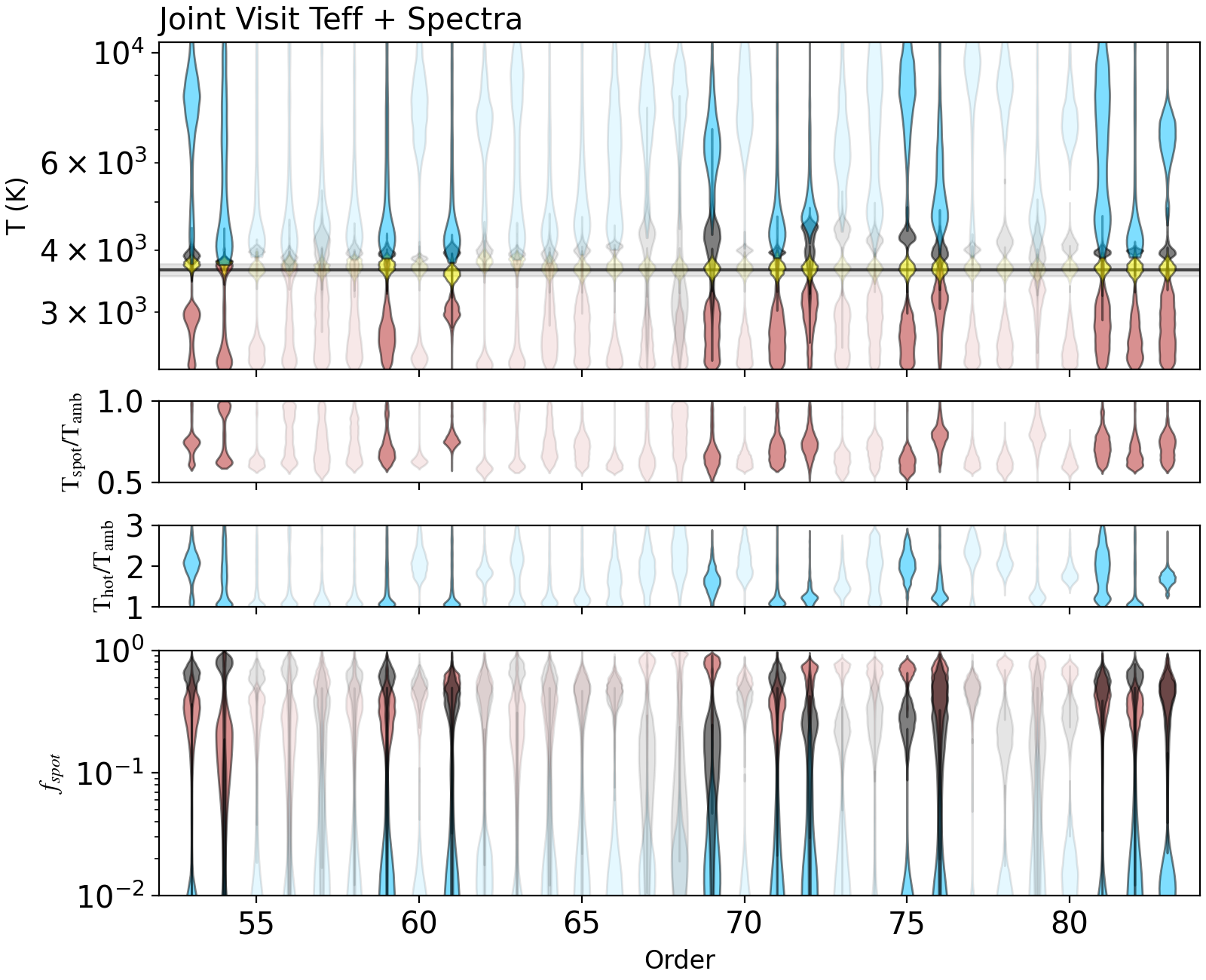}}
\caption{Combined-visit model violin plots showing the posterior distributions when modeling only the spectra and \Teff. Spectral orders shown in this plot span 0.557-0.888 $\mu$m, with wavelength decreasing to the right. Top: Temperature components for the hot (blue), middle (black) and spot (red) components. In yellow is the corresponding \Teff. Most orders exhibit a hot component between 7000-10000 K, consistent with the temperature of flares in this wavelength range. Spot temperatures are generally poorly constrained, with the 3000K spot seemingly detected in a handful of orders (53, 61, 72, 76, 79). Middle: the temperature ratio, \Tspot/\Tamb. The 5 orders showing a spot solution tend to show a temperature ratio of between 0.7-0.8. Bottom: coverage fractions for the spotted (red), ambient (black), and hot (blue) components. The hot component is very small, less than 3$\%$. There is no clear agreement with spot coverage between orders, with the largest component being the ambient photosphere in the early orders while the later orders show more spot-dominated photospheres. Orders with very poor constraints tend to overlap with orders that are contaminated by tellurics. Orders omitted from the analysis are grayed out.}
\label{f:combined_Teff_Spec_violin}
\end{figure*}

In the ensemble model described in the main text, we fit spotted spectral models to 18 orders simultaneously, but here we examine the results from modeling each order individually to check for consistency between orders and which orders might be the most informative.
In this part of the modeling, we model the data from both F21 and S22 with the same set of parameters, under the assumption that AU Mic's spot characteristics have not changed measurably between visits. 
Examining the results (Figure \ref{f:combined_Teff_Spec_violin}), we cut orders from the analysis if their solutions are poorly constrained and they satisfy any of the following criteria: there is telluric contamination, unconstrained ambient characteristics (which should be the dominant signal), or extremely cold spot temperatures (which is often coincident with heavy telluric contamination).
Spectral orders that satisfy these conditions are excluded if we suspect useful information cannot be gained from the spectral order, most frequently due to telluric contamination.
Many orders also exhibit what appears to be a hot component a few hundred Kelvin hotter than the ambient component, which could be evidence of a facular flux component.
Attempts to recover this component were unsuccessful, but perhaps worth future investigation.

\end{document}